\documentclass[
reprint,
superscriptaddress,
nofootinbib,
bibnotes,
amsmath,amssymb,amsthm,
aps,
prx,
floatfix,
]{revtex4-2}

\usepackage{graphicx}
\usepackage{dsfont, lipsum}
\usepackage{amsmath,amssymb,amsfonts,amsthm,mathtools}
\usepackage{enumitem}
\usepackage[table]{xcolor}
\usepackage{mdframed}
\usepackage{adjustbox}
\usepackage{placeins}
\usepackage{stackengine}
\usepackage{float}
\usepackage{bm}

\usepackage{booktabs}
\usepackage{array}
\newcolumntype{P}[1]{>{\centering\arraybackslash}p{#1}}
\newcolumntype{N}[1]{>{\footnotesize\arraybackslash}p{#1}}
\newcolumntype{L}[1]{>{\arraybackslash}p{#1}}

\usepackage{hyperref}
\hypersetup{
	colorlinks = true,
	linkcolor = blue,
	citecolor = blue,
	urlcolor = blue,
	filecolor = black,
	linktoc = all,
}

\usepackage[numbered,openlevel=1,open]{bookmark}

\makeatletter
\newcommand{\fmarki}{\ensuremath{\dagger}}
\newcommand{\fmarkii}{\ensuremath{\ddagger}}
\newcommand{\fmarkiii}{\ensuremath{\mathsection}}
\newcommand{\fmarkiv}{\ensuremath{\&}}
\newcommand{\fmarkv}{\ensuremath{\#}}
\newcommand{\fmarkvi}{\ensuremath{\|}}
\newcommand{\fmarkvii}{**}
\newcommand{\fmarkviii}{\ensuremath{\dagger\dagger}}
\newcommand{\fmarkix}{\ensuremath{\ddagger\ddagger}}         
\def\@fnsymbol#1{{\ifcase#1\or \fmarki\or \fmarkii\or \fmarkiii\or \fmarkiv\or \fmarkv\or \fmarkvi\or \fmarkvii\or \fmarkviii\or \fmarkix \else\@ctrerr\fi}}
\makeatother

\newcommand{\inst}[1]{{\mathcal{#1}}}

\newcommand{\vnf}[1]{{\bm{\mathsf{#1}}}}

\newcommand{\ket}[1]{|{#1}\rangle}
\newcommand{\bra}[1]{\langle{#1}|}
\newcommand{\ketbra}[1]{{\ket{#1} \! \bra{#1} }}
\newcommand{\braket}[1]{\langle #1 \rangle}

\newcommand{\tr}{\operatorname{Tr}}
\newcommand{\one}{\mathds{1}}
\newcommand{\rr}{ {\hspace{1pt} \| \hspace{1pt}} }

\newcommand{\rcg}{\rho_{{\rm cg}}}

\newcommand{\xn}{^{\otimes n}}

\DeclareMathOperator*{\argmax}{argmax}
\DeclareMathOperator*{\argmin}{argmin}

\newcommand{\lam}{\lambda}

\newcommand{\eps}{\epsilon}

\newcommand{\lamtherm}{\lambda_{\rm th}}
\newcommand{\deff}{d_{\rm eff}}

\newcommand{\tot}{{\rm tot}}
\newcommand{\Itot}{I_{\tot}}
\newcommand{\Imax}{I_{\max}}

\newcommand{\E}{\mathcal{E}}

\newcommand{\rhobar}{\overline{\rho}}
\newcommand{\Teq}{T_{\rm eq}}
\newcommand{\rcgt}{\tilde{\rho}_{\rm cg}}
\newcommand{\taut}{\tilde{\tau}}
\newcommand{\X}{\mathcal{X}}
\newcommand{\SMTR}{$S_{M}^{\tau}(\rho)$}
\newcommand{\supp}{\operatorname{supp}}
\newcommand{\chitau}{\chi_{\tau}}
\newcommand{\dchitau}{\partial \chi_{\tau}}

\newcounter{propnum}
\newcommand{\tpn}{\roman{propnum}{\stepcounter{propnum}}}

\newcommand{\AppClassical}{\hyperref[app:classical-hard-sphere-gas]{App.~\ref{app:classical-hard-sphere-gas}}}
\newcommand{\AppQuantum}{\hyperref[app:quantum-random-matrix-model]{App.~\ref{app:quantum-random-matrix-model}}}
\newcommand{\AppRE}{\hyperref[app:relative-entropies-review]{App.~\ref{app:relative-entropies-review}}}

\newcommand{\M}{M}

\newcommand{\MM}{\inst{M}}
\newcommand{\NN}{\inst{N}}

\newcommand{\bbM}{\mathbb{M}}

\newcommand{\cf}{{\textit{cf.}}}
\newcommand{\ie}{{\textit{i.e.}}}
\newcommand{\eg}{{\textit{e.g.}}}
\newcommand{\etal}{{\textit{et.~al.}}}

\theoremstyle{plain}
\newtheorem{thm}{Theorem}

\newtheorem{prop}[thm]{Proposition}
\theoremstyle{definition}

\makeatletter
\def\thmheadbrackets#1#2#3{%
  \thmname{#1}\thmnumber{\@ifnotempty{#1}{ }\@upn{#2}}%
  \thmnote{ {\the\thm@notefont{\bf [#3]}}}}
\makeatother

\newtheoremstyle{squarebrackets}
  {}
  {}
  {\normalfont}
  {}
  {\bfseries}
  {.}
  { }
  {\thmheadbrackets{#1}{#2}{#3}}

\theoremstyle{squarebrackets}
\newtheorem{proofy}{Proof}

\definecolor{bc}{cmyk}{0.05,0.05,0,0}
\definecolor{contents}{cmyk}{0.025,0.025,0,0}
\newmdenv[
    linewidth=0pt,
    backgroundcolor=bc,
    skipabove = 1pt,
    skipbelow = 3pt,
    innerleftmargin=3pt,
    innerrightmargin=2pt,
    innertopmargin=-3pt,
    innerbottommargin=2pt,
]{bbbox}

\newmdenv[
    linewidth=0pt,
    backgroundcolor=contents,
    skipabove = 1pt,
    skipbelow = 3pt,
    innerleftmargin=3pt,
    innerrightmargin=0pt,
    innertopmargin=4pt,
    innerbottommargin=1pt,
]{contentsbox}

\definecolor{proofline}{gray}{0.7}
\newcommand{\pfline}{{\color{proofline} \rule{.8\columnwidth}{.2pt}}}

\begin{document}

\title{Unification of observational entropy with maximum entropy principles}

\author{Joseph Schindler}
\email{josephc.schindler@uab.cat}
\affiliation{F\'{\i}sica Te\`{o}rica: Informaci\'{o} i Fen\`{o}mens Qu\`{a}ntics, Departament de F\'{\i}sica, Universitat Aut\`{o}noma de Barcelona, 08193 Bellaterra, Spain}

\author{Philipp Strasberg\ensuremath{^*}}
\thanks{strasberg at ifca dot unican dot es}
\affiliation{F\'{\i}sica Te\`{o}rica: Informaci\'{o} i Fen\`{o}mens Qu\`{a}ntics, Departament de F\'{\i}sica, Universitat Aut\`{o}noma de Barcelona, 08193 Bellaterra, Spain}
\affiliation{%
Instituto de F\'isica de Cantabria (IFCA), Universidad de Cantabria--CSIC, 39005 Santander, Spain}

\author{Niklas Galke}
\email{niklas.galke@uab.cat}
\affiliation{F\'{\i}sica Te\`{o}rica: Informaci\'{o} i Fen\`{o}mens Qu\`{a}ntics, Departament de F\'{\i}sica, Universitat Aut\`{o}noma de Barcelona, 08193 Bellaterra, Spain}

\author{Andreas Winter}
\email{andreas.winter@uab.cat}
\affiliation{F\'{\i}sica Te\`{o}rica: Informaci\'{o} i Fen\`{o}mens Qu\`{a}ntics, Departament de F\'{\i}sica, Universitat Aut\`{o}noma de Barcelona, 08193 Bellaterra, Spain}
\affiliation{ICREA, 
Passeig Lluís Companys, 23, 08010 Barcelona, Spain}
\affiliation{Institute for Advanced Study, Technische Universit\"at M\"unchen, Lichtenbergstra{\ss}e 2a, 85748 Garching, Germany}

\author{Michael G. Jabbour\ensuremath{^* \,}}
\email{mjabbour@telecom-sudparis.eu}
\thanks{\smallskip}
\affiliation{SAMOVAR, T\'el\'ecom SudParis, Institut Polytechnique de Paris, 91120 Palaiseau, France}
\affiliation{Centre for Quantum Information and Communication, \'Ecole polytechnique de Bruxelles, CP 165/59, Universit\'e libre de Bruxelles, 1050 Brussels, Belgium}

\date{\today}

\keywords{entropy, relative entropy, observational entropy, measured entropy, maximum entropy principle, Boltzmann entropy, Shannon entropy, second law, statistical mechanics, thermodynamics, information theory}


\begin{abstract}
We introduce a definition of coarse-grained entropy that unifies measurement-based (observational entropy) and max-entropy-based (Jaynes) approaches to coarse-graining, by identifying physical constraints with information theoretic priors. The definition is shown to include as special cases most other entropies of interest in physics. We then consider second laws, showing that the definition admits new entropy increase theorems and connections to thermodynamics. We survey mathematical properties of the definition, and show it resolves some pathologies of the traditional observational entropy in infinite dimensions. Finally, we study the dynamics of this entropy in a quantum random matrix model and a classical hard sphere gas. Together the results suggest that this generalized observational entropy can form the basis of a highly general approach to statistical mechanics.
\end{abstract}

\maketitle


\bookmark[dest=contents, level=1]{Contents}

\section{Introduction}
\label{sec:introduction}

One of the great challenges of statistical mechanics is to make sense of a diversity of successful yet seemingly incompatible approaches. This is evident from the widely different methods that are often applied to non-/equilibrium, open/isolated, pure/mixed, and classical/quantum systems, and ongoing debates about which types of ``entropy'' are fundamental to statistical thermodynamics. For a satisfactory and complete understanding of statistical physics, a central goal is the development of a single consistent framework, with a transparent information theoretic underpinning, in which these various methods can be united.

\begin{figure}[b]

\hypertarget{contents}{}

\footnotesize
\sf

\vspace*{10pt}
\footnoterule
\vspace*{-14pt}

\begin{contentsbox}

CONTENTS

\vspace{2pt}

\begin{tabular}{lr}

\begin{tabular}{r p{3.5cm}}
     \hyperref[sec:introduction]{\ref{sec:introduction}.} &
     \hyperref[sec:introduction]{Introduction \hfill \pageref{sec:introduction}}
     \\
     \hyperref[sec:informational]{\ref{sec:informational}.} &
     \hyperref[sec:informational]{Informational motivation \hfill \pageref{sec:informational}}
     \\
     \hyperref[sec:constraints]{\ref{sec:constraints}.} &
     \hyperref[sec:constraints]{Constraints and priors \hfill \pageref{sec:constraints}}
     \\
     \hyperref[sec:definition]{\ref{sec:definition}.} &
     \hyperref[sec:definition]{Definition \hfill \pageref{sec:definition}}
     \\
     \hyperref[sec:special-cases-and-limits]{\ref{sec:special-cases-and-limits}.} &
     \hyperref[sec:special-cases-and-limits]{Special cases and limits \hfill \pageref{sec:special-cases-and-limits}}
     \\
     \hyperref[sec:need-for-generalized-volumes]{\ref{sec:need-for-generalized-volumes}.} &
     \hyperref[sec:need-for-generalized-volumes]{The need for generalized \hfill \pageref{sec:need-for-generalized-volumes}}
     \\[-2pt]
     &
     \hyperref[sec:need-for-generalized-volumes]{ volumes \hfill}
     \\
     \hyperref[sec:second-laws]{\ref{sec:second-laws}.} &
     \hyperref[sec:second-laws]{Second laws \hfill \pageref{sec:second-laws}}
     \\
     \hyperref[sec:mathematical-properties]{\ref{sec:mathematical-properties}.} &
     \hyperref[sec:mathematical-properties]{Mathematical properties \hfill \pageref{sec:mathematical-properties}}
     \\
     \hyperref[sec:examples]{\ref{sec:examples}.} &
     \hyperref[sec:examples]{Physical examples \hfill \pageref{sec:examples}}
     \\
    \hyperref[sec:conclusion]{\ref{sec:conclusion}.} &
     \hyperref[sec:conclusion]{Conclusions \hfill \pageref{sec:conclusion}}
     \\
\end{tabular}

\begin{tabular}{r p{3.35cm}}
     \hyperref[app:classical-hard-sphere-gas]{\ref{app:classical-hard-sphere-gas}.} &
     \hyperref[app:classical-hard-sphere-gas]{Classical hard sphere \hfill \pageref{app:classical-hard-sphere-gas}}
     \\[-2pt]
     &
     \hyperref[app:classical-hard-sphere-gas]{gas \hfill}
     \\
     \hyperref[app:quantum-random-matrix-model]{\ref{app:quantum-random-matrix-model}.} &
     \hyperref[app:quantum-random-matrix-model]{Quantum random ma- \hfill \pageref{app:quantum-random-matrix-model}}
     \\[-2pt]
     &
     \hyperref[app:quantum-random-matrix-model]{trix model \hfill}
     \\
     \hyperref[app:relative-entropies-review]{\ref{app:relative-entropies-review}.} &
     \hyperref[app:relative-entropies-review]{Relative entropies \hfill \pageref{app:relative-entropies-review}}
     \\[-2pt]
     &
     \hyperref[app:relative-entropies-review]{review \hfill}
     \\
     \hyperref[app:proofs]{\ref{app:proofs}.} &
     \hyperref[app:proofs]{Proofs \hfill \pageref{app:proofs}}
     \\ 
     &
     \\ 
     &
     \\ &
\end{tabular}

\end{tabular}

\end{contentsbox}

\vspace{-12pt}

\flushleft{\rm \footnotesize \ensuremath{^*}MGJ and PS contributed equally to this work.}

\end{figure}

One way to approach this problem of unification is through the lens of coarse-graining, which plays a basic role in relating physical to informational quantities. Techniques based on coarse-graining are of widely accepted importance, and recently coarse-grained entropies have been of particular interest, especially since the re-invigoration of the topic by {\v S}afr{\'a}nek, Deutsch, and Aguirre~\cite{safranek2019a,safranek2019b,safranek2021brief,safranek2020classical}, who emphasized the physical relevance of coarse entropies and defined \textit{observational entropy} (OE) as a framework to analyze arbitrary coarse-grainings. Since then such entropies have been subject to increasingly systematic investigation, with studies illuminating both physical applications and information theoretic insights~\cite{schindler2020correlation,
faiez2020typical, 
strasberg2020first,riera2020finite,
hamazaki2022speed,strasberg2022book,stokes2022nonconjugate,modak2022anderson,zhou2022relations,zhou2022renyi,buscemi2022observational,sreeram2022chaos,safranek2022work,strasberg2023classicality,
safranek2020quantifying,sinha2023alpha,schindler2023continuity,safranek2023expectation,zhou2023dynamical,bai2023observational,
strasberg2024comparative,nagasawa2024generic,amadei2019unitarity,bonfill2023entropic}. Although of recent interest in its general form, OE can also be found in various forms throughout the literature~\cite{ehrenfest1912conceptual,vonNeumann1929proof,tolman1938book,vanKampen1954statistics,vonNeumann1955mathematical, wehrl1978general, PenroseRPP1979, LatoraBarangerPRL1999, GellMannHartlePRA2007, VanKampenBook2007, GemmerSteinigewegPRE2014, LeePRE2018}; the quantum version was first introduced by von Neumann~\cite{vonNeumann1929proof,vonNeumann1955mathematical}, who emphasized its role connecting micro- to macroscopic phenomena. 

Observational entropy provides a ``measurement-based'' approach to coarse-graining, where a state space is decomposed into different ``macrostates'' each corresponding to some possible observation. Meanwhile, another important approach to quantifying coarsely viewed systems is given by Jaynes' maximum entropy principle~\cite{jaynes1957informationI,jaynes1957informationII}, where states are estimated based on a limited set of given expectation values.

A hint at the necessity of combining these two approaches comes from the behavior of observational entropy in infinite-dimensional quantum systems. In such systems, the standard definition of OE is subject to several mathematical pathologies (to be described in more detail later), including discontinuity and unphysical infinities. Although these pathologies appear quite technical at first glance, attempting to repair them actually reveals subtler and more fundamental questions: How should the OE framework account for constraints and conservation laws, and how does prior information affect the entropy definition.

To resolve these issues leads one to a joint picture involving both measurement and max-entropy coarse-graining notions. The Jaynes' part determines a state~$\tau$, which derives from constraints on the system and acts as an informational prior. The OE part specifies a measurement $M$ and determines the information gained by performing this measurement on the system. In more traditional language, $\tau$ can be identified with an \textit{ensemble} (reflecting the a priori likelihood of microstates), while $M$ can be identified with the notion of \textit{macrostate} (some measurable, often coarse, property of the system), which in this framework are two separate concepts.

\begin{figure}[t]
    \centering
    \includegraphics[width=1\columnwidth, trim={10mm, 3mm, 3mm, 3mm}, clip]{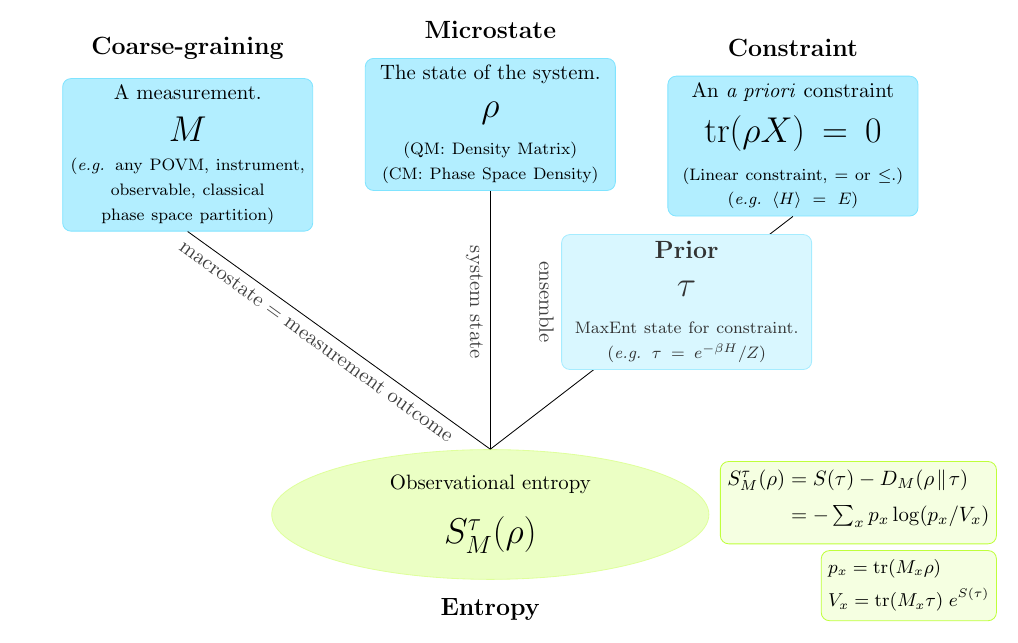}
    \caption{The (generalized) observational entropy $S_M^\tau(\rho)$ is a function of three arguments: a microstate $\rho$, a measurement~$M$, and a linear constraint encoded by a prior $\tau$. The definition unifies standard observational entropy  with maximum entropy principles, micro-/canonical entropies, and other entropies of statistical mechanics.}
    \label{fig:entropy-schematic}
\end{figure}

Based on this unified approach, we introduce in the present paper a new generalized definition of OE, the quantity $S_M^\tau(\rho)$ in Fig.~\ref{fig:entropy-schematic}. By unifying these two methods, one not only resolves the mathematical pathologies that formed the motivation, but also unites many diverse quantities within a single framework, obtains powerful and general results on entropy increase, and reveals numerous new connections in statistical mechanics.

\smallskip

The remainder of the paper is structured as follows:
We first complete the introduction with some notation and conventions. Before introducing the new generalized definition of observational entropy, in Sec.~\ref{sec:informational} we hint at the general form of the definition, and in Sec.~\ref{sec:constraints} we introduce a framework for relating physical constraints to information theoretic priors. We then give the main definition in Sec.~\ref{sec:definition} and examine its complementary physical and informational interpretations. In Sec.~\ref{sec:special-cases-and-limits} we see how the definition generalizes many other entropies in the literature by reduction to special cases. Turning to physical questions, in Sec.~\ref{sec:need-for-generalized-volumes} we consider generic entropy increase from a conceptual perspective, clarifying the role of effective macrostate volumes. We then tackle the question of second laws more precisely, in Sec.~\ref{sec:second-laws}, where we prove entropy increase theorems in various regimes and connect the results to thermodynamics. In Sec.~\ref{sec:mathematical-properties} we prove important mathematical properties of the new quantity, and resolve the infinite-dimensions pathologies of traditional observational entropy. Finally, in Sec.~\ref{sec:examples} we illustrate the theory by studying entropic dynamics in both quantum and classical examples. Concluding remarks appear in Sec.~\ref{sec:conclusion}. Apps.~\ref{app:classical-hard-sphere-gas}--\ref{app:quantum-random-matrix-model} provide further details of the physical example systems, and App.~\ref{app:relative-entropies-review} reviews properties of relative entropies. Proofs are given in App.~\ref{app:proofs}.

\medskip

\paragraph*{Notation and conventions.}

Traditionally, coarse-grainings are defined by a partition of phase space into subsets or a splitting of the Hilbert space in linear subspaces. However, from a modern perspective, a coarse-graining may be specified by any possible measurement. For quantum systems, the most general measurement is formalized as a quantum instrument, which is a collection of quantum operations whose sum is trace-preserving~\cite{nielsen2010book}. These are necessary when post-measurement states are involved. When only outcome probabilities are relevant, the most general measurement is a POVM $M$, defined by
\begin{equation}
    M = (M_x)_{x \in \X}, \qquad M_x \geq 0, \qquad \textstyle\sum_x M_x = \one,
\end{equation}
which is a collection of positive-semidefinite Hermitian operators summing to the identity. Here $x \in \X$ labels the set of possible outcomes, \ie~macrostates, and $p_x = \tr M_x \rho$ is the probability to obtain outcome $x$ from state $\rho$. 
(If not discrete, the sums become integrals and $p_x$ is a density.) 
A projective measurement is one where $M_x M_y = \delta_{xy}$, forming a set of orthogonal projectors, corresponding to the traditional ``partitioning'' definition. POVMs include measurements of Hermitian observables ($M_x$ project into eigenspaces) or measurements in a basis ($M_x = \ketbra{x}$) as special cases, but also describe cases where measurements are noisy or act only on a subsystem. We define measurement $M$ as a POVM unless otherwise stated.

A combined classical/quantum notation, where in the classical case $\tr \to \int \frac{d\Vec{x}^N d\Vec{p}^N}{(l_0 p_0)^{Nd}}$ and Hermitian operator $\to$ real phase space function, is used throughout the paper. The measure is usually fixed to $(l_0 p_0)^{Nd} = (2 \pi \hbar)^{Nd} N!$ by matching of classical/quantum limits, but can be rescaled contributing an additive constant to entropies. 
States $\rho$ may refer equally well to quantum density operators or classical phase space density distributions.

We also make use of standard informational entropies~\cite{cover2006book,wilde2011notes}. Most fundamental is the relative entropy~(RE), defined by \mbox{$D(\rho \rr \sigma) = \tr (\rho [\log \rho - \log \sigma])$} when acting on states, and $D(p \rr q) = \sum_x p_x \log (p_x/q_x)$ when acting on probability distributions.%
\footnote{
These formulae define RE when the support of $\rho$ is contained in the support of $\sigma$, or likewise for $p,q$. Otherwise, $D(\bullet \rr \bullet)=\infty$, in line with the conventions that $\log 0 = - \infty$ but $0 \log 0 = 0$. 
For further review of relative entropies, see App.~\ref{app:relative-entropies-review}.
}
Measured relative entropy, denoted
\begin{equation}
    D_M(\rho \rr \sigma) = D(p^\rho \rr p^\sigma),
\end{equation}
where $p^{\rho}_x = \tr M_x \rho$  and $p^{\sigma}_x  = \tr M_x \sigma$, is the RE between probabilities induced on $\rho, \sigma$ by a measurement~$M$. The Shannon and von Neumann entropies are denoted by $H(p) = -\sum_x p_x \log p_x$ and $S(\rho) = -\tr \rho \log \rho$.

A standard but less common informational quantity is the \textit{cross entropy} defined by
\begin{equation}
\label{eqn:cross-def}
    S(\rho; \sigma) = - \tr \rho \log \sigma
\end{equation}
for states or by $S(p;q)=-\sum_x p_x \log q_x$ for probabilities. In classical information theory, $S(p;q)$ is the number of bits needed to encode outcomes drawn from $p$ in a code optimized for $q$~\cite{cover2006book}. This relates to the minimal bits required to encode $p$ in any encoding, and the extra bits needed due to non-optimal encoding, by $S(p;q) = H(p) + D(p \rr q)$. The informational form for constraints used later takes the form $S(p;q) \leq S(q)$, which specifies the set of $p$ that are encodable at least as easily as $q$ in a $q$-optimized code.

\section{Informational Motivation}
\label{sec:informational}

We start with a simple rewriting of conventional entropies as ``entropy~+~information $=$ total information'', or
\begin{equation}
\label{eqn:missing-info}
    S = \Itot - I.
    \vspace{-2pt}
\end{equation}
For the Shannon entropy $H(p) = - \sum_x p_x \log p_x$, the von~Neumann entropy $S(\rho) = -\tr \rho \log \rho$, and the conventional observational entropy $S_M(\rho)$ (defined below),
this relation manifests as (with $N$ a number of possible outcomes and $d$ a Hilbert space dimension)
\begin{equation}
\label{eqn:missing-info-list}
\begin{split}
        H(p) &= \log N - D(p \rr 1/N), \\ 
        S(\rho) &= \log d - D(\rho \rr \one/d),  \\ 
        S_M(\rho) &= \log d - D_M(\rho \rr \one/d).
\end{split}
\end{equation}
That is, entropy measures missing~information.

In the Shannon entropy case, the information term \mbox{$I = D(p \rr 1/N)$} relates to the number of bits needed to encode outcomes drawn from $p$~\cite{shannon1948mathematical,cover2006book}. In particular, it is the informational value (in bits saved) of knowing $p$, relative to the prior guess that all outcomes are equally likely. This implicitly assumes prior ignorance about the distribution; if instead one had an initial estimate $q$, the information gained by obtaining $p$ would be $D(p \rr q)$.

The case of $S_M(\rho)$ contains a similar assumption about prior ignorance in the identification $I = D_M(\rho \rr \one/d)$. This is the number of classical bits gained from measurement~$M$, given the prior assumption that the state is maximally mixed---or in other words, the  assumption that all possible microstates are a priori equally likely. Given a different prior estimate $\tau$ of the state, the observational information would instead be $I = D_M(\rho \rr \tau)$. 

A natural candidate to extend the ``missing information'' interpretation of OE given the possibility of non-trivial prior information would thus take the form
\begin{equation}
\label{eqn:foreshadow-definition}
    S = \Imax - D_M(\rho \rr \tau),
\end{equation}
with $\tau$ being a prior estimate of the state. This would make sense if some $\Imax$ could be identified with a consistent interpretation, and if $S$ could be understood as the remaining uncertainty in the state given both prior and measured information.

One physically relevant type of prior information is the knowledge of constraints obeyed by the system. Given such knowledge, a natural prior guess $\tau$ of the state would be the Jaynes maximum entropy state for the constraints. In this situation, $\Imax$ can be identified with $S(\tau)$ leading to a consistent interpretation of the entropy.

\section{Constraints and priors}
\label{sec:constraints}

The idea is to define entropy in a way that accounts for both prior and measured information. The former comes into the definition as a state $\tau$ that represents an initial estimate, or ``Bayesian prior'', for the system state. 

The type of prior information we consider is that associated with linear constraints on a system, with $\tau$ arising from the maximum entropy principle. Basic examples of constraints include energy conservation $\braket{H} = E$ or energy bounds $\braket{H} \leq E$ (for some Hamiltonian $H$), or restriction of the system to a spatial region $\tr \Pi_X \rho = 1$ or energy shell $\tr \Pi_E \rho = 1$. Constraints may be stationary or time-dependent, and may be either equalities or inequalities. This can also include less elementary situations, such as the inclusion of multiple non-commuting charges, or constraints that apply only to the system or bath in a thermodynamic setup.

To bring such constraints into the context of entropies, we first introduce a setup that clarifies the informational properties of linearly constrained sets.

Table~\ref{tab:constraints-and-priors} collects examples of common priors $\tau$ and the constraints from which they arise. 

\subsubsection{Maximum entropy principle}

We consider arbitrary linear constraints on a system, defined by $\tr\rho X = 0$ (or $\tr\rho X \leq 0$) with $X$ a Hermitian operator. Jaynes' maximum entropy (MaxEnt) principle~\cite{jaynes1957informationI,jaynes1957informationII} states that $\tau = \argmax_{\tr \! \rho X \leq 0} S(\rho)$ is the most unbiased prior estimate of the state given the constraint, and that for sufficiently well-behaved $X$
\begin{equation}
    \tau = \argmax_{\tr \rho X = 0} \; S(\rho) = \frac{e^{-\lambda X}}{Z},
\end{equation}
where $Z=\tr e^{-\lambda X}$ and $\lambda$ is such that $\tr\tau X=0$. The meaning of well-behaved is effectively that some state of the desired form exists satisfying the constraint.

\begin{table*}[t]
    \centerline{
    \begin{tabular}{P{25mm}P{28mm}P{30mm}P{35mm}P{32mm}N{45mm}}
    \toprule \addlinespace[2pt]
    {\hypertarget{table1}{}}
          & \textsc{Prior} & \textsc{Constraint} 
         & \textsc{Volumes} & \textsc{Prior Entropy} & \centerline{\small \textsc{Notes}}
    \\[-8pt] \midrule \addlinespace[10pt]
         general 
         & $\tau$
         & $S(\rho;\tau) \leq  S(\tau)$ \linebreak $S(\rho;\tau) =  S(\tau)$
         & $ V_x = \tr M_x \tau \; e^{S(\tau)}$ 
         & $S(\tau) = -\tr \tau \log \tau$ 
         & $\deff = e^{S(\tau)}$.
    \\[20pt]
         uniform \linebreak (finite dims)
         & $\tau = \one/d$
         & $\tr \rho \leq 1$ (trivial) \linebreak $\tr \rho = 1$ (trivial) 
         & $\tr M_x$ 
         & $\log d$ 
         & Traditional OE $S_M(\rho)$. 
    \\[24pt]
         canonical \linebreak (energy)
         & $\tau = e^{-\beta H}/Z$
         & $\braket{H}\leq E$ \linebreak $\braket{H} = E$ 
         & $ \tr M_x \, e^{-\beta(H-E)}$ 
         & $\beta E + \log Z$ 
         & Average energy conservation. \hfill \linebreak $\braket{H} = \tr \rho H$,  $E = \tr \tau H$. 
    \\[16pt]
         canonical \linebreak (charges $Q_k$)
         & $\tau = e^{-\sum_k \lambda_k Q_k} /Z$
         & $\sum_k \lambda_k [\braket{Q_k} - c_k] \leq  0$ \linebreak $\sum_k \lambda_k [\braket{Q_k} - c_k] =  0$ 
         & {\footnotesize $\tr M_x \, e^{-\sum_k \lambda_k (Q_k-c_k) }$}
         & {\footnotesize $\sum_k \lambda_k c_k + \log Z$} 
         & Multiple non-commuting charges.  \hfill \linebreak $\braket{Q_k} = \tr \rho Q_k$,  $c_k = \tr \tau Q_k$. 
    \\[24pt]
         microcanonical \linebreak
         (energy or arbitrary)
         & $\tau = \Pi/W$
         & $\rho \in \Pi$ \linebreak $\rho \in \Pi$ 
         & $\tr M_x \Pi$ 
         & $\log W$ 
         & $\Pi$ a projector onto any subspace. $\rho \in \Pi$ is support containment. $V_x$~are volumes within subspace. 
    \\[12pt]
         time-averaged
         & $\tau = \rhobar$
         & $S(\rho_t;\rhobar) \leq S(\rhobar)$  \linebreak $S(\rho_t;\rhobar) = S(\rhobar)$ 
         & $\overline{p_x(t)} \; \deff$ 
         & $-\tr \rhobar \log \rhobar$ 
         & $\overline{\bullet} = \lim_{T \to \infty} \int_{-T/2}^{T/2} (\bullet/T) \, dt$. \hfill \linebreak Tightest time-independent prior.
    \\ \addlinespace[6pt]
    \bottomrule
    \end{tabular}
    }
    \caption{Constraints and priors.
    }
    \label{tab:constraints-and-priors}
\end{table*}

\subsubsection{An informational form for linear constraints}

We now introduce a useful form for writing linear constraints, based on the simple observation that any (sufficiently well-behaved, as above) constraint can be rewritten directly in terms of its associated prior.

The key observation is that the constraint $\tr \rho X \leq 0$ is equivalent to the cross entropy relation
\begin{equation}
\label{eqn:cross-ent-constraint}
    S(\rho;\tau) \leq S(\tau).
\end{equation}
Likewise, $\tr \rho X = 0$  is equivalent to $ S(\rho;\tau) = S(\tau)$. This can be called the informational form of the constraint.

The cross entropy $S(\rho ;\tau)=-\tr \rho \log \tau$ would be, in classical information theory, the number of bits needed to encode outcomes drawn from $\rho$ in a code optimized for $\tau$ (see \eqref{eqn:cross-def} and surrounding text). Borrowing this language provides a nice interpretation: the states obeying the constraint are those that are somehow ``efficiently encodable'' with respect to $\tau$. 

Any state $\tau$ gives rise in this way to a linear constraint $S(\rho ; \tau) \leq S(\tau)$, with $\tau$ automatically the MaxEnt state on the constrained set. This must be the case since for any~$\rho,\sigma$ it is true that $S(\rho) \leq S(\rho;\sigma)$. This provides a general identification between states and constraints, and it is convenient to introduce the notations \mbox{$\chitau = \{ \rho \, | \, S(\rho;\tau) \leq  S(\tau) \}$} and \mbox{$\dchitau = \{ \rho \, | \, S(\rho;\tau) = S(\tau) \}$} for the sets of states obeying the (in-/equality) constraints associated with $\tau$.

When constraints are written in this form, the elementary relation $S(\rho ;\tau) = S(\rho) + D(\rho \rr \tau)$ can be used to obtain certain properties of constrained sets that will be useful below. One is that, as already seen, $\tau$ is the \mbox{MaxEnt} state among the set $\chitau$. Another is that any $\rho \in \chitau$ obeys $D(\rho \rr \tau) \leq S(\tau) - S(\rho)$, signifying no state obeying the constraint is ``very far'' from~$\tau$. Additionally, if $\rho \in \dchitau$ then~$D(\rho \rr \tau) = S(\tau) - S(\rho)$. 

These identities help analyze the suitability of the identification $S(\tau)=\Imax$ in the earlier \eqref{eqn:foreshadow-definition}. Suppose one knows that $\rho$ obeys the constraint associated with~$\tau$, meaning $\rho \in \chitau$. Then first of all, $S(\tau)$ is by definition the prior entropy, the missing information given only prior knowledge. Second, for quantum systems, $S(\tau)$~is the maximum information extractable given the constraint, since (Proof~\ref{pf:Imax} in App.~\ref{app:proofs})
\begin{equation}
\label{eqn:Imax}
    S(\tau)  = \max_{M,\rho} D_M(\rho \rr \tau) = \max_{M,\rho} \; I,
\end{equation}
with the maximum taken over $\rho$ obeying the constraint (the latter equality is just the definition of $I$ above). And third, $S(\tau) = \max_\rho S(\rho)$ is the maximum entropy of any state obeying the constraint. Together, these show $S(\tau)=\Imax$ makes sense from several perspectives.

We remark that in the classical case, \eqref{eqn:Imax} fails because of the possibility that $S(\rho) < 0$. This occurs when classical $\rho$ is concentrated in regions smaller than the scale set by the arbitrary integration measure $l_0 p_0$ (see notation and conventions). Classically $\rho$ may be arbitrarily concentrated, but if the distribution actually arises from underlying quantum physics then it will not concentrate beyond some limit (for one way to analyze this, see~\cite{wehrl1979relation}). Then, so long as $l_0 p_0$ reflects this limit, $S(\rho) \geq 0$ and as a consequence \eqref{eqn:Imax} can be restored.

Multiple constraints can also be treated, as illustrated in Table~\ref{tab:constraints-and-priors} for charges $Q_k$. The MaxEnt $\tau$ for the joint set of constraints is a prior whose $\chitau$ contains, but is larger than, the original constraint set.

\subsubsection{Physical remarks}

We have described a general picture where any state~$\tau$ is associated with the constraint $S(\rho;\tau) \leq S(\tau)$, for which $\tau$ is automatically the \mbox{MaxEnt} prior. A given $\tau$ is a valid prior for the system, and can be used to calculate entropies, if $\rho$ obeys this constraint.

For a given physical system, many different priors may be valid in that the relevant constraint holds. Any of these can be used to calculate entropies. However, only for sufficiently ``good'' priors will second laws hold and entropy maximization be expected. Among these, a special role is played by the time-averaged state $\rhobar$, which in isolated systems is (as shown later) the tightest time-independent prior.

Later we will find that the ``good'' priors (for which second laws hold) are those $\tau$ that look like~$\rhobar$ when viewed by coarse measurements. In the non-integrable examples we will consider, the canonical $\tau = e^{-\beta H}/Z$ is a good prior in this sense. 

Either time-dependent or stationary constraints can be relevant. If $\rho$ and $\tau$ both evolve unitarily then any constraints are conserved, since $S(U \rho U^\dag; U \tau U^\dag) = S(\rho;\tau)$ and $S(U \tau U^\dag) = S(\tau)$, for any unitary $U$. On the other hand, sometimes \textit{explicitly} time-dependent constraints, where $\tau(t)$ does not involve simple unitary evolution, are also of interest, for example when $\tau(t)$ tracks the time-dependent energy expectation value of a heat bath.

Having now established the connection between constraints, priors, and missing information, we will next turn to defining the entropy.

\bookmark[dest=table1]{Table I}

\section{Definition}
\label{sec:definition}

Given a constraint for which $\tau$ is the MaxEnt prior, we define OE by analogy with~(\ref{eqn:missing-info}--\ref{eqn:foreshadow-definition}) (and justified by Secs.~\ref{sec:informational}--\ref{sec:constraints}) in the most obvious possible way:
\begin{equation}
    S_M^\tau(\rho)  = S(\tau) - D_M(\rho \rr \tau).
\end{equation}
The first term $\Imax = S(\tau)$ is the prior entropy, the missing information after the constraint is taken into account. Meanwhile, $D_M(\rho \rr \tau)$ is the informational value of measuring $M$~on~$\rho$, given the prior assumption~$\tau$. Together, this quantifies both the baseline uncertainty given $\tau$, and the distinguishability of $\rho$ from $\tau$ as viewed by $M$.

The definition was introduced in the informational form, but it also has a natural interpretation in terms of macrostate volumes, as
\begin{equation}
\label{eqn:oe-volume-form}
    S_M^\tau(\rho) = -\sum_x p_x \log \frac{p_x}{V_x},
\end{equation}
where $p_x = \tr\rho M_x$ and $V_x = \tr (\tau M_x ) \, e^{S(\tau)}$. These effective macrostate volumes can be understood as
\begin{equation}
\label{eqn:effective-volumes}
    V_x = q_x \, \deff,
\end{equation}
where $q_x = \tr M_x \tau $ are prior macrostate probabilities, and \mbox{$\deff = e^{S(\tau)}$} is the effective dimension of the space of states obeying the constraint (the dimension of a uniform distribution with equal entropy to $\tau$). This can be viewed as a refinement of the ``equal \textit{a priori} probability'' assumption of statistical mechanics. 

The above form captures OE as the average ($-\log$) probability-to-volume ratio: a large probability to be in a small macrostate is highly informative. Expanding to
\begin{equation}
\label{eqn:shan-boltz}
    S_M^\tau(\rho) = H(p) + \sum_x p_x \log V_x
\end{equation}
shows it is also the Shannon entropy over macrostates plus the mean Boltzmann entropy. This incorporates both ``which-macrostate'' uncertainty and ``which-microstate-given-macrostate'' uncertainty, interpolating Gibbs and Boltzmann entropy concepts.

The definition can be compared to the traditional OE, which we denote and define by
\begin{equation}
\label{eqn:traditional-oe}
     S_M(\rho) = -\sum_x p_x \log \frac{p_x}{W_x},
\end{equation}
where $W_x = \tr M_x$. This arises as the case $\tau \propto \one$ in finite dimensions, giving the trivial constraint $\tr \rho \leq 1$. We abuse notation by calling this case $\tau \propto \one$ also in infinite dimensions, although in that case the definitions technically are inequivalent.

To illustrate the definition we can consider two of the most common constraints. The ``canonical'' energy constraint $\braket{H}=E$ gives prior $\tau = e^{-\beta H} /Z$, and thus
\begin{equation}
\label{eqn:simple-canonical}
    S_M^\tau(\rho) = \beta E + \log Z - D_M(\rho \rr e^{-\beta H}/Z).
\end{equation}
The ``microcanonical'' constraint that $\rho$ has support only within an energy shell, gives prior $\tau= \Pi_E/W_E$ (with $\Pi_E =$ projector onto energy shell, $W_E=\tr\Pi_E$), and so
\begin{equation}
\label{eqn:simple-microcanonical}
    S_M^\tau(\rho) = \log W_E - D_M(\rho \rr \Pi_E/W_E).
\end{equation}
In this way, textbook equilibrium entropies become maxima dynamically approached from non-equilibrium.

Based on these elementary definitions, we will now see that a wide variety of commonly used physical and informational entropies arise as limits.

\section{Special cases and limits}
\label{sec:special-cases-and-limits}

Numerous entropies are used throughout the statistical mechanics literature for complementary (and/or overlapping) purposes. OE provides a simple framework in which many of these can be unified. Interestingly, even those sometimes seen as incompatible (such as Gibbs/Boltzmann~\cite{jaynes1965gibbs, goldstein2020gibbs} or surface/volume~\cite{SwendsenPRR2018} entropies), can be regarded simply as choices of $M,\tau$ or as different limits.

Here we survey some of the essential special cases. These are also summarized in Table~\ref{tab:special-cases-and-limits}, at the end of the Appendix. 

\begin{enumerate}

    \item Jaynes max entropy~\cite{jaynes1957informationI,jaynes1957informationII} arises as the case of no measurement, meaning trivial $M=(\one)$, which gives von Neumann entropy $S(\tau)$ where $\tau$ is the MaxEnt state for the constraint. It is also the upper bound~(Proof~\ref{pf:bounds})
    \begin{equation}
    \label{eqn:upper-bound}
        S(\tau) \geq S_M^\tau(\rho).
    \end{equation}
    In many cases this upper bound will be the equilibrium entropy. The bound is saturated whenever $M$ cannot distinguish $\rho$ from $\tau$ (all $\tr\rho M_x = \tr\tau M_x$), meaning that the measurement provides no additional information beyond the prior.

    \item The traditional OE~\cite{safranek2021brief} $S_M(\rho)$ is the case of $\tau \propto \one$, which is when no constraint is present. Notably, this case includes coarse-grained entropies used to describe the second law of thermodynamics by Wehrl~\cite{wehrl1978general} and von Neumann~\cite{vonNeumann2018mathematical}.

    \item Boltzmann entropies $\log V_x$ are the OE for systems with a definite macrostate (only one nonzero $p_x$). This in itself is still a generalization, as normally Boltzmann volumes would be computed from a uniform (or at best microcanonical) prior, \ie~the case $\log W_x$ with $W_x=\mbox{tr}M_x$ derived from $\tau \propto \one$. Allowing $V_x$ to account for general constraints has important consequences in both classical and quantum systems (see below).

    \item Shannon entropy of measurement outcomes~\cite{meier2024emergence} arises, up to a constant, as the case where all macrostate volumes $V_x$ are equal (equal prior probability of macrostates).

    \item von Neumann (or classical Gibbs) entropy of the microstate,  $S(\rho)$, is the lower bound (Proof~\ref{pf:bounds})
    \begin{equation}
    \label{eqn:lower-bound}
        S_M^\tau(\rho) \geq S(\rho).
    \end{equation}
    This encapsulates the fundamental statement: \textit{observations can extract no more information than the amount of information inherently available in the state itself}. For the case of no prior constraints, this is captured by $\min_M S_M(\rho) = S(\rho)$, limiting the information extractable by measurements alone. More broadly, the lower bound shows $S(\rho)$ limits the information available to extract even by a combination of constraints and measurements. The bound is not always tight: penalties occur for non-commutativity of $\rho,\tau$ and for loose constraints. Note the bound assumes the constraint is true---if not, one can seem to extract ``too much'' information, but some of that information is false.
    
\end{enumerate}

Those are the broadest cases. One can also observe reductions to more particular cases of interest. 

\begin{enumerate}[resume]

    \item Diagonal entropy~\cite{polkovnikov2011microscopic,giraud2016average,oliveira2024thermodynamic} occurs in two different ways. First, as the case $M=(\ketbra{E})_E$ of a measurement in the energy eigenbasis, with $\tau \propto \one$. Second, assuming nondegenerate energies it arises as $S(\rhobar)$, the von Neumann entropy of the time-averaged state. From the second perspective, diagonal entropy is an equilibrium entropy associated with the tightest equilibrium constraint $\tau=\rhobar$. This view aligns with the fact that it is constant in isolated systems.
    
    \item Entanglement entropy~\cite{schindler2020correlation,eisert2010area} is the minimum OE for local measurements on entangled subsystems. That is, $S_{\rm ent}(\ket{\psi_{AB}}) = \inf\limits_{ M_A, M_B} S_{M_A \otimes M_B}(\ket{\psi_{AB}})$, which equals $S(\rho_A)$ of the reduced state. This infimum can be compared to the von Neumann entropy $S(\rho) = \inf_M S_M(\rho)$, the minimum over global (non-local) measurements. For mixed or multipartite $\rho$, this local-$M$ infimum generalizes to a quantifier of quantum correlations~\cite{schindler2020correlation,rossetti2025observational}. 

    \item Canonical entropy~\cite{matty2017comparison,huang1987statistical}, an important special case of Jaynes' entropy above, arises from prior $\tau =  e^{-\beta H}/Z$, corresponding to a constraint based on the average energy. For a heat exchange scenario, this entropy naturally leads to formulations of Clausius' inequality by adapting the prior $\tau(t)$ in time to match the changing energy (as in \eqref{eqn:EP-clausius} below)~\cite{strasberg2020first, strasberg2021clausius, strasberg2024typical}. This is an example where genuinely time-dependent constraints are relevant.

    \item The stochastic thermodynamic \mbox{entropy~\cite{seifert2018stochastic,strasberg2022book,PelitiPigolottiBook2021,ShiraishiBook2023}} arises, in a weakly coupled system/bath setting, from a global microcanonical prior $\tau = \Pi_E/W_E$ (constraint to an energy shell of total system plus bath), with a projective $M=(\Pi_x \otimes \one_B)_x$ acting only on the system. This gives an entropy $S^\tau_M(\rho) = S(\tau) - S(\tau_S) + S^{\tau_S}_{M_S}(\rho_S)$, with $\tau_S,\rho_S$ reduced states in the system. For a large weakly coupled bath, under standard assumptions~\cite{goldstein2005canonical,popescu2006entanglement}, one has approximately that $\tau_S = e^{-\beta H_S}/Z$ is canonical. To make the connection to stochastic thermodynamics, define the conditional (``meso'') state $\tau_x = \frac{\Pi_x \tau_S \Pi_x}{q_x}$, with $q_x = \tr \Pi_x \tau_S$. Further let $E_S = \tr \tau_S H_S$, $E_x = \tr \tau_x H_S$,  and \mbox{$S_x = S(\tau_x)$} be the equilibrium energy, mesostate energy, and intrinsic entropy of mesostates, respectively.  Meanwhile, let $V_x = \tr (\Pi_x \tau_S) \, e^{S(\tau_S)}$ be the macrostate volumes for system OE. Interestingly, one finds $S(\tau_x) = \log V_x - \beta (E_S-E_x)$ (but see the caveat below for noncommutative case). For a state $\rho$ with probabilities $p_x = \tr \Pi_x \rho_S$, therefore, one has $S^{\tau_S}_{M_S}(\rho_S) = \sum_x p_x [ - \! \log p_x \! + S_x \! + \! \beta (E_S - E_x)]$. This means changes in the total OE are given by
    \begin{equation}
        \Delta S^\tau_M(\rho) = -\beta \Delta \braket{E_x} + \Delta \braket{S_x - \log p_x}
        \hspace*{-14pt}
    \end{equation}
    (brackets average over $p_x$), which is equivalent to stochastic thermodynamic entropy production, \eg, as in (11) of \cite{seifert2018stochastic}. Note that the above expression for $E_x$, consistent with \cite{seifert2018stochastic}, actually assumes the classical/commuting case; for quantum systems, it must be replaced by $E_x = - \beta^{-1} \tr [ \tau_x \log ( \Pi_x e^{-\beta H_S} \Pi_x)]$, which reduces in the commuting limit. For further connections with stochastic thermodynamics see \eqref{eqn:rate-matrix} below.
    
    \item Surface and volume entropies, whose incompatibilities have been the subject of some debate~\cite{SwendsenPRR2018}, here correspond simply to the choice of two different priors, $\tau \propto \Pi_{E \approx E_0}$ (projector on energy shell at~$E_0$) and $\tau \propto \Pi_{E\leq E_0}$ (projector on all energies up to~$E_0$), respectively.

    \item Wehrl entropy~\cite{wehrl1979relation,lieb1978proof} is the case of the POVM $M = \big(\frac{\ketbra{z}}{\pi}\big)_z$, where $\ket{z}$ are the over-complete basis of coherent states, with $\tau \propto \one$.

    \item A related generalization of OE  was proposed in~\cite{bai2023observational}, showing another way to include prior information in the definition. That definition is equivalent to $S = S(\rho;\tau) - D_M(\rho \rr \tau)$, coinciding with $S_M^\tau(\rho)$ if the constraint $S(\rho;\tau)=S(\tau)$ is an exact equality.

    \item Free energies~\cite{huang1987statistical} can arise in a number of ways. One is for a weakly coupled system/bath with global $\tau = e^{-\beta H_{SB}}/Z$ (total average energy constraint) and $M=M_S \otimes \one_B$ acting on the system. In the weak coupling limit, the OE becomes $S_M^\tau(\rho) \approx S_{M_S}^{\tau_S}(\rho_S) - \beta E_S + C$ (constants independent of $\rho$). Another is that for any system with canonical $\tau = e^{-\beta H}/Z$, one can write the OE as $S_M^\tau(\rho) = \beta(E-\braket{F_x})$ (bracket averages over $p_x$), with $\beta F_x = - \log Z_x + \log p_x$. Here, $Z_x = \tr(\M_x e^{-\beta H})$ and $E = \tr \tau H$. In either case, OE is maximized when free energy is minimized.

    \item R{\'e}nyi, Tsallis, and related entropies, give generalized formulae for computing informational entropies while relaxing some of Shannon's assumptions~\cite{shannon1948mathematical,renyi1961measures,rippchen2024locally,zhou2022relations,sinha2023alpha}. To make the corresponding OE generalizations, a useful prescription is to replace $D_M$ by the corresponding measured divergence. For the R{\'e}nyi case this leads to the definition $S_{M,\alpha}^\tau(\rho) = S(\tau) - D_M^\alpha(\rho \rr \tau)$. Note the lack of $\alpha$ on $S(\tau)$. In this form, OE continues to measure moments of the probability-to-volume ratio, as $S_{M,\alpha}^\tau(\rho) =-\log \big\langle \big(\frac{p_x}{V_x}\big)^{\! s}\big\rangle_{p_x}^{1/s}$, where $\alpha=1+s$. This should be compared to \eqref{eqn:oe-volume-form}. This form can be useful in proofs of OE properties, in analogy with proofs like \cite{lieb1978proof} for example.
    
    \item The coarse- and fine-grained entropies sometimes used in high energy physics~\cite{polchinski2017blackhole,almheiri2020entropy} are the Jaynes entropy $S(\tau)$ and von Neumann entropy $S(\rho)$ that were discussed earlier.

    \item The term ``thermodynamic entropy'' tends to have different meanings in different contexts, usually  associated with relations like $dS = dQ /T$. Such relations can be obtained from various microscopic entropy definitions, including many discussed in this section. This often occurs for energy constraints/measurements, as in (\ref{eqn:EP-clausius}) and (\ref{eqn:thermo}) below.
    
\end{enumerate}

The above are mainly instances of entropy at a single time. Another important class of methods are those describing entropy production (EP), the creation of entropy during the course of a process~\cite{potts2019introduction,strasberg2022book,bai2024fully}. While this includes a large class of methods whose full analysis is beyond the present scope, we remark that in a number of cases entropy production quantities may be reinterpreted as
\begin{equation}
\label{eqn:EP-general}
    \Delta S_M^{\tau} = S_{M_t}^{\tau_t}\big(\rho(t)\big) - S_{M_0}^{\tau_0}\big(\rho(0)\big),
\end{equation}
for some time-dependent measurements and constraints. The connection may at first appear obscure, because EP is often written with a single relative entropy term. However, this typically involves some state $\tau_t$ defined by a constraint equivalent to the equality $S(\rho(t); \tau_t) = S(\tau_t)$. The constraint equation can then be used to turn the RE into a difference of observational entropies. We give one such example here.

\begin{enumerate}[resume]

    \item Entropy production for a system plus bath setup arises in a standard form from $M(t) = M_S(t) \otimes \one_B$ with prior $\tau(t) = \one_S/d_S \otimes \tau_B(t)$, where $M_S(t)$ are optimal measurements on the system at each time and $\tau_B(t) = e^{-\beta(t) H_B}/Z_B$ is a canonical state of the bath, whose temperature $\beta(t)$ is determined by $\tr H_B \rho(t) = \tr H_B \tau(t)$. In other words, optimal measurements are able to be performed on the system, and the constraint is given by a known bath energy expectation $\braket{H_B} = E_B(t)$ at each time. This yields as the entropy $S_M^\tau(\rho) = S(\rho_S) + S(\tau_B)$, where all the quantities are time dependent. With $T_B = \beta^{-1}$ and using $\frac{d}{dt}S(\tau_B)  = \beta(t) \frac{d}{dt} E_B(t)$, this leads to entropy production
    \begin{equation}
    \label{eqn:EP-clausius}
        \Delta S_M^{\tau} = \Delta S(\rho_S) + \int_{0}^{t} \frac{dE_B(t')}{T_B(t')} \, ,
    \end{equation}
    where $\Delta S(\rho_S) = S(\rho_S(t)) - S(\rho_S(0))$ is the change in von Neumann entropy of the system. To relate this to the relative entropy form common in EP literature, define the state $\gamma(t) = \rho_S(t) \otimes \tau_B(t)$. Note that it is true that the constraint equation $S(\rho(t); \gamma(t)) = S(\gamma(t))$ holds (this is tighter than the earlier constraint which defined~$\tau$). Therefore, one has $D(\rho(t) \rr \gamma(t)) = S(\gamma(t)) - S(\rho(t)) = S(\gamma(t)) - S(\rho(0))$, with the first equality from the cross entropy equation and the second because $\rho(t)$ evolves unitarily. If one further assumes that at the initial time the system is decorrelated from a perfectly thermal bath, meaning $\rho(0) = \rho_S(0) \otimes \tau_B(0)$, then it follows that $\Delta S_M^{\tau} = D(\rho(t) \rr \gamma(t)) \geq 0$. This can be straightforwardly extended to include several baths and/or grand canonical contributions, and can be compared to EP as it is developed in references such as~\cite{potts2019introduction,strasberg2021clausius,strasberg2020first}. 
\end{enumerate}

Many of the cases seen above make use of either \mbox{$M$ or $\tau$} nontrivially, but not both simultaneously. To harness the full power of OE to prove second laws in isolated systems it is especially useful to include both together. This will be evident below, but remarkably, it is also apparent from historical examples such as some of the famous~H-theorems. While a broader discussion of entropy increase theorems is deferred until later, in Sec.~\ref{sec:second-laws}, we will now show how the entropies used in several well known historical theorems fit neatly within  the framework.

\begin{enumerate}[resume]

    \item  Boltzmann's first H-theorem~\cite[Eqs. (17), (17a)]{boltzmann1909further} is equivalent to the statement \mbox{$\frac{d}{dt} S^\tau_M(\rho_t) \geq 0$}, with $\tau = e^{-\beta H}/Z$ and $M=M_{P(E)}$, as in~\eqref{eqn:empirical-boltzmann-energy} of App.~\ref{app:classical-hard-sphere-gas} (which treats a classical hard-sphere gas in detail). This $M$ measures the distribution of kinetic energy among the particles. To show equivalence, one uses \eqref{eqn:sanov-general} in the \mbox{large-$N$}/\mbox{small-$\Delta$} limit. Then Boltzmann's $E$ function is equivalent to $D_M(\rho \rr \tau)$ up to terms constant by total energy conservation.

    \item Boltzmann's second H-theorem~\cite[No-number equation between (44) and (45)]{boltzmann1909further} (see also~\cite[(10)]{ehrenfest1912conceptual}) is equivalent to the statement \mbox{$\frac{d}{dt} S^\tau_M(\rho_t) \geq 0$}, with $\tau = e^{-\beta H}/Z$ and $M=M_{P(\vec{x},\vec{v})}$ as in \eqref{eqn:empirical-boltzmann}. This $M$ measures the distribution of particles over single-particle phase cells. Equivalence is proved as above. Note that strict non-negativity in Boltzmann's theorems is true only because of his idealizing assumptions (``Stosszahlansatz'') and many-particle limit.

    \item Gibbs' H-theorem~\cite[XII]{gibbs1902book} is equivalent to $S_M(\rho_0) \leq S_M(\rho_{t \to \infty})$ (see (66--67) of \cite{ehrenfest1912conceptual}), where $M$ is the measurement dividing the full phase space into a grid of equal cells.

    \item von Neumann's H-theorem~\cite{vonNeumann1929proof} is equivalent to a statement about the time-average $\overline{S_M^\tau(\rho)}$ for certain choices of $M,\tau$. In particular, von Neumann gives an inequality equivalent to $S(\tau) - \overline{S_M^\tau(\rho)} < \eps$ (and considers when this $\eps$ must be small), ensuring under certain conditions that the time average is near the maximum value. This is a special case of the equilibration inequality \eqref{eqn:eq-eps} studied below, and can be rewritten as $\overline{D_M(\rho \rr \tau}) < \eps$. The relevant prior is his ``microcanonical'' state, $\tau = \sum_E \tr(\rho\Pi_E) \frac{\Pi_E}{\tr\Pi_E}$, which is a mixture of microcanonical energy shells, essentially a coarse version of the time-averaged state~$\rhobar$. The relevant $M$ is any measurement coarser than his ``quantum phase cell'' measurement (effectively, anything coarser than an $M$ commuting with the $\Pi_E$ shells). Comparing to von Neumann's notation, $\tau = \vnf{U}_\psi$ and  $M=(\vnf{E}_{\nu,a})_{\nu,a}$, with $\Pi_E \to \vnf{\Delta}_a$. The entropies $S(\vnf{U}_\psi)$ and $S(\psi)$ he defines (see (34--35)~of~\cite{vonNeumann1929proof}) most directly translate to standard OEs, which has been the usual interpretation. But in fact, one finds $S(\vnf{U}_\psi)-S(\psi) = \sum_{\nu,a} p_{\nu,a} \log \frac{p_{\nu,a}}{t_{\nu,a}}$, where $p_{\nu,a} = (\vnf{E}_{\nu,a} \psi, \psi)$ and $t_{\nu,a} = (\vnf{\Delta}_a \psi, \psi) s_{\nu,a}/S_a$ (here making use of von Neumann's notation). This $t_{\nu,a}$ is the probability for $\vnf{E}_{\nu,a}$ on $\vnf{U}_\psi$. Thus, his entropy difference is in fact $S(\vnf{U}_\psi)-S(\psi) = D_M(\psi \rr \vnf{U}_\psi)$, providing a new interpretation of the theorem. Recasting it this way, as $\overline{S_M^\tau(\rho)} \approx S(\tau)$ where $\tau$ is a coarse version of $\rhobar$, sheds more light on its fundamental meaning, and it can be seen as an early result in the same direction as (\ref{eqn:eq-general}--\ref{eqn:eq-rhobar}).
    
\end{enumerate}

A number of well known Clausius inequalities, capturing thermodynamic entropy increase, also fit in neatly. Here we mention connections to several more common entropy increase results that arise as special cases.

\begin{enumerate}[resume]

    \item A canonical Clausius inequality of the form \mbox{$\Delta S_M^{\tau} = \Delta S(\rho_S) + \int \frac{dE_B}{T_B} \geq 0$}, where $E_B = \braket{H_B}$ is the bath energy expectation value, $T_B = \beta^{-1}_B$ the bath canonical temperature, and $\Delta S(\rho_S)$ change in system von Neumann entropy, derives from the setup of \eqref{eqn:EP-clausius} above. The relevant prior is $\tau = \one_S/d_S \otimes e^{-\beta_B(t) H_B}/Z_B$ constraining the bath energy, and $M = M_{S}(t) \otimes \one_B$ is an optimal measurement on the system. Under \eqref{eqn:EP-clausius} this inequality was shown for the special initial state $\rho(0) = \rho_S(0) \otimes \tau_B(0)$. This can be extended to other initial states using typicality methods~\cite{strasberg2024typical}.

    \item A microcanonical Clausius inequality of the form \mbox{$\Delta S_M^{\tau} = \int \frac{dE_S}{T_S} + \int \frac{dE_B}{T_B} \geq 0$} derives from the setup of \eqref{eqn:thermo} below. Here $T_S^{-1} = \frac{\partial}{\partial E_S} \log W_{E_S}$, with $W_{E_S} = \tr \Pi_{E_S}$ the volume of a microcanonical shell at $E_S$, is the standard Boltzmann temperature of the system; bath quantities are defined analogously, and the system and bath each have definite energies $E_S, E_B$. The relevant prior is $\tau = e^{-\beta H}/Z$ constraining the total energy ($H$ is a weakly coupled Hamiltonian), and $M=M_{E_S} \otimes M_{E_B}$ are coarse measurements of local energy in the system and bath. Conditions for the inequality to hold are studied around \eqref{eqn:thermo}.

    \item The second law from stochastic thermodynamics~\cite{strasberg2022book} arises as the case of \eqref{eqn:rate-matrix} below. This is equivalent to the statement that $dS_M^\tau /dt \geq 0$ assuming $p_x$ evolves by a master equation $dp_x/dt = \sum_x' R_{xx'} p_x$ whose rate matrix obeys a local detailed balance condition.
    
\end{enumerate}

Each of the special cases throughout this section has good motivations for its use in some context. At the same time, each has its limits and domain of applicability, and none alone explains all elementary statistical phenomena. This is evidenced by endless debates about ``correct'' entropy definitions~\cite{ehrenfest1912conceptual, jaynes1965gibbs, goldstein2020gibbs, SwendsenPRR2018}, which often dismiss some widely used entropy or another as misguided or non-fundamental. 

In general, one would like to explain diverse phenomena like why a gas expands to fill its container, why heat flows from hot to cold bodies in both classical and quantum systems, why open quantum systems thermalize, why pure states in isolated systems also thermalize, the role of chaotic mixing in equilibration, and other related questions, all within the same language. By unifying all the above quantities within a more general setting, OE provides a way to do so, without the need to argue against the foundation of existing methods. And below we will see that, beyond special cases, in its general form OE allows one to obtain a number of useful new results.

\section{The need for generalized volumes}
\label{sec:need-for-generalized-volumes}

We now begin to consider the question of why entropy increases. In this section, we explain why the properly defined effective Boltzmann volumes $V_x$ are necessary in order for entropy increase to be generic, and observe how Boltzmann and Shannon terms work together to produce entropy increase. We begin by here highlighting some conceptual aspects, with more detailed entropy increase theorems to be derived in later sections.

One effect of including a prior $\tau$ that encodes the constraints on a system, is the appearance of effective macrostate volumes $V_x = \tr (\tau M_x) \, e^{S(\tau)}$ (see \eqref{eqn:effective-volumes}) in the entropy definition. These generalize the traditional macrostate volumes $W_x = \tr M_x$ that are calculated uniformly on the system's state space.

\begin{figure}[t]
    \centering
    \begin{tabular}{ll}
       \includegraphics[width=.45\columnwidth]{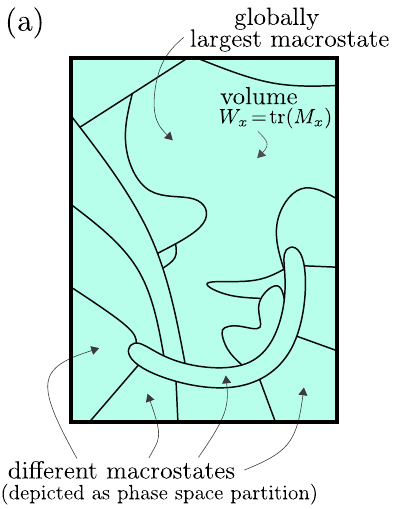}
       &
       \includegraphics[width=.45\columnwidth]{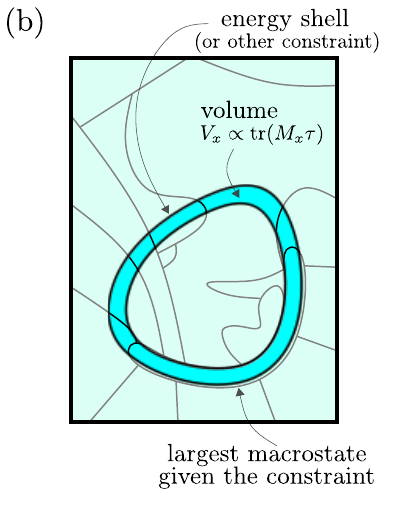}
    \end{tabular}
    \caption{The most likely macrostates are the ones with largest volumes $V_x$ that account for constraints on the system.}
    \label{fig:conserved-quantities}
\end{figure}

In the context of Boltzmann entropies $\log V_x$ (which contribute to OE as in \eqref{eqn:shan-boltz} and as a limiting case), the conceptual need for this generalization is clear from Fig.~\ref{fig:conserved-quantities}, which compares global and effective volumes. If a system's dynamics are constrained to an energy shell, it is the volumes on the shell, and not the global volumes, that determine which macrostate is more likely. 

While it is already common to (often implicitly) assume microcanonical energy shell constraints, the extension to general constraints is crucial if one expects Boltzmann entropies to increase in generic systems. For instance, the example in Fig.~\ref{fig:quantum-example} admits no microcanonical constraint, but exhibits increasing Boltzmann entropies only when appropriate $\tau$ is used to compute volumes.

If constraints are \textit{not} taken into account, then for $M$ of the type depicted in Fig.~\ref{fig:conserved-quantities}, the typical behavior is that Boltzmann entropy will equilibrate to a value \textit{smaller} than the initial value, contradicting the usual second law where entropies increase.

However, this problem only occurs for $M$ like in Fig.~\ref{fig:conserved-quantities}, where constrained and uniform volumes have opposite behaviors. One might think such $M$ were unusual or pathological, but they are actually quite common.

As a pedagogical example, we explain how this applies to a simple system: an isolated box of gas in a uniform gravitational field. 

In the standard story, a box of gas uniformly fills its container because of microstate counting---more microstates correspond to ``uniformly filled'' than any other distribution. This microstate counting corresponds to the traditional volumes $W_x$.

The same is still true in the presence of the gravitational field. More microstates correspond to uniformly filled than otherwise. However, in actuality the gas will accumulate at the bottom of the container---not fill it uniformly. The explanation is simple: if one only counts microstates on the allowed energy shell, most microstates actually correspond to ``exponentially accumulates at the bottom''. This counting corresponds to volumes $V_x$ given a microcanonical energy constraint.

Thus the new OE definition, with properly computed volumes, sees the equilibrium state of the gas as a high entropy state, while more naive entropy definitions would see the equilibrium as a low entropy state.

Furthermore, the easiest way to prove the above analysis highlights the usefulness of generalized constraints. While the microcanonical $V_x$ for this ``spatial distribution'' measurement are difficult to compute directly, the canonical $V_x$ derived from $\tau = e^{-\beta H}/Z$ are straightforward to compute as in \eqref{eqn:sanov-general}. For this classical system the canonical and microcanonical constraints are the same, and either can be applied to the problem. Further, in the many particle limit where equivalence of ensembles holds, they must also yield about the same volumes.  

This example provides a perspective on cases of ``apparent entropy decrease'' such as oil separating from water, which is highly analogous to the above example. Another prototypical example for $M$ of this type are measurements of ``energy on the hot side'' during heat exchange---for which we explicitly see this phenomenon occur in Figs.~\ref{fig:quantum-example}--\ref{fig:ideal-gas-1}.

This section has suggested the intuition that with properly chosen $\tau$, entropy increase laws can be expected. While this is true, in quantum systems such laws will not hold for the Boltzmann terms alone. 

Consider a system initially with probability 1 to be in the macrostate of largest volume $V_x$. During typical dynamics, $\ket{\psi(t)}$ (the system's state vector) will later equilibrate to a configuration overlapping many macrostates, which necessarily lowers the Boltzmann part of the entropy. But this decrease is compensated by an increase in the Shannon entropy of the distribution over macrostates, leading to an increase in total OE.

Similarly, consider a system with macrostate probabilities initially evenly distributed over all possibilities. During typical dynamics, $\ket{\psi(t)}$ will later equilibrate to a configuration with probability more concentrated at the largest volume $V_x$, necessarily lowering the Shannon entropy. This decrease is compensated by the increasing Boltzmann contributions, for increasing total OE.

In this way, with properly chosen $\tau$, the Boltzmann and Shannon terms conspire leading OE to approach its maximum. In the next section, we consider the conditions where this intuition can be made quantitative via entropy increase theorems.

\section{Second laws}
\label{sec:second-laws}

In this section we turn to the more precise treatment of entropy increase theorems and second laws. By virtue of its reduction to the numerous special cases above, OE neatly incorporates previously proven entropy increase theorems within its framework. Here, however, we are concerned with second laws that can be derived for OE from a completely general point of view.

\subsubsection{Overview}

The central question is under what circumstances it can be proved that $S_M^\tau(\rho)$ increases, and what this increase implies for physics.

We focus here on equilibration under fixed constraints. This means we consider $S_M^\tau(\rho(t))$ increasing towards its maximum $S(\tau)$, where $\tau$ is fixed. In that case the difference $S(\tau) - S_M^\tau(\rho(t)) = D_M(\rho(t) \rr \tau)$ becomes the central quantity of interest.

One can say that $\rho(t)$ ``equilibrates to $\tau$'' if, for essentially any coarse $M$, this quantity $D_M(\rho(t) \rr \tau)$ becomes small and remains small for long times (equivalently, $S_M^\tau(\rho) \approx S(\tau)$ becomes nearly maximal). One example is thermalization, \ie~equilibration to $\tau \propto e^{-\beta H}$.

More broadly, $\rho(t)$ ``equilibrates to $\tau$ as viewed by~$M$'' if $D_M(\rho(t) \rr \tau)$ becomes small for that particular $M$. The $M$-dependence of the idea of equilibration gives the method flexibility---to prove something like Clausius relations, one only needs to show equilibration occurs for some particular $M,\tau$, without a full analysis of the ergodic hypothesis at the microstate level. 

Note that we use the terminology ``equilibration'' in a stronger sense than usual in pure state statistical mechanics~\cite{gogolin2016equilibration}, where equilibration only refers to reaching a time-independent steady value. Here, we go beyond a steady entropy and demand that it \textit{maximizes}, which is not always true for other entropies (\cf~\cite{meier2024emergence}). While well known theorems for expectation value dynamics (such as~\cite{linden2009evolution}) ensure that essentially any function of measurement outcome probabilities (including most entropy definitions) will obtain such a steady state, maximization is an additional requirement with stronger consequences.

The increase of a quantity called entropy, on its own, is not satisfactory to establish a second law. For such a law to be useful, it must imply something about physics, such as that heat flows from hot to cold bodies.

To establish such connections to physics as done below, the fact that OE increases/maximizes (not just equilibrates in the weaker sense), and that equilibration to generic $\tau$ (and not just $\rhobar$) can be analyzed, both play important roles, letting statistical laws reduce into simpler thermodynamic ones.

Here, we focus on fixed constraints. Second laws involving time-dependent constraints  $\tau = \tau(t)$, like the Clausius relation for expectation values proved in \eqref{eqn:EP-clausius} above (and its extension to pure states via dynamical typicality~\cite{strasberg2024typical}), also play a crucial role complementary to the present results. For instance, after removing a partition, a gas not only equilibrates to a new $\tau$, but to one with a higher equilibrium entropy---which involves both the approach to equilibrium captured by $D_M(\rho \rr \tau)$ and the changing equilibrium $S(\tau(t))$ as contributions.

\subsubsection{Open vs. isolated systems}

We consider equilibration in both open and isolated systems. These two cases have a basic difference. In open systems, the actual microstate $\rho_S(t)$ can (and typically does) approach an equilibrium state $\tau_S$. This obviously also implies equilibration for any particular $M$. In isolated systems, it is impossible for $\rho(t)$ to approach any stationary $\tau$, since the usual distance measures like  $D(\rho \rr \tau)$  or the trace distance are unitary invariant. So in isolated systems, equilibration completely relies on being viewed by $M$. Nonetheless, equilibration is still generic in isolated systems, because chaotic mixing implies $M$ would need very high resolution to ``see'' the difference between $\rho$ and $\tau$ after long times. The two cases are also directly connected---an open system can by viewed as a subsystem of a larger isolated one, in which case any $M$ with access only to system degrees of freedom is necessarily in some sense coarse.

For present purposes we define isolated systems as those evolving unitarily under an undriven time-independent Hamiltonian $H$, and open means not isolated.

\subsubsection{Open systems}

In the theory of open systems, there is a significant body of research establishing when $\rho_S(t)$ approaches an equilibrium state~$\tau_S$, including results based on  typicality~\cite{goldstein2005canonical,popescu2006entanglement,GemmerMichelMahlerBook2004}, open system dynamics~\cite{spohn1978entropy,DeVegaAlonsoRMP2017, LandiPolettiSchallerRMP2022, TrushechkinEtAlAVSQS2022},  master equations~\cite{spohn1978entropy}, resource theory of thermal operations~\cite{brandao2013resource,lostaglio2019review}, and other approaches~\cite{wehrl1978general,strasberg2022book}. From these methods, it can often be established as a corollary that $D(\rho_S(t) \rr \tau_S) \to 0$ with time, reflecting equilibration of the system state.

The connection of OE to these approaches therefore essentially follows from \eqref{eqn:qre-bound-props} below, which in this case provides the bound
\begin{equation}
\label{eqn:qre-bound}
    S_{M_S}^{\tau_S}(\rho_S) \geq S(\tau_S) - D(\rho_S \rr \tau_S).
\end{equation}
While this inequality can be applied directly to the open system, it may be the case that $\rho_S$ does not obey the constraint associated with equilibrium state $\tau_S$. This is because usually, the relevant constraint actually holds in the larger system + environment setting. This can be remedied by considering total entropy where $M$ acts only on the system, in which case one finds (see Proof~\ref{pf:open-system-bound})
\begin{equation}
\label{eqn:open-system-bound}
    S_{M_S \otimes \one_E}^{\tau_{SE}} (\rho_{SE}) \geq S(\tau_{SE}) - D(\rho_S \rr \tau_S).
\end{equation}
These bounds show that, if for any reason one can show that $D(\rho_S(t) \rr \tau_S) \to 0$ with time, then the entropy $S_M^\tau(\rho) \to S(\tau)$ also maximizes for all $M$. That is, equilibration of  $\rho_S(t)$ to $\tau_S$ at the microstate level also implies OE equilibration with respect to system measurements.

This relation, when it applies, takes the form of an ever increasing lower bound on $S_M^\tau$. This means that transient decaying fluctuations in entropy are allowed along the way. This has the realistic feature that entropy of a specific observable may fluctuate temporarily downward even while generally increasing. 

In this way, the existing theories of open system equilibration already imply OE maximization laws. We now turn to the subtler case of isolated systems.

\subsubsection{Time-averaged state}

As for many equilibration theorems~\cite{reimann2008foundation,linden2009evolution,short2011equilibration,gogolin2016equilibration}, the time-averaged state $\rhobar$ will play an important role below. Time average is denoted $\overline{\, \bullet \,} = \lim_{T \to \infty} \int_{-T/2}^{T/2} (\bullet/T) \, dt$.  

Here we mention several useful facts that hold assuming an isolated system (see Proof~\ref{pf:time-average}). First, for any stationary~$\tau$, the cross-entropy $S(\rho(t);\tau)$ is time-independent, so the constraint~\eqref{eqn:cross-ent-constraint} is conserved. Second, the constraint
\begin{equation}
\label{eqn:time-average}
    S(\rho(t);\rhobar)=S(\rhobar)
\end{equation}
always holds, so that $\tau=\rhobar$ is always a valid prior. And third, $S(\rhobar) \leq S(\rhobar; \tau) = S(\rho(t);\tau) \leq S(\tau)$ for any stationary $\tau$, implying $\rhobar$ is the minimum-entropy (\ie~tightest) stationary prior for which the constraint holds. Note that for a quantum system with no degenerate energies, one has $\rhobar = \sum_E \ketbra{E}\rho \ketbra{E}$ ($H$~eigenbasis), and the above results can be checked easily.

\subsubsection{Isolated systems}
 
\begin{figure}
    \centering
    \includegraphics[width=\linewidth]{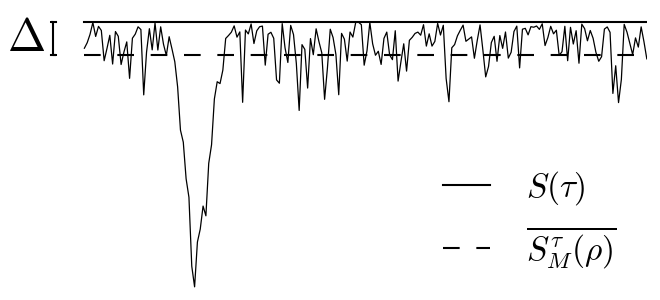}
    \caption{If time-averaged entropy is near the maximum $S(\tau)$, then fluctuations to low entropies must be rare, short-lived, and compensated by long times near the maximum.}
    \label{fig:fluctuations}
\end{figure}

One class of second laws in isolated systems derives from showing that the time-averaged entropy is nearly equal to its maximum possible value~\cite{ehrenfest1912conceptual,vonNeumann1929proof,gogolin2016equilibration}. This type of time-averaged law is the type we now consider here, and the basic principle of the laws we will derive is sketched in Fig.~\ref{fig:fluctuations}.  

In the present context, time-averaged entropy being near the maximum means that \mbox{$\overline{S^\tau_M(\rho)} \approx S(\tau)$}. The central quantity is therefore the deviation from maximal of the time-averaged entropy (compare Fig.~\ref{fig:fluctuations}),
\begin{equation}
\label{eqn:eq-bigdelta}
    \Delta = \overline{D_M(\rho \rr \tau)} = S(\tau) - \overline{S_M^\tau(\rho)}.
\end{equation}
The probability of observing entropy fluctuations of order~$\delta$ below $S(\tau)$ decays as $\Pr(\delta) \propto \Delta/\delta$, (see Proof~\ref{pf:eq-bigdelta}). Thus, $\Delta$ sets the scale of expected entropy fluctuations. Here, second laws take the form of bounds on $\Delta$.

To see why bounds on $\Delta$ can be viewed as second laws, consider the bound $\Delta < \eps$ where $\eps$ is some sufficiently small parameter. If such a bound holds, it ensures that (see Proof~\ref{pf:eq-eps})  (i) entropy is nearly maximal at almost all times, (ii) large entropy fluctuations are rare, and (iii)~when entropy does happen to be low, it will subsequently very likely increase. 
This follows from the one-sidedness of the bound: upward entropy fluctuations above the maximum are impossible, so downward entropy fluctuations must be suppressed.

Therefore, to prove second laws, we wish to prove bounds of the form
\begin{equation}
\label{eqn:eq-eps}
    \overline{D_M(\rho \rr \tau)} \leq \eps.
\end{equation}
To prove such bounds, the basic requirement is that $\tau$ be a ``good enough prior'' and that $M$ be ``coarse enough'', in a sense seen more precisely below.

To make progress, a useful first step is the decomposition (see Proof~\ref{pf:eq-general})
\begin{equation}
\label{eqn:eq-general}
    \overline{D_M(\rho \rr \tau)} = \overline{D_M(\rho \rr \rhobar)}  + D_M(\rhobar \rr \tau).
\end{equation}
The first term is the ``equilibration term'', describing whether the system equilibrates to its time average as viewed by $M$. The second term is the ``to $\tau$'' term, describing whether $\tau$ looks like $\rhobar$ to $M$. Similarity of $\tau$ to~$\rhobar$ (``good prior'') helps the second term become small. Meanwhile, ``coarseness of $M$'' helps with both terms, since $\rho(t), \rhobar, \tau$ may appear similar as viewed by $M$, even if not at the microstate level. Together, these two factors determine the validity of second laws for a given $M,\tau$.

We now consider the conditions under which bounds on these two terms may be proved, and present partial entropy increase theorems that hold under different sets of assumptions.

We begin with the former term, which can be proved small for quantum systems in several important regimes, even without a detailed specification of the system. 

First, we consider the case of a macroscopic system where many energy eigenstates are occupied. In this case, $\Delta$ is small if the number of macrostates $m$ is small compared to the effective dimension $d_2(\rhobar) = (\tr \rhobar^2)^{-1}$. In particular, consider a quantum system with non-degenerate energies and energy gaps~\cite{linden2009evolution,short2011equilibration}. With only this assumption, one obtains the bound (see Proof~\ref{pf:eq-rhobar})
\begin{equation}
\label{eqn:eq-rhobar}
    \overline{D_M(\rho \rr \rhobar)} \leq \eps \log m + g(\eps),
\end{equation}
with $\eps = m/4\sqrt{d_2(\rhobar)}$ and \mbox{$g(\eps)\! =\! -\eps\log \! \eps \! + \! (1\! +\!\eps)\! \log (1\! +\! \eps)$}. For macroscopic systems and realistic finite-precision measurements, the effective dimension of $\rhobar$ is typically much bigger than the number $m$ of possible measurement outcomes, making this a generic and powerful result. 

On the opposite end of the spectrum is the case when only a handful of energy eigenstates are occupied, in the sense that $S(\rhobar) \ll S(\tau)$. Here, we have a bound because for any quantum system, \eqref{eqn:time-average} implies that
\begin{equation}
\label{eqn:eq-eigenstate}
    \overline{D_M(\rho \rr \rhobar)} \leq S(\rhobar).
\end{equation}
For this to be useful, it must hold that $S(\rhobar) \ll S(\tau)$, which means $\tau$ is not very similar to $\rhobar$ at the microstate level. However, this theorem will be useful in situations where, for instance, a canonical $\tau \propto e^{-\beta H}$ looks similar to an energy eigenstate $\ketbra{E}$ of equal energy when viewed by the measurement $M$, which is not so unrealistic. In fact, this situation is a central consequence of the eigenstate thermalization hypothesis to which we will turn shortly.

The intermediate regime between these two cases will be useful for continued study: in the example in Fig.~\ref{fig:quantum-example}, $\Delta$ is empirically small even though neither \eqref{eqn:eq-rhobar} nor \eqref{eqn:eq-eigenstate} give useful bounds there (the system is too mesoscopic), suggesting a search for more proof regimes.

Now we move on to the second term. This term is more difficult to make progress on in a general setting, because it depends more on details of the measurement $M$ and how it relates to the system's Hamiltonian. 

One general remark is that it suffices for the time-averaged measurement probabilities $\overline{p_x(t)} = \overline{\tr M_x \rho(t)}$ to be similar to $q_x = \tr M_x \tau$ derived from the prior. This is essentially a statement of the ergodic hypothesis as viewed by $M$, avoiding the more difficult question of ergodic hypotheses at the microstate level. Indeed, $D_M(\rhobar \rr \tau)$ is really shorthand for the underlying $D(\overline{p} \rr q)$, so that this term can be analyzed even in classical systems where $\rhobar$ may not be well-defined.

To illustrate how closeness of the probabilities ensures high entropy, it is demonstrative to consider the bound
\begin{equation}
    D_M(\rhobar \rr \tau) \leq \log \; \sup_x \; \frac{\overline{p_x(t)}}{q_x}
\end{equation}
with the $\sup$ taken over $x$ such that $\overline{p}_x >0$. Evidently it suffices that no time-averaged macrostate occupancy is much greater than the prior occupancy. Whether this bound will be small, however, depends on further details of the system.

One case where a bound can be obtained is under the eigenstate thermalization hypothesis (ETH)~\cite{DAlessioEtAlAP2016, deutsch2018eigenstate}. Specifically, for energy eigenstates $\psi_E = \ketbra{E}$, the diagonal part of the ETH asserts that $|\tr(M_x(\psi_E-\psi_{E'}))|$ is negligibly small for all $x$, large systems, and reasonable~$M$, if $E,E'$ belong to an energy window with subextensive size $\Delta E$. It follows that $D_M(\psi_E \rr \psi_{E'}) \leq \eps_{\textsc{eth}}$ for some $\eps_{\textsc{eth}}\geq 0$ for all $E,E'$ in the energy window. Since $\tau$~(by assumption) and $\rhobar$ are stationary, they are diagonal in an energy eigenbasis, so if they are supported in such an energy window, convexity implies
\begin{equation}
    \label{eqn:ETH}
    D_M(\rhobar \rr \tau) \leq \eps_{\textsc{eth}}.
\end{equation}
In this way ETH provides one sufficient condition for smallness of \eqref{eqn:eq-general}'s ``to $\tau$'' term.

Together, (\ref{eqn:eq-rhobar}--\ref{eqn:ETH}) show a few situations where entropy increase theorems can be proved under minimal further assumptions. This provides a basis for continued investigation of bounds like~\eqref{eqn:eq-eps}, both in general and in particular systems. In the examples in Figs.~\ref{fig:quantum-example}--\ref{fig:ideal-gas-dynamics}, one can observe that $\Delta$ is empirically small in all cases; but in none of those cases was the smallness proved here analytically.

This class of second laws, based on time-averaged entropy being near a maximum value, are the type of law argued for by von Neumann~\cite{vonNeumann1929proof}, who emphasized its compatibility with the time-reversibility and recurrence problems of statistical mechanics~\cite{ehrenfest1912conceptual}. In fact, as was shown earlier, von Neumann's quantum H-theorem is a special case of \eqref{eqn:eq-eps} for a particular choice of $M,\tau$. In tandem with complementary classes of laws, such as fluctuation theorems and laws based on time-dependent constraints, this class provides a piece of the puzzle in the story of equilibration, thermalization, and entropy increase.

We remark that equilibration theorems based on infinite time averages are sometimes criticized on the grounds that the relevant timescale may be too long. We emphasize that, as discussed in Proof~\ref{pf:eq-eps}, what is shown by bounds like \eqref{eqn:eq-eps} is that some finite equilibration timescale $T_{\rm eq}$ exists, and on this timescale entropy increase can be expected. Determining this timescale is a separate problem, and it is obvious that without specifying more about the Hamiltonian, the existence of some $T_{\rm eq}$ is the best one can hope for. Given additional specificity, the above can be improved using methods like those reviewed in~\cite{gogolin2016equilibration, WilmingEtAlBook2018}. And in general, any of a number of results on equilibration under restricted sets of POVMs, such as those appearing in~\cite{gogolin2016equilibration}, can be extended to analogs of~\eqref{eqn:eq-rhobar}.

\subsubsection{Connections to thermodynamics}

So far we have considered the question of when entropy can be shown to increase. But supposing entropy $S_M^\tau(\rho)$ does increase, what does this imply for physics? We now turn to showing that if bounds like \eqref{eqn:eq-eps} do hold, then generalized thermodynamic laws follow. 

To illustrate this, consider the example of two weakly coupled systems exchanging heat, with $\tau=e^{-\beta H}/Z$ corresponding to total energy conservation, and $M=M_{E_A}\otimes M_{E_B}$ being a coarse measurement of the local energy in each subsystem. This is precisely the setup for the examples shown in Figs.~\ref{fig:quantum-example}--\ref{fig:classical-example}.

Suppose that a bound like \eqref{eqn:eq-eps} holds in this system, as is empirically true in both of those examples. Further, suppose the system always has a definite macrostate (perhaps because it is classical, or perhaps because quantum fluctuations in local energy are smaller than coarse energy windows of the measurement). If this is true, then~\eqref{eqn:eq-eps} implies equilibration to (see Proof~\ref{pf:thermo}) 
\begin{equation}
\label{eqn:thermo-eq}
    T_A = T_B,
\end{equation}
where $T_A^{-1} = \frac{\partial}{\partial E_A} \log W_{E_A}$ (and similarly for~$T_B$) are traditional Boltzmann temperatures. Moreover, starting from a non-equilibrium state, during time $\Delta t$ it implies that with high probability the change in entropy obeys (see Proof~\ref{pf:thermo}) 
\begin{equation}
\label{eqn:thermo}
    \Delta S_M^\tau(\rho) =  \int \frac{dE_A}{T_A} + \int \frac{dE_B}{T_B} > 0,
\end{equation}
thus implying that heat flows from hot to cold bodies. 

The volumes $W_{E_{A,B}}$ defining $T_{A,B}$ are the traditional ones, recovering textbook thermodynamics. But the fact that (\ref{eqn:thermo-eq}--\ref{eqn:thermo}) were derived from $\tau \propto e^{-\beta H}$ (with volumes $V_{E_A,E_B}$ later reducing to $W_{E_{A,B}}$ and other terms in the proof) is relevant, since it provides the possibility of proving this relation using the results of previous sections. For instance, an everyday system like hot tea, which is classical for all macroscopic purposes but actually quantum, obeys both the definite-macrostate assumption of \mbox{(\ref{eqn:thermo-eq}--\ref{eqn:thermo})}, and the conditions for \eqref{eqn:eq-rhobar} to provide a useful bound, making it plausible (\eg~by establishing \eqref{eqn:ETH} or similar) to prove \eqref{eqn:eq-eps} directly from first principles in such a system. What we showed here is that such a proof would as a corollary constitute a proof of the thermodynamics.

Further insight is revealed if one repeats the above calculation, but now with $M'=M_{E_A} \otimes \one_B$ measuring energy in only one subsystem. In that case, one can quickly find (see Proof~\ref{pf:thermo}) a new equilibrium condition, which is that the Boltzmann temperature $T_A$ becomes equal to the Gibbs temperature $\beta^{-1}$ appearing in the global prior $\tau \propto e^{-\beta H}$. This gives a somewhat novel connection between distinct temperature notions (note that the equality was not derived from equivalence of ensembles). Interestingly, if a bound like \eqref{eqn:eq-eps} holds for~$M$, it must also hold for the coarser $M'$, and so not only implies $T_A$ and $T_B$ equilibrate, but to the value $\beta^{-1}$.

This provides a simple first example of how, by considering arbitrary $M$, one could obtain a variety of generalized thermodynamic laws.

An immediate question is how \eqref{eqn:thermo} generalizes if the single-macrostate assumption is discarded. Interestingly, this generalization does not take the form of a Clausius relation, but becomes a relation on time derivatives of macrostate probabilities \mbox{$p_x(t) = \tr M_x \rho(t)$}, \mbox{$q_x = \tr M_x \tau$}. This relation takes the form (see Proof~\ref{pf:generalized-thermo})
\begin{equation}
\label{eqn:generalized-thermo}
    \frac{d}{dt} S^\tau_M(\rho) = \sum_x \frac{dp_x}{dt} \, \big(\log q_x - \log p_x \big).
\end{equation}
This describes entropy increase in terms of a stochastic relation on macrostate populations. Further progress can be made if the probabilities evolve through a master equation $dp_x/dt = \sum_{x'} R_{xx'}p_{x'}$. If the rate matrix $R_{xx'}$ satisfies local detailed balance $R_{xx'}/R_{x'x} = q_x/q_{x'}$, then entropy increase follows, as (see~Proof~\ref{pf:rate-matrix})
\begin{equation}
\label{eqn:rate-matrix}
    \frac{d}{dt} S^\tau_M(\rho) = \sum_{x,x'} R_{xx'}p_{x'}\log\frac{R_{xx'}p_{x'}}{R_{x'x}p_{x}} \ge 0,
\end{equation}
further cementing the connection to stochastic thermodynamics~\cite{strasberg2023classicality, PelitiPigolottiBook2021, strasberg2022book, ShiraishiBook2023,FalascoEspositoRMP2025}. In this light, Clausius relations like~\eqref{eqn:thermo} can be viewed as the single-macrostate (\eg~classical or macroscopic) limit of the more general stochastic relation \eqref{eqn:generalized-thermo}. Meanwhile, with different $\tau$ and $M$ we found, in \eqref{eqn:EP-clausius}, other Clausius relations that can instead be established for expectation values, further highlighting the flexibility of the generalized OE framework.

Another interesting question is how \eqref{eqn:thermo} generalizes if the systems are strongly coupled, but this is left open for the moment. More broadly, further clarifying the structure of what generalized thermodynamic laws can be derived for general $M,\tau$ in various systems will be a useful topic for further study.

\section{Mathematical properties}
\label{sec:mathematical-properties}

In this section, we discuss some useful formal properties, and resolve the pathologies of the traditional OE definition. 

Proofs are given in Proof~\ref{pf:properties}. Most of the results follow from the structural properties of relative entropies: monotonicity, joint convexity, chain rule, and non-negativity~\cite{cover2006book,wilde2011notes}. Relative entropy properties are reviewed in \AppRE.

\subsubsection{General structure}

As a function of states and measurements, the OE has many natural properties concerning its basic structure. Many such properties were shown previously for $S_M(\rho)$ (see \eg~\cite{safranek2021brief,buscemi2022observational}) and similar ones continue to hold here.

\smallskip
(\tpn) Bounds are $S(\tau) \geq S_M^\tau(\rho) \geq S(\rho)$, assuming the constraint holds. The upper bound holds independently of the constraint.

\smallskip

(\tpn) \SMTR \ is a jointly concave function of states $\rho,\tau$, and a concave function of the POVM $M$. In $\rho$ it is bounded-concave~\cite{schindler2023continuity}. In $M$ it is linear under disjoint convex combinations. See \AppRE \ for definitions. 

\smallskip

(\tpn) \SMTR \  is monotonic under coarser/finer measurements, defined in the standard sense of the postprocessing partial order on POVMs~\cite{leppajarvi2021postprocessing,bonfill2023entropic}. Namely, if $N,M$ are POVMs such that $N_j = \sum_i \Lambda_{j|i} M_i$, with $\Lambda$ a stochastic matrix ($\Lambda_{j|i} \geq 0$, $\sum_j \Lambda_{j|i}=1$), then $N$ is called coarser than $M$, and it holds that $S^\tau_N(\rho) \geq S^\tau_M(\rho)$.

\smallskip

(\tpn) \SMTR \  is monotonic under sequential measurements. This is best formalized in terms of quantum instruments (and their classical analogues), as in \AppRE. Namely, if $\NN,\MM$ are composable  instruments, and $\NN \circ \MM$ their composition, then $S^\tau_{\NN \circ \MM}(\rho) \leq S^\tau_{\MM}(\rho)$. Hence, additional measurements only provide more information. This extends to a concrete chain rule as in \eqref{eqn:mre-chain-rule}.

\smallskip

(\tpn) A fundamental bound is given by
\begin{equation}
\label{eqn:qre-bound-props}
    S_M^\tau(\rho) \geq S(\tau) - D(\rho \rr \tau).
\end{equation}
This shows that similarity of $\rho$ to $\tau$ at the microstate level puts a bound on the measured entropy.

\smallskip

(\tpn) For a convex combination $\rho = \sum_k \lambda_k \rho_k$,
\begin{equation}
    S_M^\tau(\rho) = \textstyle\sum_k \lambda_k S_M^\tau(\rho_k) + I(M:\mathcal{E}),
\end{equation}
where $I(M \! : \! \E)$ is the mutual information between outcomes of POVM $M$ and the ensemble $\E = \{\lambda_k,\rho_k\}_k$~\cite{buscemi2024note}. This is defined by $I(M \! : \! \E) = D(P_{xk} \rr P_x P_k)$ where $P_{xk} = \lambda_k \tr(M_x \rho_k)$ and $P_x,P_k$ are its marginals. Thus
\begin{equation}
    S_M^\tau(\rho) - S(\rho) \leq \textstyle\sum_k \lambda_k  \big( S_M^\tau(\rho_k) - S(\rho_k) \big)
\end{equation}
is implied by Holevo's bound on $I(M \! : \! \E)$~\cite{wilde2011notes}.

\subsubsection{Continuity}
 
The bound for continuity in $\rho$ of $S_M(\rho)$ that appeared in~\cite{schindler2023continuity} fails in infinite dimensions. Here we identify a new bound that applies to the present case, at the cost of sacrificing $M$-independence of the bound.

\smallskip

(\tpn) For any measurement with an effectively finite number of outcomes (meaning a finite number of nonzero values of $V_x = \tr M_x \tau \; e^{S(\tau)}$), one obtains the bound
\begin{equation}
\label{eqn:continuity-bound}
    |S_M^\tau(\rho) - S_M^\tau(\sigma)| \leq h(s) + s \, (\log A + B),
\end{equation}
with $s = \frac{1}{2} \| \rho -\sigma \|_1$, $A$ the number of nonzero $V_x$, $B$~the maximum finite value of $|\log V_x|$, and $h(s)$ binary Shannon entropy. This bound ensures continuity in $\rho$.

\smallskip

(\tpn) $S_M^\tau(\rho)$ is a continuous function of $M$ in the simulation distance topology, by the same reasoning as in~\cite{schindler2023continuity}.

\subsubsection{Coarse-grained states}

Several notions of coarse-grained state arise in the context of OE. These are related to the estimation or retrodiction of microstates based on knowledge only of  measurement outcomes and priors~\cite{buscemi2022observational,bai2023observational}.

\smallskip

(\tpn) One way to obtain a coarse-grained state is based on Bayes-like inference about the true state $\rho$, based on outcomes of $M$ given prior $\tau$. This is obtained by applying the Petz recovery map with prior $\tau$ to the quantum-classical channel (\cf~\eqref{eqn:MRE-quantum-channel}) implementing the measurement $M$~\cite{buscemi2022observational}. This leads to the coarse-grained state 
\begin{equation}
\label{eqn:rcg}
    \rcg = \sum_x p_x \frac{\sqrt{\tau} M_x \sqrt{\tau}}{\tr\tau M_x}
\end{equation}
with $p_x=\mbox{tr}M_x\rho$. This is essentially a convex sum of POVM elements weighted by their probabilities and the prior.

\smallskip

(\tpn) Because quantum Bayesian inference is not unique, other related coarse-grained states are obtained from rotated Petz recovery maps~\cite{parzygnat2023axioms}. This leads to the generalization
$\rcg^s = \sum_x p_x (\tau^{(1+is)/2} M_x \tau^{(1-is)/2})/(\tr\tau M_x)$. A special linear combination of these, given by
\begin{equation}
\label{eqn:rcgt}
    \rcgt = \int_{-\infty}^{\infty} \beta(s) \rcg^s \, ds
\end{equation}
with $\beta(s) = \frac{\pi/2}{1+\cosh(\pi s)}$, is of technical importance due to its appearance in strong RE monotonicity theorems~\cite{sutter2017multivariate}. If $\tau$ is proportional to a projector, or if everything commutes (including the classical case), then $\rcg = \rcg^s = \rcgt$. 

\smallskip

(\tpn) Another notion of coarse-grained state is given by the MaxEnt state $\taut_M$ compatible with the fixed measurement probabilities $p_x = \tr M_x \rho$. This takes the form $\taut_M = e^{-\sum_x\lambda_x M_x}/Z$, if such a state exists with the correct probabilities. Prior constraints could also be included, yielding $\taut_{M,X} = e^{-\sum_x\lambda_x M_x-\lambda' X}/Z$.

\smallskip

(\tpn) For the case of $\tau \propto \one$ and projective $M = (\Pi_x)_x$,
\begin{equation}
     \taut_M = \rcg = \rcg^s = \rcgt =  \sum_x \, \tr (M_x \rho) \, \frac{\Pi_x}{\tr \Pi_x},
\end{equation}
so the coarse-grained state is essentially unique, and is obtained by naively smearing $\rho$ over macrostates. In this case $S(\rcg) = S_M(\rho)$ so that OE is the von Neumann entropy of the coarse-grained state---but this does not extend to general POVMs or general $\tau$. 

\smallskip

(\tpn) For general $M,\tau$ and $\rcg$ as in \eqref{eqn:rcg}, one has $D(\rho \rr \tau) \geq D_M(\rho \rr \tau) \geq D(\rcg \rr \tau) \geq D_M(\rcg \rr \tau)$, which follow from RE monotonicity. With $\tau \propto \one$ and projective~$M$, some of these become equalities, and the chain reduces to $S(\rho) \leq S_M(\rho) = S(\rcg) = S_M(\rcg)$.

\subsubsection{Recovery bound}

We obtain a recovery bound for OE with general $M,\tau$ analogous to the one found for $\tau \propto \one$ in~\cite{buscemi2022observational}. This bound is a consequence of the powerful property of strong RE monotonicity~\cite{sutter2017multivariate}. 

\smallskip

(\tpn) Given the constraint $S(\rho ; \tau) \leq S(\tau)$, and $\rcgt$ as in \eqref{eqn:rcgt}  above, it can be shown that
\begin{equation}
\label{eqn:recovery-bound}
    S_M^\tau(\rho) - S(\rho) \geq \textstyle\sup_M D_{M}(\rho \rr \rcgt).
\end{equation}
This shows how OE bounds the closeness between the true state and an inferred coarse state. The bound is also useful for technical purposes.

\smallskip

(\tpn) As a corollary, assuming the constraint holds, it follows that if $S_M^\tau(\rho) = S(\rho)$ then $\rho = \rcgt$, giving conditions for optimal measurements.

\subsubsection{Resource theories}

OE minimized over restricted sets of POVMs connects naturally to resource theories. Denote $S_{\Omega}^\tau(\rho) = \inf_{M\in \Omega} S_M^{\tau}(\rho)$ for any set $\Omega$ of measurements. In resource theory, a function can be considered a resource measure when it changes monotonically, \eg~$F(\Phi(\rho)) \leq F(\rho)$ for all $\rho$, under all ``free operations'' of the theory.

\smallskip

(\tpn) If $\Omega$ contains all $M$, $S^\tau_{\Omega}$ is a monotone of \mbox{$\tau$-preserving maps}, connecting in case $\tau \propto e^{-\beta H}$ to resource theories of thermodynamics~\cite{lostaglio2019review}. 

\smallskip

(\tpn) For other $\Omega$ this will be a monotone of a subclass of (\eg~separable) \mbox{$\tau$-preserving} maps, opening possible relations to joint (\eg~thermo+entanglement~\cite{oppenheim2002thermodynamical} or thermo+complexity~\cite{munson2024complexity}) multi-resource theories~\cite{sparaciari2020firstlaw}.

\subsubsection{Fluctuation theorems}

To address this important topic is beyond the scope of this article, but we remark that fluctuation theorems involving the observational entropy production \eqref{eqn:EP-general} can be obtained and will be reported in future work. Observational entropy fluctuation theorems for the case $\tau \propto \one$~\cite{strasberg2020first}, and for the case of time-dependent constraints with trivial $M$~\cite{strasberg2024typical}, have also been studied previously. Other fluctuation theorems are automatically incorporated by the reduction of OE to special cases noted earlier.

\subsubsection{Comparison to traditional OE}

The traditional definition of OE ($S_M(\rho)$ as in \eqref{eqn:traditional-oe}) is subject to several issues that motivated the present work. One is the question of apparent entropy decrease, which has already been discussed above and will be further illustrated in the next section. The others are mathematical pathologies that arise in infinite-dimensional systems---namely those systems where \mbox{$\tr \one = \infty$} and no maximally mixed state exists (which includes classical systems).

In such systems, the Boltzmann volumes $W_x = \tr M_x$ may be infinite. Not only is this possible, but for any $M$ with a finite number of outcomes (which includes any realistic finite-precision $M$), at least one $W_x$ \textit{must} be infinite. This has the consequence that \mbox{$S_M(\rho)=\infty$} for almost all states, including for instance any $\rho$ mixed with~$\eps>0$ of thermal noise ($\rho' = (1-\eps) \, \rho + \eps \, e^{-\beta H}/Z$), which is clearly problematic. This implies every $\rho$ is arbitrarily close to a state with infinite entropy, which means  $S_M(\rho)$ is a highly discontinuous function of the state.

In contrast, we can see from the upper bound \eqref{eqn:upper-bound} and the continuity bound \eqref{eqn:continuity-bound} that both of these issues are resolved by the generalization, whenever a constraint is present that ensures $S(\tau) < \infty$.

Note that $S_M(\rho)=\infty$ is not really problematic in principle; it only indicates that a finite amount of information is available in the coarse description while an infinite amount is missing. However, the problem is that traditional OE is formally infinite \textit{too often}, even when it physically should not be. Including $\tau$ generalizes straightforward approaches to remedying this issue, such as insisting that $\rho$ has energy below some cutoff based on prior knowledge of the system, in a conceptually clean and mathematically precise way.

We also remark that discontinuity due to infinite entropies is a familiar problem, occurring also for von Neumann entropy. In that case continuity can be restored using energy bound techniques~\cite{winter2016tight}, which provided the inspiration for the present approach.

Earlier studies of $S_M(\rho)$ avoided problems with these issues by restricting focus to measurements $M$ that explicitly include conserved quantities, like $S_{xE}$ \mbox{in~\cite{safranek2019a,safranek2019b,safranek2020classical}}. The generalization allows one to consider completely generic $M$ without pathologies. Ideally $M$ can model arbitrary experimental capabilities, and need not be tailored to the system.

\section{Physical examples}
\label{sec:examples}

We consider two numerical example systems, one quantum and one classical.

\begin{figure}[t!]
    \centering
    \stackinset{r}{41mm}{b}{8mm}{\includegraphics[height=21mm]{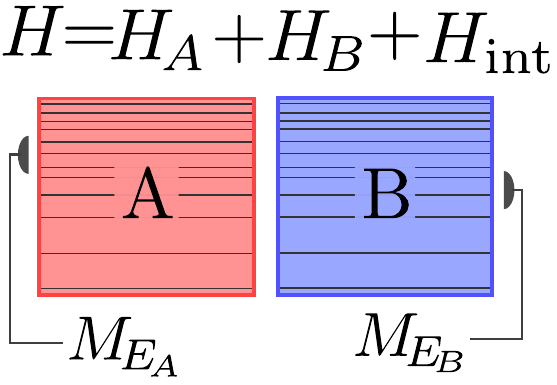}}{\includegraphics[width=\columnwidth]{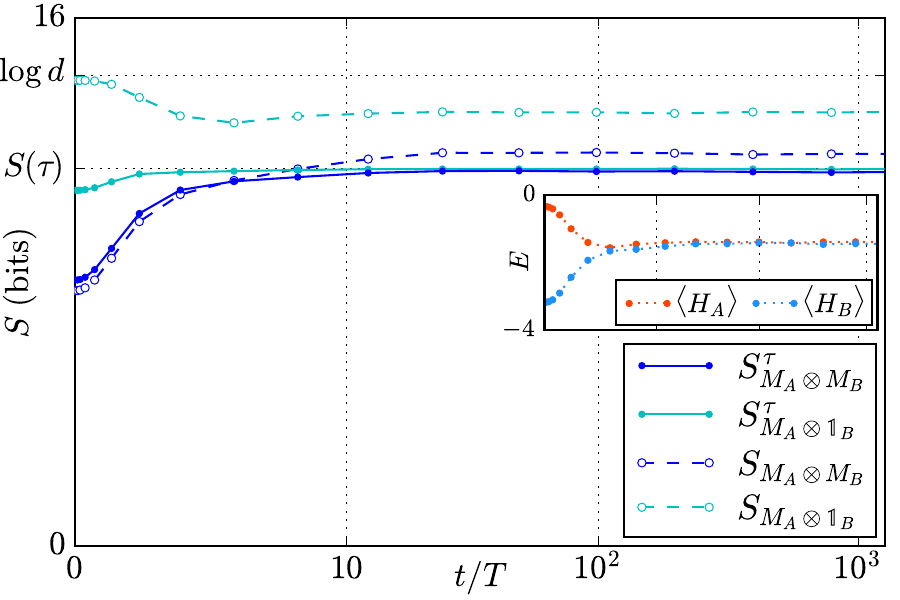}}
    \caption{Entropies during heat exchange in a quantum random matrix model. Depicted are $S_M^\tau(\rho)$ with \mbox{$\tau = e^{-\beta H}/Z$} (solid), traditional OE $S_M(\rho)$ (dashed), and energies (inset). $M$ measure coarse local energies, either jointly~(blue) or only in system $A$ (cyan). See text for details.}
    \label{fig:quantum-example}
\end{figure}

The quantum system, depicted in Fig.~\ref{fig:quantum-example}, is a random matrix model of heat exchange between two weakly interacting subsystems. The initial state is a product of locally thermal-like pure states with different temperatures in each subsystem. During the dynamics, initially hot/cold systems $A$/$B$ come to thermal equilibrium. For details of the model see \AppQuantum.

\begin{figure}[t!]
    \centering
    \includegraphics[width=\columnwidth]{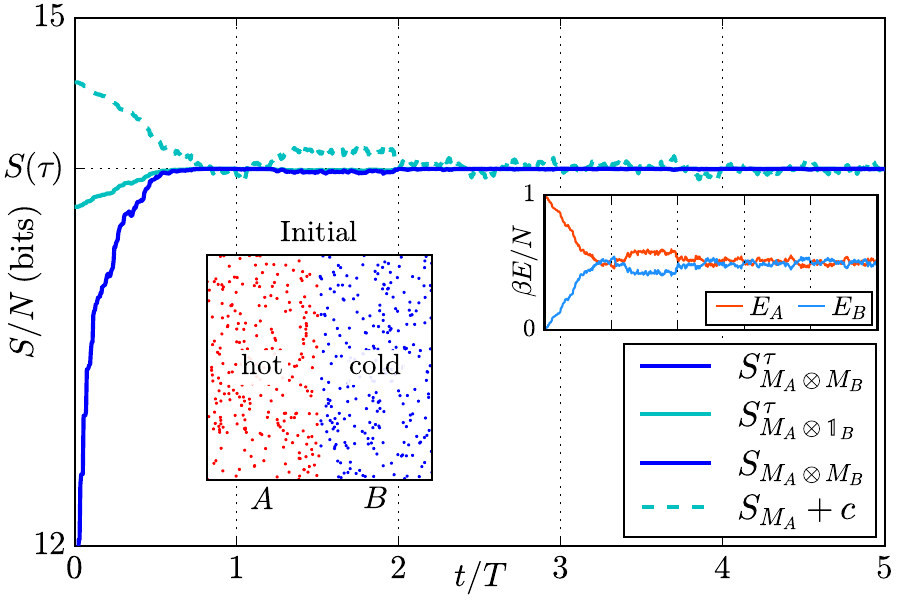}
    \caption{Classical analog of Fig.~\ref{fig:quantum-example}, in a 2d hard sphere gas. Systems A/B consist of the red/blue species of particles, initially separated and at different temperatures, which then intermix and equilibrate. Coarse measurements of subsystem energy $M_{E_A}$, $M_{E_B}$ capture heat exchange between the two systems. 
    Animations and code are available at~\cite{schindler2024animations}.}
    \label{fig:ideal-gas-1}
    \label{fig:classical-example}
\end{figure}

\begin{figure*}[t]
    \centerline{
    \begin{tabular}{llll}
         (a) IC 1 & (b) IC 2 & (c) IC 3 & (d) IC 4
         \\[2pt]
         \includegraphics[width=.25\textwidth]{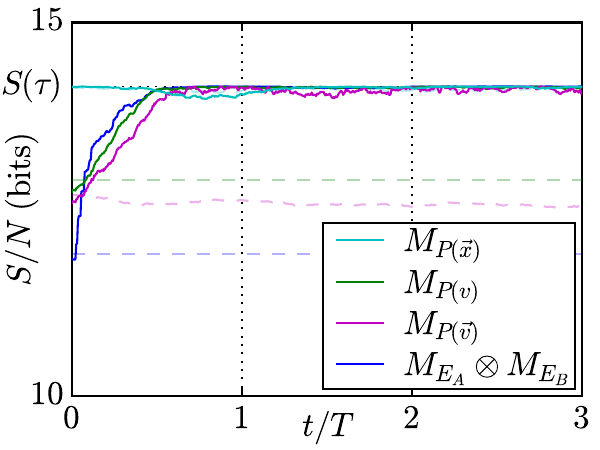}
         &
         \includegraphics[width=.25\textwidth]{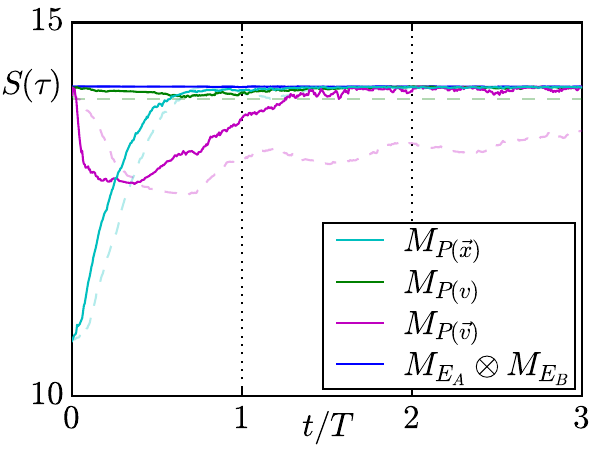}
         &
         \includegraphics[width=.25\textwidth]{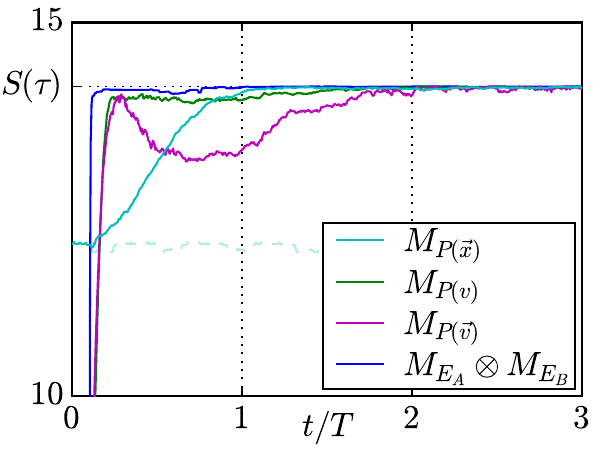}
         &
         \includegraphics[width=.25\textwidth]{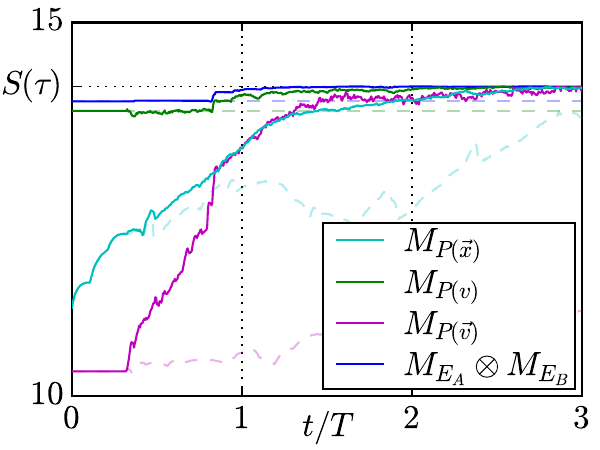}
         \\[2pt]
         \begin{tabular}{ll}
             \includegraphics[width=.125\textwidth, trim={0 2in 2in 0}, clip]{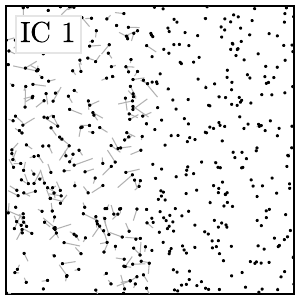}
             &
         \includegraphics[width=.125\textwidth, trim={0 2in 2in 0}, clip]{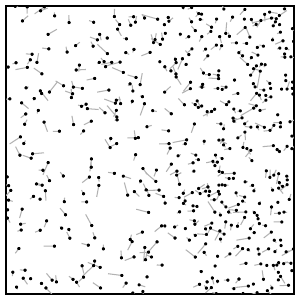}
         \\
         $t/T=0$ & $t/T=0.5$
         \end{tabular}
         &
         \begin{tabular}{ll}
             \includegraphics[width=.125\textwidth, trim={2in 2in 0 0 }, clip]{figs/006y/fig5a.pdf}
             &
         \includegraphics[width=.125\textwidth, trim={ 2in 2in 0 0}, clip]{figs/006y/fig5b.pdf}
         \\
         $t/T=0$ & $t/T=0.5$
         \end{tabular}
         &
         \begin{tabular}{ll}
             \includegraphics[width=.125\textwidth, trim={0 0 2in 2in }, clip]{figs/006y/fig5a.pdf}
             &
         \includegraphics[width=.125\textwidth, trim={ 0 0 2in 2in }, clip]{figs/006y/fig5b.pdf}
         \\
         $t/T=0$ & $t/T=0.5$
         \end{tabular}
         &
         \begin{tabular}{ll}
             \includegraphics[width=.125\textwidth, trim={ 2in 0 0 2in }, clip]{figs/006y/fig5a.pdf}
             &
         \includegraphics[width=.125\textwidth, trim={ 2in 0 0 2in }, clip]{figs/006y/fig5b.pdf}
         \\
         $t/T=0$ & $t/T=0.5$
         \end{tabular}
         \\[4pt]
    \end{tabular}
    }
    \caption{Dynamics of the entropy $S^\tau_M(\rho)$ for several different $M$'s and IC's (initial conditions) of a 2d hard sphere gas. The prior $\tau \propto e^{-\beta H}$ is the canonical state~\eqref{eqn:tau-ideal-gas} corresponding to the energy constraint $\braket{H}=E$, so the maximum value $S(\tau)$ is the canonical ideal gas entropy. Each IC is low entropy in a different way. This is reflected by various $M$ that each capture entropy of different degrees of freedom. In particular shown are $M_{P(\vec{x})}$~(spatial distribution),  $M_{P(v)}$~(speed distribution), $M_{P(\vec{v})}$~(velocity distribution), and $M_{E_A} \otimes M_{E_B}$ (division of energy between two subsystems, with systems $A$/$B$ as defined by Fig.~\ref{fig:ideal-gas-extra}). Light dashed lines indicate the free case (interactions turned off) for similar ICs. Animations and code are available at~\cite{schindler2024animations}.}
    \label{fig:ideal-gas-dynamics}
\end{figure*}

The classical system, depicted in Figs.~\ref{fig:classical-example}--\ref{fig:ideal-gas-dynamics} (see also Fig.~\ref{fig:ideal-gas-extra}), is a two dimensional hard sphere gas. Various initial conditions and measurements are studied in the figures. We also study the free case with interactions turned off. For more details see \AppClassical.

In both systems we consider prior $\tau=e^{-\beta H}/Z$ corresponding to total energy conservation $\braket{H}=E$. In Figs.~\ref{fig:quantum-example}--\ref{fig:classical-example}, we study coarse measurements of local energy in each subsystem, capturing heat exchange between two subsystems. In Fig.~\ref{fig:ideal-gas-dynamics}, we study several different $M$, with $M_{P(\bullet)}$ capturing the spatial, speed, and velocity distributions among particles, and $M_{E_{A,B}}$ again being coarse local energy measurements.

We now briefly elaborate on the definitions of these $M$ and how the entropies are calculated. We then proceed to explain various observations and remarks about the numerical results.

\subsubsection{Definitions of $M$ and calculations}

The first class of $M$ in the examples are coarse energy measurements.
These are defined as collections $M_E = (\Pi_E)_E$ of projectors $\Pi_E$, labeled by a discrete set of energies $E$, which project onto energy windows of interval $E \pm \Delta E/2$.
In the quantum case, these projectors are defined by $\Pi_{E} = \sum_{E_k \in E \pm \Delta E/2} \ketbra{k}$ in terms of the Hamiltonian eigenbasis $\ket{k}$ and spectrum $E_k$. In the classical case, $\Pi_E = 1$ on the part of phase space with energy in the window $E \pm \Delta E/2$, and $\Pi_E=0$ otherwise.

These definitions were applied to the local Hamiltonians $H_A,H_B$. Then either one-sided, $M_{E_A} \otimes \one_B =$ $ (\Pi_{E_A} \otimes \one_B)_{E_A}$, or joint $M_{E_A} \otimes M_{E_B} =$ $ (\Pi_{E_A} \otimes \Pi_{E_B})_{E_A,E_B}$ measurements were formed from the local projectors.
Entropy was evaluated in the quantum case by direct numerical evaluation of probabilities and volumes in \eqref{eqn:oe-volume-form}. For the classical case, the volumes can be approximated analytically, and entropy was evaluated by \eqref{eqn:numerical-oe-thermo-joint} and \eqref{eqn:numerical-oe-thermo-single} in \AppClassical, with system $A$ as defined in Fig.~\ref{fig:ideal-gas-extra}. Widths~$\Delta E$ are given in the Appendix.

A second class of $M$ appear only in the classical examples, measurements $M_{P(\bullet)}$ that measure the distributions of single-particle properties.
These are defined by first specifying any measurement $M^{(1)}$ that acts on a single particle, with discrete outcomes labeled by $j$. This $M^{(1)}$ is performed on each particle individually, and one collects the observed probability $P(j)$ of each outcome $j$. These probabilities are then themselves discretized, into bins of width $\delta P$, defining the macrostates. Thus, each macrostate corresponds to a (discretized) distribution $P(j)$ of the values $j$ of a single-particle property. The typical example is $M_{P(\vec{x})}$ where $M^{(1)}$ measures ``which spatial bin'', and $P(\vec{x})$ is the fraction of particles in each bin.
Entropies for $M_{P(\bullet)}$ are evaluated by \eqref{eqn:sanov-general} in \AppClassical, which employs a trick based on Sanov's theorem~\cite{cover2006book} to approximate the volumes. This approximation becomes exact as the number of particles $N \to \infty$. 

Additional details of both classes of $M$ are given in \AppClassical \ and \AppQuantum, with illustrations in Fig.~\ref{fig:ideal-gas-extra}c.

\subsubsection{Discussion of the results}

The examples in Figs.~\ref{fig:quantum-example}--\ref{fig:ideal-gas-dynamics} illustrate various aspects of the theory discussed earlier. We highlight the following observations:
\begin{itemize}

    \item Different $M$ reflect different ways a state may have low entropy---for instance, in Fig.~\ref{fig:ideal-gas-dynamics}b, the initial state has high entropy in terms of the distribution of velocity among particles, but low entropy in terms of configuration. Any $M$ can be considered, defining a valid entropy. Which one is physically relevant depends on the problem under consideration.
    
    \item All entropies $S_M^\tau(\rho)$ in the non-integrable systems are seen to equilibrate to their maximum value~$S(\tau)$, which here is the canonical entropy. In the classical case, this canonical entropy is the Sackur-Tetrode ideal gas entropy~\cite{huang1987statistical}.

    \item Figs.~\ref{fig:quantum-example}--\ref{fig:classical-example} capture the type of entropy increase that occurs when two systems exchange heat, with thermodynamic entropy identified with the energy measurement $M_{E_A} \otimes M_{E_B}$, as also in (\ref{eqn:thermo-eq}--\ref{eqn:thermo}). In both examples, this thermodynamic entropy increases and maximizes while the energies equilibrate. 

    \item For $M_{E_A} \otimes M_{E_B}$ acting on weakly coupled systems, the OE $S_M^\tau(\rho)$ with canonical $\tau$, and the traditional OE $S_M(\rho)$, are equal to first order (usually differing by sub-extensive terms). In Fig.~\ref{fig:classical-example}, they are only calculated to the order where they are equal. In Fig.~\ref{fig:quantum-example}, they are calculated exactly, and can be seen to agree up to lower order effects. 

    \item Meanwhile, looking at  $M_{E_A} \otimes \one_B$ (solid vs. dashed cyan curves), Figs.~\ref{fig:quantum-example}--\ref{fig:classical-example} both demonstrate the necessity of generalized volumes for entropy increase as discussed at length earlier. While $S_M^\tau(\rho)$ with canonical $\tau$ is seen to increase in all examples, this is an $M$ for which $S_M(\rho)$ generically dynamically decreases during equilibration, and equilibrates far from its maximum of $\log d$. This occurs because this measurement of ``energy on the hot side'' is an example of the type depicted in Fig.~\ref{fig:conserved-quantities}. Note that no microcanonical energy constraint is possible in this quantum case, since the state overlaps all energy subspaces, leaving a canonical prior as the most obvious way to account for energy conservation.

    \item There is a striking similarity between Figs.~\ref{fig:quantum-example}~and~\ref{fig:classical-example}, even though one is a classical system always in a definite macrostate, and one is a pure quantum state always overlapping many macrostates. From the coarse-grained statistical mechanics perspective, they both simply look like weakly coupled systems exchanging heat (but the quantitative similarity of the axes is only a coincidence; the classical case is entropy per particle).

    \item In Fig.~\ref{fig:classical-example}, one will note a surprising transient bounce in the energies. This bounce is also evident in the entropies when more zoomed in. The effect is caused by the cold side effectively colliding first with the hot side, then the wall, then the hot side again returning some energy, before the particles fully intermix, as visible in animations~\cite{schindler2024animations}.

    \item In Fig.~\ref{fig:ideal-gas-dynamics}, each example starts in a different type of initial low entropy state. Eventually, each one evolves to a state that has high entropy with respect to every $M$ considered. This parallels the fact that at late times, one cannot tell which IC each final state came from without access to detailed microscopic information. In this way, the final states appear to be ``high entropy'' states in a very generic sense (even though the microscopic Gibbs/von Neumann entropy $S(\rho)$ is invariant). 

    \item Reassuringly, entropies for $M_{P(\bullet)}$ and $M_{E_A} \otimes M_{E_B}$ are all quantitatively on the same scale and are mutually intelligible, even though these quantities are evaluated using quite unrelated methods.

    \item Comparing the interacting and free cases (solid vs. dashed curves) in Fig.~\ref{fig:ideal-gas-dynamics} helps clarify the role of integrability. For the highly chaotic interacting case, entropy increases and maximizes for all $M$ considered. In the free case, entropy increase is observed for some $M$ but not others, depending on the IC. This suggests that, while one should expect generic entropy increase for chaotic systems, for integrable ones it may be highly measurement and state dependent. In the latter case, if entropy increase does occur, it occurs much slower and with larger fluctuations. This conclusion is somewhat different than what has been recently suggested in the literature~\cite{chakraborti2022entropy,garrido2024time}. Note that the IC's we consider, while atypical in the sense that they start low entropy, are still fairly generic, and the results are reasonably stable under perturbing the IC's.

    \item In Fig.~\ref{fig:ideal-gas-dynamics}(b,c) one observes a transient decrease in the entropy associated with velocity degrees of freedom. This occurs as the particles with initially down/left velocities bounce out of the lower left corner, temporarily aligning many velocities before the later equilibrium is reached, as visible in animations~\cite{schindler2024animations}.
    
    \item In the classical case, since only Boltzmann terms appear (see discussion below), the calculation of entropies differs from traditional methods only by the use of effective volumes $V_x$ corresponding to canonical $\tau$. The result is that OE increases to the maximum value $S(\tau)$ for generic coarse $M$, unlike what occurs for traditional volumes $W_x$.

\end{itemize}

\noindent These observations clarify the main aspects of the theory and demonstrate some of the flexibility of the OE approach.

Turning to more speculative matters, these observations suggest a somewhat observer-dependent picture of ``high entropy'' vs. ``low entropy'' states. About this we remark the following:
\begin{itemize}

    \item A ``high entropy'' state can be viewed as one which has nearly maximal entropy with respect to essentially any coarse $M$ (given its constraints are accounted for by $\tau$). Conversely, a state may be ``low entropy'' if some reasonable $M$ reveals a low entropy. Conceived this way, ``essentially any'' and ``reasonable'' are observer-dependent---this reflects the fact that the ability to experimentally implement some $M$ revealing low entropy allows the possibility to extract resources from the system, by manipulating that degree of freedom (\eg~as in Jaynes' Whifnium example~\cite{jaynes1992gibbs}). 
    
    \item Despite this observer dependence, final states in Fig.~\ref{fig:ideal-gas-dynamics} do appear to be high entropy in a non-arbitrary sense---even though there certainly exists some $M$ that does reveal a low entropy. However, such $M$ are special and may be difficult to realize physically (\eg~one might imagine $M_x=\ketbra{\psi}$ projecting on the actual microstate of the system). Roughly speaking, any not-specially-chosen coarse $M$ would yield a high entropy.

    \item The entropies appearing in the famous H-theorems, viewed as instances of OE, are closely tied to the question of how to make this picture more precise, \ie~how ``high entropy'' can be defined in an observer-independent way. This is visible in Gibbs' H-theorem, where his phase cell measurement can be cast (with the freedom to choose cell sizes appropriately) as the finest possible $M$ that cannot resolve $\hbar$ scale details in phase space. Therefore, due to coarser/finer monotonicity, high OE under Gibbs' $M$ is sufficient to ensure high OE under any physically reasonable $M$ in classical systems. Similarly, Boltzmann's $M$ (the one we call $M_{P(\vec{x},\vec{v})}$ as in \eqref{eqn:empirical-boltzmann}, also recently studied in~\cite{chakraborti2022entropy,garrido2024time}) can be cast as the finest possible $M$ that measures distributions of single-particle properties, \ie~that does not distinguish between the individual particles. Again, by coarse/finer monotonicity, high OE for Boltzmann's $M$ ensures high OE for any $M$ in the $M_{P(\bullet)}$ class. Meanwhile, the construction of ``quantum phase cell'' and ``macroscopic'' measurements used by von Neumann in his quantum H-theorems appear to be based on precisely this kind of reasoning~\cite{vonNeumann1929proof}. Recasting these entropies in the OE forms (which changes signs and constants relative to original forms) clarifies that, beyond having intuitive conceptual meanings, these $M$ actually quantitatively correspond to minimum entropies over physically motivated classes of coarse-grainings.
    
\end{itemize}

From this discussion it is evident that defining the notion of ``high entropy states'' is subtle. Clearly, if the inherent informational entropy $S(\rho)$ of a state is large, then measured entropies are necessarily also large for every $M$. But this is not the only way to have a \mbox{high entropy state}; pure states and classical systems also appear to become more generic during equilibration. We have seen that this occurs in an $M$-independent sense that seems to be evident empirically, but will require additional future clarification to state precisely.

\subsubsection{Further remarks about the classical case}

The classical cases studied here are pure states (definite phase point, no mixtures) with projective measurements. Thus, only one macrostate is occupied at a time, and the entropy reduces to a Boltzmann entropy $\log V_x$ with generalized volume $V_x$. In this case, dynamical maximization of OE is less robust compared to mixed or quantum systems, both because of the inequality $S_M^\tau(\rho) \leq \log V_{\rm max} < S(\tau)$, and because the Boltzmann and Shannon terms cannot ``conspire'' towards maximization as discussed earlier. 

In light of this $V_{\rm max}$ bound, $S(\tau)$ can only be approached if one macrostate is overwhelmingly more likely than the others---which is indeed the case for the coarse $M$'s considered here. One may observe that in the plots, the maximum is at certain times even obtained exactly. This is possible because we calculate entropy per particle with an approximation assuming the $N \to \infty$ limit. For mixed classical states (or any quantum states), entropy maximization will be generic for a much larger class of $M$, including $M$ where all macrostates have small and/or equal volumes.

Despite these differences from the more general case, OE still has a tendency towards maximization in the classical case for the usual reason of Boltzmann entropies, so long as volumes are computed from appropriate $\tau$, as seen in the figures. And indeed, despite reducing to Boltzmann terms, pure classical cases form an important part of the OE framework.

\section{Conclusions}
\label{sec:conclusion}

In the present paper, we introduced the definition $S_M^\tau(\rho)$ as a framework for defining coarse-grained entropies in physical systems. 

One benefit of this framework is its dual interpretation in terms of information theory and statistical \mbox{mechanics---it} was derived first in terms of informational quantities like relative entropies and Bayesian priors, but shown to have equally valid meaning in terms of macrostate volumes and the outcomes of observables.

Another is that the new generalization provides a simple framework where many entropies can be unified, and was shown to connect to nearly all entropies commonly used in physics---including in classical/quantum, open/isolated, pure/mixed, and non-/equilibrium contexts. In doing so, it helps clarify the relation of many entropies to each other, and helps disambiguate the roles of microstate (the system state $\rho$), macrostate (an arbitrary, often coarse, property $M$ of the system), and ensemble (the $\tau$ w.r.t.~to which microstates are \mbox{``\textit{a priori} equally likely''}), in statistical mechanics.

Of particular interest was the fact that it includes as special cases the entropies from the historically important H-theorems of Boltzmann, Gibbs, and von Neumann, with transparent relations to Jaynes' maximum entropy and von Neumann's informational entropy, as well as to numerous more modern quantities like the entanglement entropy and quantum entropy production.

Beyond conceptual purposes, the new definition was shown to be useful in obtaining technical results relating to the second law of thermodynamics---it was shown that in terms of $S^\tau_M(\rho)$, one can obtain strong and general entropy increase theorems, and that these theorems can in turn be seen to imply physical thermodynamic laws. To do so we leveraged powerful results from various parts of the equilibration literature, including theories of microstate equilibration in open systems, and equilibration of expectation values in isolated systems, finding that these take a natural expression within the framework.

To demonstrate the theory, we investigated entropic dynamics in classical and quantum physical examples from a unified point of view. In these examples we found that, by virtue of its flexibility, the framework allowed us to formalize intuitive naive statements about entropies that can otherwise be difficult to state precisely.

We often restricted the use of the definition $S_M^\tau(\rho)$ to states $\rho$ obeying the constraint associated with $\tau$, namely $S(\rho;\tau) \leq S(\tau)$; the necessity of $\tau$ to be a ``valid'' prior played a central role in motivating the informational form of the  definition. However, from a mathematical standpoint, this can be relaxed to apply the definition to arbitrary $\rho,\tau$ without any catastrophic problems. The main technical loss is the lower bound $S_M^\tau(\rho) \geq S(\rho)$ (see~\eqref{eqn:lower-bound}) relating OE to the von Neumann entropy. Conceptually speaking, the relaxation does somewhat disrupt the informational motivation---however, as discussed below~\eqref{eqn:lower-bound}, this disruption also has an intuitive explanation. Meanwhile, the statistical mechanical interpretation in terms of macrostate volumes, \textit{\`a la} (\ref{eqn:oe-volume-form}--\ref{eqn:effective-volumes}), is not affected. Overall, if care is taken not to use any inapplicable results (throughout the work the constraint assumption is explicitly stated when it is used), there is no real harm in allowing $\tau$ to represent a generic prior guess---where constraints may be unknown---of the system state.

One question that often arises in relation to physical entropies is whether the entropy amount is an objective or a subjective quantity. In the present context it is an objective quantity, but one which takes a measurement~$M$ and a prior~$\tau$ as arguments. The values of these arguments might be identified with the subjective state of some observer, but need not be. What \textit{is} subjective is the question of which entropy (\ie~which $M,\tau$) is relevant in a given context; but this is entirely dependent on the physical questions that one is analyzing. The types of second laws derived above do not state that ``the'' entropy increases, but rather that entropy increase is generic for many $M$ and appropriate $\tau$, so that in many cases if some entropy is relevant it will obey a second law.

We primarily considered two types of systems in this work: isolated systems evolving under a fixed time-independent Hamiltonian, and open systems that are subsystems of an isolated one. Untouched so far were the case of time-dependent/driven Hamiltonians and thermodynamic cycles, whose treatment would be a useful continued step in extending the framework.

It is useful to mention connections of the present work to several other recent studies that have appeared with related goals. One is a recently proposed principle of maximum observable entropy~\cite{scarpa2023observable}. This is similar in spirit to the statement that at equilibrium $D_M(\rho \rr \tau)\approx 0$ (\ie~$S_M^\tau(\rho)$ maximizes) for $\tau$ determined by the constraint of fixed average energy, with the main difference being that in that work a constrained maximization is performed after (instead of before, as here) evaluating outcome probabilities. We considered the question of when such equilibrium conditions hold in the section on second laws. We also note in that study that constrained maximization is performed over their quantity $\tilde{S}(A)$, which is equivalent to an observational entropy, raising further interesting connections. Another recent study proposed an entropy maximization principle based on Shannon entropy of observables~\cite{meier2024emergence}, with results closely related to \eqref{eqn:eq-rhobar} above, and questioned the motivation for including volume terms in the observational entropy. We remark that for the observable Shannon entropy alone, equilibration of expectation values may ensure the entropy has an equilibrium value, but does not ensure that it \textit{maximizes}. In that work, entropy \textit{increase} was only obtained by assuming a very strong past hypothesis, which excludes many physically realistic situations. The inclusion of volume terms in OE not only allows entropy increase/maximization to be proved, but also allows OE to reduce to thermodynamic entropies, and connects the statistical properties of entropy to physics such as heat transfer. Meanwhile, another recent study proposed a new definition of entropy production that similarly makes use of MaxEnt principles to describe a reference state~\cite{varizi2024entropy}. According to their definition, entropy production is given by the term that we called $D(\rho \rr \tau)$, which in our framework provides a bound on OE via \eqref{eqn:qre-bound-props}, and through that bound contributes to entropy production. Last here we highlight the work of Bai \etal~in~\cite{bai2023observational}, which first raised the question of including prior information in the observational entropy definition, arriving at definitions coinciding with the one here in the case of equality constraints.

Numerous additional paths remain for future study.
One important branch is continuing the extension of second laws into the realm of generalized thermodynamics, where consequences of \eqref{eqn:eq-eps} (\ie~analogous to \eqref{eqn:thermo}) in strongly coupled systems, generic quantum systems where~\eqref{eqn:generalized-thermo} holds, and for general $M$, $\tau$, will be of special interest. Equally important is the task of proving inequalities like \eqref{eqn:eq-eps} both in particular systems and under minimal general assumptions. This includes searching for bounds like (\ref{eqn:eq-rhobar}--\ref{eqn:eq-eigenstate}) in new regimes, and also, for instance, identifying assumptions on the Hamiltonian and $M$ that make $D_M(\rhobar \rr e^{-\beta H}/Z)$ (or generalizations) small. With these improvements one hopes to prove second laws from first principles in a wide class of systems.

Another branch is the pursuit of translations where specialized results can be brought into the present more general context, which can provide new insights. Typicality theorems \textit{\`a la}~\cite{GemmerMichelMahlerBook2004, popescu2006entanglement, goldstein2005canonical, muller2011concentration, nagasawa2024generic, strasberg2024typical}, equilibration theorems such as~\cite{reimann2008foundation,linden2009evolution,short2011equilibration,gogolin2016equilibration, WilmingEtAlBook2018}, and fluctuation theorems~\cite{EspositoHarbolaMukamelRMP2009, campisi2011fluctuation, strasberg2022book, PottsArXiv2024}, will be particularly interesting cases, as well as the treatment of driven thermodynamic cycles. Also on the theoretical side, one would like to further investigate and formalize the notions of high and low entropy states from the OE point of view.

In terms of applications, deeper studies of entropic dynamics in other quantum systems will be of interest, especially beyond the case of simple energy measurements and constraints. In particular, this work opens the door to detailed studies of coarse-grained entropy in infinite-dimensional systems such as field theories and gravity, where improved disambiguation among entropy concepts may help clarify subtle aspects of horizon~\cite{jacobson2003horizon,bousso1999covariant} and black hole~\cite{bekenstein1981universal,wallace2019case,schindler2021unitarity,carlip2014blackhole} entropies.

\bigskip
\begin{acknowledgments}
 
The authors thank F.~Buscemi, G.~Gasbarri, and F.~Meier for useful discussions.
J.S. acknowledges support through a Beatriu de Pinós fellowship of the Catalan Agency for Management of University and Research Grants (AGAUR) and EU Horizon 2020 MSCA grant 801370. 
J.S., N.G., P.S. and A.W. acknowledge support by the Spanish MICIN (project PID2022-141283NB-I00) with the support of FEDER funds, by the Spanish MICIN with funding from EU NextGenerationEU (PRTR-C17.I1) and the Generalitat de Catalunya, and by the Spanish MTDFP through the QUANTUM ENIA project: Quantum Spain, funded by the EU NextGenerationEU within the framework of the `Digital Spain 2026 Agenda'.
M.G.J. acknowledges support by the Belgian Fund for Scientific Research (F.R.S.--FNRS).
P.S. and A.W. acknowledge support by the European Commission QuantERA grant ExTRaQT (Spanish MICIN project PCI2022-132965). 
P.S. acknowledges support from ``La Caixa'' Foundation (ID 100010434, fellowship code LCF/BQ/PR21/11840014) and from the Ram\'on y Cajal program RYC2022-035908-I.
A.W. furthermore acknowledges support by the Alexander von Humboldt Foundation, and by the Institute for Advanced Study of the Technical University Munich. 
\end{acknowledgments}

\bibliography{biblio}

\begin{thebibliography}{126}
\bibitem [{bus()}]{buscemi2024note}%
  \BibitemOpen
  \href@noop {} {}\bibinfo {note} {{T}his identity was first brought to the
  attention of the authors by Francesco Buscemi.}\BibitemShut {Stop}%
\end{thebibliography}

\clearpage
\appendix

\section{Classical hard sphere gas}
\label{app:classical-hard-sphere-gas}

\begin{figure*}[t]
    \centerline{
    \begin{tabular}{lll}
         (a) $t/T = 0$ & (b) $t/T = 0.5$  & (c) Four coarse-grainings.
         \\
         \includegraphics[width=.3\textwidth]{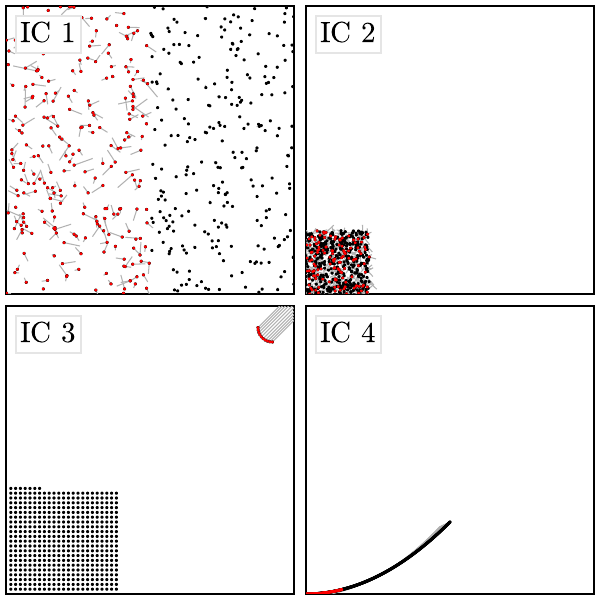}
         &
         \includegraphics[width=.3\textwidth]{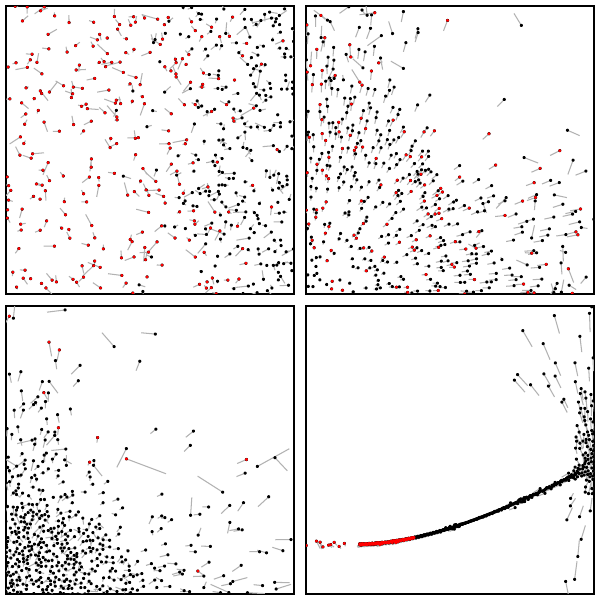}
         &
         \includegraphics[width=.3\textwidth]{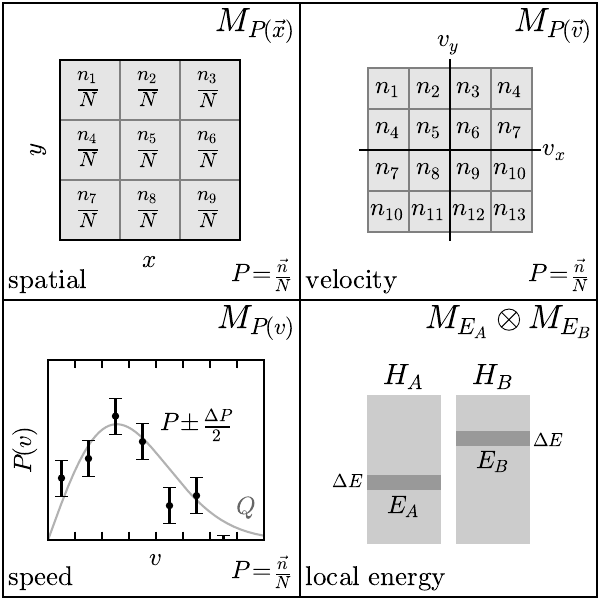}
    \end{tabular}
    }
    \caption{(a,b) More detailed view of the IC's from Fig.~\ref{fig:ideal-gas-dynamics}, with system $A$ marked in red. (c) Illustration of the $M$'s.}
    \label{fig:ideal-gas-extra}
\end{figure*}

As an elementary example of OE dynamics, we consider various coarse-grainings applied to a classical ideal gas with hard sphere interactions. Results appear in Figs.~\ref{fig:classical-example}--\ref{fig:ideal-gas-extra}.

The system is an $N=500$ particle gas in $d=2$ dimensions, with Hamiltonian $H = \sum_{k = 1}^{Nd} H_k = \sum_k p_k^2/2m$ and evolving via hard sphere collisions. Simulation parameters are roughly based on a room temperature nitrogen gas, with hard sphere radius $r_0 = 100$ pm, box size $L/r_0 = 200$, mass $m = 30$ GeV/$c^2$, temperature $\beta^{-1}=25$ meV ($\sim$ 300K), box-crossing timescale $T\approx 52$~ps, and thermal wavelength $\lamtherm \approx 18$ pm. Normalization is w.r.t.~$\tr = \int \frac{dx^{Nd}dp^{Nd}}{(l_0 p_0)^{Nd}}$, where the free integration measure is fixed to $(l_0 p_0)^{Nd} = (2 \pi \hbar)^{Nd} N!$ (but sometimes left free for clarity) by CM/QM limit.

To fix the prior we assume the constraint of average energy conservation $\braket{H}=E$ and that all particles remain inside the box. Letting $\Theta$ be the spatial projector onto $0 \leq x \leq L$, and $Z = \sqrt{2 \pi m \beta^{-1}}$, with $\beta^{-1} = 2E/Nd$, this implies
\begin{equation}
\label{eqn:tau-ideal-gas}
    \tau = \prod_{k=1}^{Nd} \frac{\Theta(x_k)}{L/l_0} \, \frac{e^{-\beta H_k}}{Z/p_0}.
\end{equation}
The prior entropy evaluates to
\begin{equation}
\label{eqn:Stau-ideal-gas}
    e^{S(\tau)} = \bigg[\tfrac{L}{l_0} \sqrt{\tfrac{E/N}{E_0}} \sqrt{\tfrac{2 \pi e}{d}} \bigg]^{Nd} \!\!\! \approx  \bigg[\tfrac{(L/N^{1/d})}{\lamtherm} \sqrt{e} \bigg]^{Nd}
\end{equation}
where $\lamtherm = \sqrt{2 \pi \hbar^2 \beta/m}$ and $E_0 = p_0^2/2m$, and the $\approx$ is Stirling's approximation, which gives the Sackur-Tetrode ideal gas entropy~\cite{huang1987statistical}. 

We consider projective measurements, meaning $M$ can be thought of as a partition of phase space into subsets (``macrostates'') each labelled by $j$, with $\Pi_j$ phase space functions projecting onto each subspace. $M=(\Pi_j)_j$ is formally the indexed collection of such projectors. The probability for $\rho$ to be in macrostate $j$ is $p_j = \tr\rho \Pi_j$, and prior probabilities of each macrostate are given by $q_j = \tr\tau\Pi_j$, with effective volumes $V_j = q_j \, e^{S(\tau)}$.  

We consider a pure system with an exactly known phase point, so only one $p_j$ is nonzero at any given time, and OE can be evaluated by $S_M^\tau(\rho) = \log V_j(t)$ at each time. Thus in order to calculate entropy one simply needs the volumes $V_j$ associated with each macrostate for the relevant measurements. 

With only Boltzmann terms appearing, this differs from traditional methods only by the appearance of effective volumes $V_j = q_j \, e^{S(\tau)}$ instead of the traditional $W_j = \tr\Pi_j$. This modification ensures the entropy has a tendency to increase for generic $M$, as discussed throughout the text.

Several types of $M$ appear in Figs. \ref{fig:classical-example}--\ref{fig:ideal-gas-extra}. These have been roughly described in the main text. We now give formal definitions here, and derive the formulas by which the entropies are calculated.

\subsubsection{Empirical frequency coarse-grainings.}

The class of measurements denoted $M_{P(\bullet)}$ describe the observed distribution (or ``empirical frequency'') of some single-particle property in a many particle system.

To define such a measurement, one first chooses any single-particle measurement $M^{(1)}$ that will be applied to each particle. We denote this by $M^{(1)} = (\tilde{\Pi}_j)_j$, a collection of single-particle projectors (denoted with tilde) whose possible outcomes are are labeled by index $j$. 

For example, in particular we consider $M^{(1)} = M_{\vec{x}}$ which measures the position $\vec{x}$ within bins of size $\Delta x$. In this case, the projectors $\tilde{\Pi}_{\vec{x}}$ are functions on single-particle phase space (four-dimensional $x,y,p_x,p_y$), labeled by a discrete set of vectors $\vec{x}$ corresponding to the center of each spatial bin. Each $\tilde{\Pi}_{\vec{x}}=1$ for the parts of phase space in the bin $(x,y) \in \vec{x} \pm \Delta x/2$, and 0 otherwise. This effectively splits the box into a grid of spatial subregions as in Fig.~\ref{fig:ideal-gas-extra}c. Similar constructions of $M^{(1)}$ are used for velocity $M_{\vec{v}}$ and speed $M_v$.

Next, one wishes to mathematically describes the measurement $M_{P(j)}$ that describes applying $M^{(1)}$ to each particle and counting the fraction $P(j)$ of particles obtaining the outcome $j$. In the above example, this counts the fraction $P(\vec{x})$ of particles in each spatial bin. This fraction will also be binned discretely, to within a width~$\Delta P$, obtaining a discrete set of macrostates.

Thus, denote $M_{P(j)} = (\Pi_{P(j)})_{P(j)}$, which is a collection of projectors on the full phase space. The label $P(j)$ is a vector of probability values indexed by $j$, and $\Pi_{P(j)}$ projects on the part of phase space where, for every $j$, acting $M^{(1)}$ leads to the fraction $P(j) \pm \Delta P/2$ of outcomes. For example, if $M^{(1)}$ has three outcomes and $\Delta P = 0.1$, then one of the $P(j)$ macrostates would be labeled by $P(j) = (.05, .65, .25)$, \ie~a possible configuration of a histogram. A state is in this macrostate if the number of particles $(n_1,n_2,n_3)$ obtaining each $j$, converted to a fraction, gives the vector $P(j)$ within $\pm \Delta P/2$.

Expressed more formally, this means each projector $\Pi_{P(j)} = \sum_{\vec{j}\in P(j)} \tilde{\Pi}_{j_1} \otimes \ldots \otimes \tilde{\Pi}_{j_N}$, where $\vec{j}\in P(j)$ indicates that sum is taken over all strings of $j_1, \ldots, j_N$ such that the number of times each $j$ value appears is consistent with the fraction $P(j) \pm \Delta P/2$.

To evaluate entropy one needs to calculate the probabilities $q_{P(j)} = \tr \Pi_{P(j)} \tau$ that determine the volumes. In general this can be difficult to calculate. However, in the special case that $\tau$ is IID (independent and identically distributed) with respect to property $M^{(1)}$, it is possible to approximate them analytically using a trick based on Sanov's theorem~\cite{cover2006book}.

By IID, we mean that acting $M^{(1)}$ on any one-particle reduced state $\tau_{0}$ of $\tau$ gives the same outcome distribution $Q_j = \tr \tilde{\Pi}_j \tau_{0}$, and that applying $M^{(1)}$ to all particles jointly leads to $tr(\Pi_{j_1} \otimes \ldots \otimes \Pi_{j_N} \tau ) = Q_{j_1} \ldots Q_{j_N}$, the IID product of this same $Q_j$ distribution. (It is sufficient but unnecessary that $\tau = \otimes^N \tau_0$.) For our example $M$'s, this holds for the canonical but not microcanonical $\tau$.

Sanov's theorem concerns the probability of drawing atypical strings of IID outcomes. Suppose one draws $n$ samples each IID from $Q$, yielding a string $j_1,\ldots, j_n$. Denote by $P$ the empirical sample distribution obtained from counting outcomes in the string. Let $\mathcal{P}$ denote a closed set of probability distributions over the same index as $P,Q$, and denote by $q(\mathcal{P})$ the probability that the sample distribution $P$ lies in this set. Sanov's theorem states that~\cite{cover2006book}
\begin{equation}
    \lim_{n\to \infty} \tfrac{1}{n} \, \log q(\mathcal{P}) = -D(P^* \rr Q),
\end{equation}
where $P^* = \argmin_{P' \in \mathcal{P}} D(P' \rr Q)$ is the dominant element of the set in the asymptotic limit.

If $\tau$ is IID in the sense above, then $q_{P(j)}$ has precisely the character of this theorem. In particular, $\mathcal{P}$ is the set of distributions in the range $P(j) \pm \Delta P/2$, and $q_{P(j)}$ is the probability that $n=N$ samples drawn IID from $Q_j$ lie in this set.

To approximate $q_{P(j)}$ using the theorem, we must estimate $P^*$ in the set. This is somewhat tricky practically, because if one allows distributions that do not have the same energy as the actual state, or are not properly normalized, the wrong values are obtained. In other words, $\mathcal{P}$ must really also be restricted to those distributions that are physically possible for the system.

In order to make an estimation of $P^*$ respecting all physical constraints, we note that the actual exact observed distribution $P^{**}$ and reference distribution $Q$ must both be physically admissible. Therefore, we convexly mix $P^{**}$ with $Q$ by the largest amount allowed by the bin width $\Delta P$ to obtain an estimate of $P^*$. This brings $P^*$ closer to the reference $Q$ while conserving any constraints linear in the distributions.

Putting this all together, suppose $\rho$ is a state whose observed empirical distribution is $P^{**}$, which is in the macrostate defined by $P(j) \pm \Delta P/2$. Estimate $P^*$ by convexly mixing the observed distribution with $Q$ as much as allowed by the macrostate bin. Then take the estimate $\log q_{P(j)} = - n D(P^* \rr Q)$ from Sanov's theorem, which becomes exact in the $n \to \infty$ limit. Since only a single macrostate is occupied, in general one finds $S_M^\tau(\rho) = S(\tau) + \log q_{P(j)}$. Therefore, one arrives at
\begin{equation}
\label{eqn:sanov-general}
    S^\tau_{M_{P(j)}}(\rho) \approx S(\tau) - n \,  D(P^* \rr Q),
\end{equation}
for the entropy of $M_{P(\bullet)}$ given IID $\tau$, which is used in the numerics with $\Delta P = 0.02$. In the limit $\Delta P \to 0$, the per-particle difference from maximum entropy becomes simply the relative entropy $D(P \rr Q)$ between the one-particle distributions derived from $\rho$ and from $\tau$.

In the numerics, this formula is applied for canonical~$\tau$ as in \eqref{eqn:tau-ideal-gas}, with several different $M^{(1)}$ that are discussed further below. This is always valid because the prior is a product $\tau = \otimes^n \tau_0$ of single-particle states, and thus automatically also IID in the sense above.

An important instance of this is the case $M_{P(\vec{x},\vec{p})}$, in which $M^{(1)} = M_{\vec{x},\vec{p}}$ splits the single-particle phase space into a grid of equal phase cells, in the limit of small cell widths $\Delta x, \Delta p \to 0$. Using $\tau$ from \eqref{eqn:tau-ideal-gas}, one finds $Q(\vec{x},\vec{p}) \propto \Theta(\vec{x}) \, e^{-\beta H(\vec{p})}$, abusing earlier notation slightly to allow the vector form. Meanwhile, suppose $P(\vec{x},\vec{p})$ is the actual observed distribution of particles over phase space. If also $\Delta P \to 0$, then $P^*=P$ and the entropy directly contains the term $D(P \rr Q)$. Due to vanishing cell widths, this becomes an integral $D(P \rr Q) = \int P (\log P - \log Q) d\vec{x} d \vec{p}$. But the second term of the integral is constant and independent of $P$ as it depends only on the conserved total energy. Therefore for a gas whose macrostate is described by one-particle phase distribution $P(\vec{x},\vec{p})$, one finds
\begin{equation}
    \label{eqn:empirical-boltzmann}
    S^\tau_{M_{P(\vec{x},\vec{p})}}(\rho) = C - n \int P(\vec{x},\vec{p}) \log P(\vec{x},\vec{p}) \; d\vec{x} d\vec{p},
\end{equation}
where $C$ is a constant depending only on total energy. The latter term is quickly recognized as Boltzmann's entropy from his (second) $H$-theorem (the equation between (44) and (45) of \cite{boltzmann1909further}, also (10) of~\cite{ehrenfest1912conceptual}), which was also recently studied in~\cite{chakraborti2022entropy,garrido2024time}.

This $M_{P(\vec{x},\vec{p})}$ is of general importance because, in the $\Delta x, \Delta p \to 0$ limit, it is finer than any other measurement in the $M_{P(\bullet)}$ class, and therefore has the minimal entropy of any such measurement. Note that one needs \mbox{$N \to \infty$} first in order to take such a limit~\cite{boltzmann1909further}. We remark that Sanov's theorem provides a general perspective on how Boltzmann's $\log W$ entropies ultimately take on an apparently different $- \int P \log P$ form.

Another instance of historical interest is the case $M_{P(E)}$, in which $M^{(1)} = M_{E}$ measures single-particle kinetic energies, in the limit $\Delta E \to 0$. Here one finds $Q(E) \propto E^{-\frac{d}{2}+1} e^{-\beta E}$ and, by the same steps as above,
\begin{equation}
    \label{eqn:empirical-boltzmann-energy}
    S^\tau_{M_{P(E)}}(\rho) = C' - n \int P(E) \log \frac{P(E)}{E^{-\frac{d}{2}+1}} \; dE.
\end{equation}
The latter term can be recognized as the entropy from Boltzmann's first H-theorem ((17), (17a) in~\cite{boltzmann1909further}). Note this is the distribution of \textit{single-particle} energies, and thus need not be constant in time. When calculated in OE forms, \eqref{eqn:empirical-boltzmann-energy} must be larger than~\eqref{eqn:empirical-boltzmann} because it is coarser.

It remains to comment on $M_{\vec{x}}$, $M_{\vec{v}}$, and $M_v$ that are used as $M^{(1)}$ in the numerics, corresponding to spatial, velocity, and speed measurements. Reference distributions $Q$ derive from applying measurements to the one-particle reductions $\tau_0$. 

$M_{\vec{x}}$ splits the box into spatial cells of side length $\Delta x$. This is labeled by a discrete set of vectors $\vec{x}$ at the center of each bin. There are $b = (L/\Delta x)^2$ total bins, over which the prior is uniform, so that $Q(\vec{x}) = 1/b$ is a uniform constant value. There is also a macrostate ``outside the box'' which has zero probability.

Similarly, $M_{\vec{v}}$ splits velocities into a grid of width~$\Delta \vec{v}$. The reference is a discretization of $Q(\vec{v}) \propto e^{-(v/v_0)^2} dv$, where $v_0 = \sqrt{2/\beta m}$. Meanwhile, $M_v$ splits the interval $[0,\infty)$ of possible speeds into bins of width $\Delta v$. The reference is a discretization of $Q(v) \propto v^{d-1} e^{-(v/v_0)^2} \, dv$. Discretization is by integration over bins. 

In the numerics, $\frac{\Delta x}{L}=\frac{1}{6}$, $\frac{\Delta \vec{v}}{L/T}=\frac{3 \Delta v}{L/T} = 0.3$ give the bin widths, and $\Delta P = 0.02$ the meta bin width. These measurements are illustrated in Fig.~\ref{fig:ideal-gas-extra}c.


\subsubsection{Thermodynamic coarse-grainings}
\label{app:sec:thermodynamic-coarse-grainings}

The class of measurements denoted $M_{E_A}$, $M_{E_B}$, and their combinations, describe measurements of the energy in a system, and are therefore of special importance in thermodynamics. These are part of the larger class of ``coarse or fine measurements of an observable'', which we describe first.

Consider any observable $F$, which is a real function on phase space. A coarse measurement of the observable, denoted $M_{F,\Delta F}$, consists of a discrete set of projectors onto level sets of width $\Delta F$ of the observable. Meanwhile, a fine measurement of the observable, denoted $M_{F,dF}$, consists of a continuous set of projectors onto level sets of infinitesimal width, which is the $\Delta F \to 0$ limit.

Analogous $M$ can be defined in the quantum case for any observable $Q$, a Hermitian operator. A fine measurement $M_Q$ is the collection of $Q$ eigenprojectors (if nondegenerate $M_Q$ measures in the eigenbasis). Meanwhile for a coarse measurement $M_{Q,\Delta Q}$, the projectors defining the measurement consist of sums of these eigenprojectors over eigenvalue ranges of width~$\Delta Q$.

Thermodynamics is generally associated with coarse measurements of energy. In particular, when two systems exchange heat, one is interested in the distribution of energy between the two systems. This is captured by coarse measurements $M_{H_A,\Delta H_A}$ and $M_{H_B,\Delta H_B}$ associated with the local Hamiltonians. We relabel these as $M_{E_A}$ and $M_{E_B}$ respectively. We study both the joint measurement $M_{E_A} \otimes M_{E_B}$ and the one-sided measurement $M_{E_A} \otimes \one$ of subsystem energies.

Let system $A$ consist of the first $N_A$ particles, with $H_A$ their kinetic Hamiltonian. Let $\Pi_{E_A}$ be a projector onto a microcanonical shell of energy $E_A$ and width $\Delta E_A$. And likewise for $B$, with the remaining $N_B$ particles. Let $\E_A = (E/N) N_A$ (and likewise $\E_B$) denote the equilibrium subsystems energies.

We need volumes like \mbox{$V_{E_A,E_B} = \tr (\tau \Pi_{E_A}\Pi_{E_B}) \, e^{S(\tau)}$}. These are closely related to traditional microcanonical volumes $W_{E_A} = \tr_A \Pi_{E_A} \Pi_X$, where $\Pi_X$ is a spatial projector keeping all particles inside the box. Note $\tr_A \Pi_{E_A}$ alone is infinite due to the space integral and $\Pi_{E_A} \rho = \Pi_X \Pi_{E_A} \rho $ on any state obeying the constraints. Alternately, we could have treated the spatial constraint as an infinite potential in the Hamiltonian.


We first calculate $W_{E_A}$. Let $\tau(E_A) \propto e^{-\beta(E_A) H_A}$ be the MaxEnt state [like~\eqref{eqn:tau-ideal-gas}] satisfying $\tr_A  H_A \rho = E_A$, which has temperature $\beta(E_A)^{-1} = 2 E_A/N_A d$. With the usual approximation that $E$ is about constant in the $\Delta E_A$ window~\cite{huang1987statistical}, and Stirling's approximation, integrating over the shell yields
\begin{equation}
\label{eqn:ideal-gas-WE}
    W_{E_A} \approx \frac{\Delta E_A}{\sqrt{2 \pi} \; \sigma(E_A)} \;  e^{S(\tau(E_A))} , 
\end{equation}
with $\sigma(E_A) \! = \! \sqrt{\frac{2}{N_A d}} \, E_A$ root energy variance of~$\tau(E_A)$. We choose widths $\Delta E_{A} = \sqrt{2 \pi} \, \sigma(\E_{A})$ (likewise $B$) determined by the equilibrium energies, giving prefactor $\E_A/E_A$ in \eqref{eqn:ideal-gas-WE}. Note $\Delta E/ E \leq \sqrt{4\pi/Nd}$ for both $A,B$, and $(\frac{d}{dE_A} \log W_{E_A})^{-1} \approx 2E_A/N_A d$.

We now evaluate entropy for $M = M_{E_A} \otimes M_{E_B}$. We make the lowest order approximation that $E \approx$ const in the $\Pi_E$ window, to connect to textbook $W_E$, although a better calculation using incomplete $\Gamma$-functions is possible. With this approximation, plugging in \eqref{eqn:tau-ideal-gas} to $V_{E_A,E_B} = \tr (\tau\Pi_{E_A}\Pi_{E_B}) \, e^{S(\tau)}$ yields
\begin{equation}
\label{eqn:ideal-gas-VE}
    V_{E_A, E_B} \approx W_{E_A} W_{E_B} \, e^{-\beta (E_A + E_B - E)}
\end{equation}
which is $\approx W_{E_A} W_{E_B}$ using energy conservation, since for any occupied macrostate $E_A + E_B \approx E$. With \eqref{eqn:ideal-gas-WE} then
\begin{equation}
    S^\tau_M(\rho) \approx S\big(\tau(E_A)\big) + S\big(\tau(E_B)\big) + \log \tfrac{\E_A \E_B}{E_A E_B}.
\end{equation}
(Note that the two first terms, being extensive, dominate the per particle entropy.) 
With \eqref{eqn:Stau-ideal-gas} this can be rearranged to
\begin{equation}
\label{eqn:numerical-oe-thermo-joint}
    S^\tau_{M_{E_A} \otimes M_{E_B}}(\rho) \approx S(\tau) - \log  \Big(\frac{\E_A}{E_A}\Big)^{\alpha_A} \Big(\frac{\E_B}{E_B}\Big)^{\alpha_B}
\end{equation}
where $\alpha_A = \frac{N_A d}{2}-1$ and likewise $\alpha_B$, which we use to perform the numerical calculation.

Next we calculate entropy for $M = M_{E_A} \otimes \one_B$. Proceeding as earlier one finds, with $\tau_B = \tr_A \tau $,
\begin{equation}
    V_{E_A} \approx W_{E_A} \, e^{-\beta (E_A - \E_A)} \, e^{S(\tau_B)}
\end{equation}
and eventually
\begin{equation}
\label{eqn:numerical-oe-thermo-single}
    S^\tau_{M_{E_A} \otimes \one_B}(\rho) \approx S(\tau) - \log  \Big(\frac{\E_A}{E_A}\Big)^{\alpha_A}  -\beta (E_A - \E_A)
\end{equation}
as used in the numerical calculation.

It remains to mention how the traditional OE $S_M(\rho)$ was calculated. For $M_{E_A} \otimes M_{E_B}$ acting on weakly coupled systems, the traditional OE $S_M(\rho)$ and the OE with canonical $\tau$ given by $S_M^\tau(\rho)$ are equal to first order. Here, we have calculated only up to the order where they are equal, as $S_M^\tau(\rho) \approx \log W_{E_A} W_{E_B} = S_M(\rho)$. Thus, the traditional $S_{M_{E_A} \otimes M_{E_B}}(\rho)$ also comes from \eqref{eqn:numerical-oe-thermo-joint} in the numerics. One expects at higher order they will differ by a sub-extensive term. Meanwhile, for $M_{E_A} \otimes \one_B$, the traditional OE $S_{M_{E_A} \otimes \one_B}(\rho)=\infty$ is infinite due to the integral over $B$'s phase space. Therefore, we instead calculate the related quantity $S_{M_{E_A}}(\rho_A) + S(\tau_B)$, which is the marginal in system $A$ plus a suitable constant. This quantity is equal to $\log W_{E_A} + S(\tau_B)$ which can be seen is approximately equal to \eqref{eqn:numerical-oe-thermo-single} but with the final $\beta E$ terms ignored, which is what was used in the numerics.

\subsubsection{Further remarks}

We comment on how initial states were generated.

\smallskip

\textit{IC 1:} Half the particles consist system $A$ and half~$B$. The $A$ ($B$) particle positions are chosen uniformly random on the left (right) half of the box. The $A$ and $B$ particle velocities are drawn randomly from Gaussian distributions at two different temperatures.

\textit{IC 2:} The particles are uniformly randomly distributed in a corner of the box, with uniformly random velocities. Within the first several timesteps the velocities equilibrate (becoming Gaussian random) due to an immense number of immediate collisions. System $A$ consists of 1/4 of the particles, chosen randomly.

\textit{IC 3:} The initial state was chosen to simulate several projectiles impinging on a stationary block, but was not fine tuned in any particular way. System $A$ consists of the projectiles.

\textit{IC 4:} The initial state was chosen in a somewhat arbitrary shape that simulates a rightward-moving chain of particles that collides with the wall. The distribution is $x=n/2N$, $y=x^2$, $v_x=e^{3x}$, $v_y=e^{3y}$, where $n$ is a label of each of the particles, before velocities are normalized. System $A$ is the initially leftmost 1/4 of particles.

\smallskip

Note that system $A/B$ definitions follow the particles, so \eg~even though in IC 1 system $A$ is initially in the left half, later system $A$ particles leave the left half and intermix with the $B$ particles. The system $A/B$ distinction only matters for the local energy coarse-grainings.

\section{Quantum random matrix model}
\label{app:quantum-random-matrix-model}

In this section we specify details of the model studied numerically in Fig.~\ref{fig:quantum-example}.

The model consists of systems $A,B$ with dimension $d_A=d_B=140$. The total Hamiltonian is given by
\begin{equation}
    H = H_0 \otimes \one + \one \otimes H_0 + H_{\rm int},
\end{equation}
with the first two terms identified as $H_A$ and $H_B$. Time evolution is performed by exact numerical diagonalization of the Hamiltonian. Entropies are calculated by numerical evaluation of measurement outcome probabilities for $\rho$ and $\tau$.

The local Hamiltonian is defined by $H_0 = \sum_k E_k \ketbra{k}$, where the spectrum $E_k$ is randomly sampled from a zero-mean unit-variance Gaussian, and $\ket{k}$ is by definition the eigenbasis. Positive temperatures correspond to negative average energies, as seen from the density of states, and we consider states in this regime.

The interaction $H_{\rm int}$ is a banded weak coupling interaction of the type studied in \cite{strasberg2024comparative}, with the parameters $\Delta_E \approx 5.2$, $\delta_V = 0.5$, $\lambda = \Delta_E^2 / 20 d_A \delta_V$, $T\approx 1/50 \lambda$. Note that the $x$-axis in Fig.~\ref{fig:quantum-example} is given by $\log(1+t/T)$ in terms of this timescale.

The initial state used in Fig.~\ref{fig:quantum-example} is a locally thermal-like pure state, defined by $\ket{\psi} = e^{-\beta_A H_0/2} \otimes e^{-\beta_B H_0/2} \ket{\phi}$ where $\ket{\phi}$ is a Haar-random state, $\beta_A=0.25$, $\beta_B=7.0$. Other initial states can also be considered, for which the qualitative behavior of $S_M^\tau(\rho)$ is the same. For traditional OE, whether entropy increase/decrease is observed for one-sided $M$ is initial state dependent as either Boltzmann or Shannon terms may dominate the evolution.

As explained in the text, we consider $\tau = e^{-\beta H}/Z$, and study coarse local energy measurements. The width of the energy measurements is $\Delta E_A = \Delta E_B = 0.5$.

\section{Relative entropies review}
\label{app:relative-entropies-review}

Here we briefly pedagogically review basic properties of relative entropies, since most properties of OE follow from the underlying RE structures.

\subsubsection{Classical, quantum, and measured relative entropies}

$D(\bullet \rr \bullet)$ is taken to refer to either classical relative entropy~(CRE) or quantum relative entropy (QRE), depending on its arguments. These are defined by%
\footnote{
More strictly, these are the definitions for $\supp(\rho) \subset \supp(\sigma)$ or $\supp(p) \subset \supp(q)$, denoting containment of the support. If these conditions are not met then by definition $D(\bullet \rr \bullet) = \infty$. This definition is natural as it can be derived from the convention that $\log 0 = - \infty$ but $0 \log 0 = 0$, if one expands $\log(p_x/q_x)=\log p_x - \log q_x$ in the classical case, or evaluates in a $\sigma$ eigenbasis in the quantum case. See \eg~\cite{wilde2011notes}.
}
\begin{align}
    D(p \rr q) &= \textstyle\sum_x p_x \log(p_x/q_x),\\
    D(\rho \rr \sigma) &= \tr (\rho \log \rho - \rho \log \sigma).
\end{align}
The measured relative entropy (MRE) is defined as the classical relative entropy of outcome statistics induced by a measurement. It is defined by
\begin{equation}
    D_M(\rho \rr \sigma) = D(p^\rho \rr p^\sigma)
\end{equation}
where $M$ is a POVM and $p^\rho_x = \tr M_x \rho $ is the probability distribution induced by on $\rho$ by measuring $M$ (likewise for $\sigma$). The MRE can also be written as a quantum relative entropy (for finitely indexed measurement outcomes)
\begin{equation}
\label{eqn:MRE-quantum-channel}
    D_M(\rho \rr \sigma) = D(\Phi_M(\rho) \rr \Phi_M(\sigma))
\end{equation}
where $\Phi_M(\bullet) = \tr(M_x \bullet) \ketbra{x}$ is the quantum-classical channel implementing the measurement. The fact that $D_M$ can be viewed as both a classical and quantum relative entropy  gives it a lot of useful structure.

\vspace{4pt}

The main structure of relative entropies is captured by the following properties of the CRE.

\vspace{4pt}

\begin{bbbox}
\begin{prop}[CRE properties]
    The following hold.
    \begin{enumerate}
        \item Non-negativity: 
        \begin{align}
            D(p \rr q) &\geq 0
        \end{align}
        with equality if and only if $p=q$.
        \item Joint convexity:
        \begin{align}
            D\big(\textstyle\sum_k \lambda_k p_k \rr \textstyle\sum_k \lambda_k q_k\big) \leq \textstyle\sum_k \lambda_k D(p_k \rr q_k)
        \end{align}
        for any probability distribution $\lambda_k$.
        \item Monotonicity:
        \begin{align}
            D\big(\Lambda p \rr \Lambda q \big) \leq D(p \rr q)
        \end{align}
        for any classical channel $\Lambda$. A classical channel is a stochastic matrix $\Lambda_{y|x}$ such that $\Lambda_{y|x} \geq 0$ and $\sum_y \Lambda_{y|x}=1$, which can also be viewed as a conditional probability distribution.
        \item Chain rule:
        \begin{align}
            D(p_{xy} \rr q_{xy}) = D(p_x \rr q_x) + \textstyle\sum_x p_x \, D(p_{y|x} \rr q_{y|x}),
        \end{align}
        where indices are being used to designate joint, conditional, and marginal distributions, related by $p_x = \sum_y p_{xy}$ and  $p_{xy} = p_x \, p_{y|x}$.
    \end{enumerate}
\end{prop}
\end{bbbox}

Non-negativity, convexity, and monotonicity, have direct quantum equivalents, while the chain rule does not.

\begin{bbbox}
\begin{prop}[QRE properties]
    The following hold.
    \begin{enumerate}
        \item Non-negativity: 
        \begin{align}
            D(\rho \rr \sigma) &\geq 0
        \end{align}
        with equality if and only if $\rho=\sigma$.
        \item Joint convexity:
        \begin{align}
            D\big(\textstyle\sum_k \lambda_k \rho_k \rr \textstyle\sum_k \lambda_k \sigma_k\big) \leq \textstyle\sum_k \lambda_k D(\rho_k \rr \sigma_k)
        \end{align}
        for any probability distribution $\lambda_k$.
        \item Monotonicity:
        \begin{align}
            D\big(\Phi(\rho) \rr \Phi(\sigma) \big) \leq D(\rho \rr \sigma)
        \end{align}
        for any quantum channel $\Phi$. A quantum channel is a completely positive trace-preserving map.
    \end{enumerate}
\end{prop}
\end{bbbox}

When the $M$ argument of MRE is given as a set of measurements, we define it with an implied supremum. For the set $\bbM$ consisting of all POVMs, this gives the supremum MRE over all measurements,
\begin{equation}
\label{eqn:supremum-MRE-bbM}
    D_{\bbM}(\rho \rr \sigma) \equiv \sup_M \; D_M(\rho \rr \sigma).
\end{equation}
The supremum MRE captures the amount of information extractable by the best possible measurements, and asymptotically attains the QRE.

\begin{bbbox}
\begin{prop}[MRE and QRE]
    The following hold.
\begin{enumerate}
    \item $D_\bbM(\rho \rr \sigma) \leq D(\rho \rr \sigma)$, with equality iff $[\rho,\sigma]=0$.
    \item $\lim_{n \to \infty} \frac{1}{n}  D_\bbM(\rho\xn \rr \sigma\xn) = D(\rho \rr \sigma)$.
\end{enumerate}
\end{prop}
\end{bbbox}
In the classical case there is a unique most informative measurement, which measures the value of states at each phase space point. Applying this measurement reveals that $D_\bbM(\rho \rr \sigma)=D(\rho \rr \sigma)$ for classical systems.

We thus have the following hierarchy. $D_M(\rho \rr \sigma)$ is the information obtained from a particular measurement~$M$. $D_\bbM(\rho \rr \sigma)$ is the most possible information available to any single measurement. And $D(\rho \rr \sigma)$ is the most information that can be extracted per copy even using joint measurements on many copies of a state.

We refer to standard references, such as for CRE~\cite{cover2006book}, for QRE~\cite{wilde2011notes,nielsen2010book}, and for MRE/QRE connection~\cite{hiai1991proper,vedral1997statistical,hayashi2001asymptotics}, for more detailed definitions and for proofs of all these properties.

\subsubsection{MRE properties}
\label{app:sec:mre-props}

From the CRE/QRE properties above follow many useful properties of MRE.

\smallskip

The basic bounds are as follows.

\begin{bbbox}
    \begin{prop}[Bounds] \label{thm:MRE-bounds}
        MRE is in the range
        \begin{equation}
            0 \leq D_M(\rho \rr \sigma) \leq D(\rho \rr \sigma).
        \end{equation}
        Lower equality when $M$ cannot distinguish $\rho$ from $\sigma$, namely when all $\tr(M_x \rho)=\tr(M_x \sigma)$.
    \end{prop}
    \pfline
    \begin{proof}
       CRE non-negativity, MRE bound above.
    \end{proof}
\end{bbbox}

MRE is jointly convex in the states, and convex in the measurement.

\begin{bbbox}
    \begin{prop}[Convexity] \label{thm:MRE-convexity}
    For any probility distribution $\lambda_k$ the following hold.
    
    MRE is joint convex over states,
        \begin{equation}
            D_M\big(\textstyle\sum_k \lam_k \rho_k \rr \textstyle\sum_k \lam_k  \sigma_k\big) \leq  \textstyle\sum_k \lam_k \, D_M(\rho_k \rr \sigma_k).
        \end{equation}
        Equality iff $\tr(M_x\rho_k)/\tr(M_x \sigma_k) = c_x$ are constant.
        
    MRE is convex over measurement,
    \begin{equation}
        D_{M'}(\rho \rr \sigma) \leq \textstyle\sum_k \lam_k \, D_{M^{k}} (\rho \rr \sigma),
    \end{equation}
    where $M' = \sum_k \lam_k M^{k}$ and $M^k=(M_{x|k})_x$ are a collection of POVMs over the same outcome set. Equality iff $\tr(M_{x|k}\rho)/\tr(M_{x|k} \sigma) = c_x$ are constant.
    \end{prop}
    \pfline
    \begin{proof}
        CRE joint convexity. Equality conditions by~\cite[Theorem~2.7.1]{cover2006book}.
    \end{proof}
\end{bbbox}

POVMs can also be joined by \textit{disjoint} convex combination, denoted for any $M=(M_x)_x$ and $M'=(M'_y)_y$ (with number of outcomes $m$ and $m'$) by
\begin{equation}
\label{eqn:disjoint}
    N = \lambda M \oplus (1-\lambda) M' ,
\end{equation}
with $\lambda \in [0,1]$ and $N = (N_z)_z$ defined by
\begin{equation}
    \begin{array}{rclll}
         N_x &=& \lambda M_x, & & x=1\ldots m, \\[2pt]
         N_{y+m} &=& (1-\lambda) M'_y, & \quad & y=1\ldots m'.
    \end{array}
\end{equation}
This is simply convex combination with outcome sets kept separate. MRE is linear under this operation.

 \begin{bbbox}
     \begin{prop}[Disjoint convex linearity] \label{thm:MRE-disjoint-convexity}
         MRE is linear under disjoint convex combination,
         \begin{equation}
             D_{\lambda M \oplus (1-\lambda) M'}(\rho \rr \sigma) = \lambda D_M(\rho \rr \sigma) + (1-\lambda) D_{M'}(\rho \rr \sigma).
         \end{equation}
     \end{prop}
     \pfline
     \begin{proof}
         Direct calculation, \cf~\eg~\cite[Lemma 10]{bonfill2023entropic}.
     \end{proof}
 \end{bbbox}

POVMs may be finer or coarser than one another, or unrelated. We say $N$ is coarser than $M$ if the elements are related by a stochastic map as $N_y = \sum_x \Lambda_{y|x} M_x$, which means the statistics of $N$ can be obtained from~$M$ by classical post-processing~\cite{martens1990nonideal,buscemi2005clean,guff2021resource}.  

For a stochastic map $\Lambda$ and POVM $M$, define $\Lambda M$ as the POVM obtained by postprocessing,
\begin{equation}
\label{eqn:coarser}
    (\Lambda M)_y = \sum_x \Lambda_{y|x} M_x.
\end{equation}
Recall that a stochastic map $\Lambda_{y|x}$ (\ie~classical channel) is a matrix such that $\Lambda_{y|x} \geq 0$ and $\sum_y \Lambda_{y|x}=1$. This is also a conditional probability distribution (read ``$|$'' as ``given''), \ie~a probability distribution for each $x$.

MRE is a monotone of the coarser/finer partial order: finer measurements can only reveal more information.

\begin{bbbox}
    \begin{prop}[Monotonicity] \label{thm:MRE-monotonicity}
    MRE is a monotone of stochastic postprocessing,
        \begin{equation}
            D_{\Lambda M}(\rho \rr \sigma) \leq D_M(\rho \rr \sigma)
        \end{equation}
        for all $\rho,\sigma$, where $\Lambda$ is a stochastic map.
    \end{prop}
    \pfline
    \begin{proof}
        CRE monotonicity.
    \end{proof}
\end{bbbox}
It is known that also $D_M=D_N$ for all $\rho,\sigma$ if and only if $N,M$ are post-processing equivalent (\ie~$M=\Lambda N$ and also $N = \Lambda' M$ for some $\Lambda,\Lambda'$)~\cite{bonfill2023entropic}.

We next consider sequential measurements. This requires describing measurement outcome states, so we must go beyond POVMs and describe measurements by quantum instruments~\cite{davies1970operational,wilde2011notes}. A quantum instrument is a collection $\MM = (\MM_x)_x$ of quantum operations (completely positive maps~\cite{wilde2011notes,nielsen2010book}) whose sum is trace-preserving. Probability to obtain outcome $x$ is given by
\begin{equation}
\label{eqn:inst1}
    p_x = \tr \MM_x(\rho) ,
\end{equation}
resulting in the normalized post-measurement state
\begin{equation}
\label{eqn:inst2}
    \rho_x = \MM_x(\rho)/p_x .
\end{equation}
To every instrument is associated a POVM that determines the probabilities (but not outcome states).

The most general sequential measurement scenario is one where the choice of a subsequent measurement can be conditioned on the outcome of the first, sometimes known as \textit{quantum post-processing}~\cite{leppajarvi2021postprocessing}. Suppose first one performs instrument $\MM = (\MM_x)_x$. They can then perform another instrument $\NN^x = (\NN_{y|x})_y$, chosen from a collection $\NN$ of available ones, conditional on the earlier outcome. The \textit{quantum post-processing} of instrument $\MM$ by the collection $\NN$ is denoted
\begin{equation}
\label{eqn:inst-compose}
    \NN : \MM = (\NN_{y|x} \circ \MM_x)_{xy},
\end{equation}
which is itself an instrument whose outcome set is labelled by $x,y$. (Note that the symbols $\MM$, and $\NN : \MM$, both denote instruments, even though $\NN$, which is a collection of instruments, does not). When $\NN$ is independent of $x$ this is a simple sequential measurement.

MRE is a monotone of quantum post-processing: performing additional measurements can only reveal more information. The amount is captured by a chain rule.

\begin{bbbox}
    \begin{prop}[Chain rule for sequential measurement] \label{thm:MRE-sequential}
        MRE obeys the chain rule
        \begin{equation}
        \label{eqn:mre-chain-rule}
            D_{\NN:\MM}(\rho \rr \sigma) = D_{\MM}(\rho \rr \sigma) + D_{\NN | \MM}(\rho \rr \sigma),
        \end{equation}
        with conditional MRE
        \begin{equation}
            D_{\NN | \MM}(\rho \rr \sigma) = \sum_x p_x \, D_{\NN^x}(\rho_x \rr \sigma_x)
        \end{equation}
        equal to the average MRE of the (conditional) second measurement $\NN^x$ performed on outcome states $\rho_x,\sigma_x$ of the first. The average is taken over the outcome probabilities $p_x = \tr(\MM_x(\rho))$ from the first measurement.

        In particular additional measurements are more informative,
        \begin{equation}
            D_{\NN:\MM}(\rho \rr \sigma) \geq D_{\MM}(\rho \rr \sigma),
        \end{equation}
        since $D_{\NN|\MM}(\rho \rr \sigma) \geq 0$.
    \end{prop}
    \pfline
    \begin{proof}
        CRE chain rule and non-negativity.
    \end{proof}
\end{bbbox}

We next consider the MRE \textit{recovery inequality}~\cite{buscemi2022observational}. Unlike the previous properties, this makes essential use of the quantum side of MRE captured by~\eqref{eqn:MRE-quantum-channel} above. It derives from strengthened monotonicity of quantum relative entropy~\cite{wilde2011notes,sutter2017multivariate}, and reduces to the classical strengthened monotonicity~\cite{li2014relative,wilde2011notes} in the classical case. 

\begin{bbbox}
    \begin{prop}[Recovery inequality] \label{thm:MRE-recovery}
        MRE obeys
        \begin{equation}
            D(\rho \rr \sigma) - D_M(\rho \rr \sigma) \geq D_{\bbM}(\rho \rr \rcgt),
        \end{equation}
        where $D_{\bbM}$ is the supremum MRE as in \eqref{eqn:supremum-MRE-bbM} above, 
        and $\rcgt$ is the coarse-grained state from \eqref{eqn:rcgt}.
    \end{prop}
    \pfline
    \begin{proof}
        See Proof~\ref{pf:properties}.
    \end{proof}
\end{bbbox}

Finally we mention that MRE is a continuous function of the measurement. To state this theorem requires a topology on the space of measurements, which is not trivial when comparing POVMs with different outcome sets. An appropriate topology is induced by the \textit{simulation distance} $d(M,N)$, which quantifies how well two POVMs can approximately simulate each other with classical postprocessing. For more details consult~\cite{schindler2023continuity}.

\begin{bbbox}
    \begin{prop}[Continuity] \label{thm:MRE-continuity}
        For any two fixed states $\rho,\sigma$ such that $D(\rho \rr \sigma) < \infty$, the MRE $D_M(\rho \rr \sigma)$ is a continuous function of the measurement $M$ (in the simulation distance topology defined by~\cite{schindler2023continuity}).
    \end{prop}
    \pfline
    \begin{proof}
        Corollary~15 of~\cite{schindler2023continuity}.
    \end{proof}
\end{bbbox}

\section{Proofs}
\label{app:proofs}

\begin{proofy}[of $\Imax=S(\tau)$]
\label{pf:Imax}
    The max is taken over states $\rho \in \chitau$ obeying the constraint, and over all $M$. The constraint \eqref{eqn:cross-ent-constraint} rearranges to $D(\rho \rr \tau) \leq S(\tau) - S(\rho)$. Since $D_M(\rho \rr \tau) \leq D(\rho \rr \tau)$ and $S(\rho) \geq 0$, $\Imax \leq S(\tau)$. This is saturated by the state $\ket{\psi} = \sum_k \sqrt{\tau_k} \ket{k}$, where $\tau = \sum_k \tau_k \ketbra{k}$, since measuring $M = (\ketbra{k})_k$ reveals $D_M(\psi \rr \tau) = S(\tau)$. Thus $\Imax = S(\tau)$.
    \qed
\end{proofy}

\begin{proofy}[of Eqs.~\eqref{eqn:upper-bound}, \eqref{eqn:lower-bound}]
\label{pf:bounds}
    Upper bound: $D(p \rr q) \geq 0$ with equality if and only if all $p_x = q_x$~\cite{cover2006book}. Lower bound: By RE monotonicity, one has $D_M(\rho \rr \sigma) \leq D(\rho \rr \sigma)$, with equality possible only in the commuting case~\cite{hiai1991proper,vedral1997statistical}. So $S^\tau_M(\rho) \geq S(\tau) - D(\rho \rr \tau) = S(\tau) + S(\rho) - S(\rho;\tau) \geq S(\rho)$. The final step applied the constraint \eqref{eqn:cross-ent-constraint}, and is tight only if the constraint is an equality. If $[\rho,\tau]=0$, then $M$ in mutual eigenbasis yields $S^\tau_M(\rho) = S(\tau) - D(\rho \rr \tau)$.
    \qed
\end{proofy}

\begin{proofy}[of Eqs.~\eqref{eqn:open-system-bound}]
\label{pf:open-system-bound}
    Let $M = M_S \otimes \one_E$. Then $p_x = \tr (\rho M_x \otimes \one)) = \tr \rho_S M_x$ and similarly $q_x = \tr\tau_S M_x$. Therefore, $D_{M_S \otimes \one_E}(\rho_{SE} \rr \tau_{SE})= D_{M_S}(\rho_S \rr \tau_S)$ since they both equal $D(p \rr q)$. Therefore we have $S_{M_S \otimes \one_E}^{\tau_{SE}} = S(\tau_{SE}) - D_{M_S}(\rho_S \rr \tau_S)$. But $D_{M_S}(\rho_S \rr \tau_S) \leq D(\rho_S \rr \tau_S)$, which gives the result. This last step follows from relative entropy monotonicity, with the same proof as \eqref{eqn:qre-bound-props}.
    \qed
\end{proofy}

\begin{proofy}[of Eq.~\eqref{eqn:time-average} and $\rhobar$ properties]
\label{pf:time-average}

    Assume $\rho(t) = U \rho(0) U^\dagger$ undergoes Hamiltonian unitary evolution, and that $\tau$ is stationary meaning $[\tau,H]=0$.
    
    The first point is true since cross entropy is invariant under joint unitaries, giving $S(\rho(t);\tau) = S(U \rho(0) U^\dagger ; U \tau U^\dagger)=S(\rho(0);\tau)$.

    For the second point we wish to prove that the constraint $S(\rho(t) ; \rhobar) = S(\rhobar)$ holds at all times. The proof is elementary, only using that $\rhobar$ is the time average and that $\rhobar$ is time-independent. The latter holds because a time evolution applied to $\rhobar$ can be taken inside the integral, which only shifts the bounds of integration, which are then taken to infinity. Therefore one argues as follows. Since $\rhobar$ is stationary, from the previous point, $S(\rho(t);\rhobar)$ is constant. And the time average $\overline{S(\rho(t);\rhobar)} = \overline{-\tr \rho(t) \log \rhobar} = -\tr \overline{\rho(t)} \log \rhobar) = S(\rhobar)$ by linearity. But since the value is constant, it must be equal to the time average at all times.

    The third point is proved similarly. $\overline{S(\rho(t) ; \tau)} = S(\rhobar; \tau)$ by linearity. But $S(\rho(t);\tau)$ must be constant, thus equals the time average at all times. Thus $S(\rhobar;\tau) \leq S(\tau)$ if the same holds for $\rho(0)$. Thus by the properties of the constraint set, $S(\rhobar) \leq S(\tau)$. 

    A caveat is that $\rhobar$ may not always be well defined, particularly for classical systems. In this case, one works directly with time-averaged probabilities $\overline{\tr M_x \rho(t)}$, which are all that really appear in the entropies.
    \qed
\end{proofy}

\begin{proofy}[of Eq.~\eqref{eqn:eq-bigdelta}]
\label{pf:eq-bigdelta}
    We first consider finite times. Let~$\overline{\; \bullet \;}^T$ denote a finite time average, and suppose that  $\Delta_T = \overline{D_M(\rho(t) \rr \tau)}^T = S(\tau) - \overline{S_M^\tau(\rho)}^T$. We will show $\Delta_T$ sets the scale of entropy fluctuations away from the maximum value $S(\tau)$.

    The reason the average value bounds fluctuations (without \eg~calculating a variance) is that values are one-sided: entropy values above the maximum are impossible, so large downward fluctuations far below the average must be rare. Any large fluctuation that occurs must be balanced by long times near the maximum value.

    To make this quantitative, let $\delta$ be the magnitude of an entropy fluctuation. Let $T(\delta)$ be the amount of time in the interval $-T/2 < t < T/2$ for which $D_M(\rho(t) \rr \tau) > \delta$. Then $f(\delta) = \frac{T(\delta)}{T}$ is the fraction of time for which OE is at least $\delta$ below the maximum.
    
    If the system spends time fraction $f(\delta)$ at values $D_M(\rho(t) \rr \tau) > \delta$ then clearly $\Delta_T \geq \delta f(\delta)$, which would be saturated by the best case (least fluctuations) scenario. Thus for any $\delta$ one has $f(\delta) \leq \Delta_T/\delta$. 

    This simple observation can be summarized by
    \begin{equation}
    \label{eqn:fluctuation-bound}
        \Pr \big[S(\tau) - S^\tau_M(\rho(t)) \geq \delta \big] \leq \frac{\Delta_T}{\delta}.
    \end{equation}
    This bounds the probability (defined by randomly sampling times $t \in [-T/2,T/2]$) to experience entropy fluctuations of size $\delta$ below the maximum, an instance of Markov's inequality.

    Now suppose $\Delta = \overline{D_M(\rho(t) \rr \tau)}$ and let $\Delta' > \Delta$. As we mainly care about the order of magnitude of equilibration time, let us somewhat arbitrarily set $\Delta' = 2 \Delta$. Convergence of the limit defining the infinite time average ensures there exists a time $\Teq$ such that $\Delta_T \leq \Delta'$ for all $T \geq \Teq$.
    
    We can call this $\Teq$ the equilibration time. While the value of this $\Teq$ is both arbitrary (given the freedom in choosing $\Delta'$) and unknown, the important thing here is that a finite equilibration time exists. A more physically meaningful $\Teq$ cannot be deduced without further specification of the system.
    
    On timescales $T \ll \Teq$, we cannot say something about equilibration; obviously, this is physically reasonable. However, the above reasoning shows that (with probability defined as before), for any $T \geq \Teq$
    \begin{equation}
        \Pr \big[S(\tau) - S^\tau_M(\rho(t)) \geq \delta \big] \leq \frac{2 \Delta }{\delta}.
    \end{equation}
    The prefactor $2$ could be changed to any $\alpha >1$ adjusting $\Teq(\alpha)$ accordingly. In this way the infinitely time averaged $\Delta$ sets the scale of entropy fluctuations even on finite timescales, with the penalty of a small prefactor.
    \qed
\end{proofy}

\begin{proofy}[of Eq.~\eqref{eqn:eq-eps}]
\label{pf:eq-eps}

    The question here is why \eqref{eqn:eq-eps} has the character of a ``second~law''. Here we show it implies that initially low entropies will very likely increase, that entropy will have a tendency to maximize, and that large entropy fluctuations will be rare. These statements are true relative to timescales defined by any $T \geq \Teq$ in Proof~\ref{pf:eq-bigdelta} and on entropy scales $\eps$ as in \eqref{eqn:eq-eps}.

    It was already shown in Proof~\ref{pf:eq-bigdelta} that large entropy fluctuations are rare. In particular, during $t \in (-T/2, T/2)$, the fraction of time during which $S_M^\tau(\rho(t))$ is at least $\delta$ below maximal is bounded by
    \begin{equation}
        f(\delta) \leq \frac{\alpha \, \eps }{\delta}
    \end{equation}
    where $\alpha$ is an order 1 prefactor. The majority of time is spent near maximal, viewed relative to any scale $\delta \gg \eps$. Any large fluctuations must be very brief. Meanwhile, entropy fluctuations of order $\eps$ are perfectly likely, consistent with the stochastic nature of statistical mechanics.

    If one chooses a time randomly, entropy is very likely already near maximized. However, second laws concern situations where initial entropy is low. This bound also limits the amount of time the entropy can remain low. 
    
    Suppose that a fluctuation at least $\delta$ below maximum entropy has persisted during an interval $\Delta t$. By the same logic used to obtain the bound above, the probability that, at a subsequent time, entropy remains equally as low or lower, must be  $\leq \big(\frac{\alpha \eps}{\delta} - \frac{\Delta t}{T} \big)$. Thus the fluctuation has a maximum duration $\alpha \eps T/\delta$, and the longer a fluctuation lasts, the more likely entropy is to subsequently increase. It is in this probabilistic sense that entropy increase is ensured by these theorems.

    We note the stochastic nature of these arguments: entropy decreases are not impossible, just unlikely. We also note time symmetry: given a low entropy state, entropy is most likely higher both before and after it.
    
    This type of reasoning has been used either implicitly or explicitly at least as early as von~Neumann's H-theorem~\cite{vonNeumann1929proof} (see also \cite{gogolin2016equilibration}). We have elaborated it again here for clarity and completeness.

    We remark that Fig.~\ref{fig:classical-example} empirically rules out any formulation of the second law that involves strictly non-negative entropy changes or forbids fluctuations on arbitrarily short timescales, at least without further simplifying assumptions.     
    \qed
\end{proofy}

\begin{proofy}[of Eq.~\eqref{eqn:eq-general}]
\label{pf:eq-general}
    Let $M$ induce $p_x, q_x, t_x$ on $\rho, \rhobar, \tau$. The only time dependent quantities are \mbox{$\rho = \rho(t)$} and the related $p_x = p_x(t)$, and one has \mbox{$\overline{p_x} = q_x$}. It therefore follows that
    $
    S(\tau)-\overline{S^\tau_M(\rho)} = \overline{D_M(\rho \rr \tau)} = \overline{\sum_x p_x \log \big(\frac{p_x}{q_x}\frac{q_x}{t_x} \big)}
    = \overline{D_M(\rho \rr \rhobar)} + \overline{\sum_x p_x \log \big(\frac{q_x}{t_x} \big) }
    = \overline{D_M(\rho \rr \rhobar)} + \sum_x \overline{p_x} \log \big(\frac{q_x}{t_x} \big) 
    = \overline{D_M(\rho \rr \rhobar)} + D_M(\rhobar \rr \tau).$ Even if $\rhobar$ is not well-defined, this is in fact a useful decomposition, as it only involves the induced probabilities.
    \qed
\end{proofy}

\begin{proofy}[of Eq.~\eqref{eqn:eq-rhobar}]
\label{pf:eq-rhobar}
    This theorem is an application of the powerful Cor.~1 of \cite{short2011equilibration}, which demonstrates time-averaged equilibration of a POVM's outcome probabilities in terms of trace distance. We use this result along with the continuity of Shannon entropy (\cf~\cite{winter2016tight}) to obtain the bound (see also \cite{meier2024emergence} with related methods). With notation like Proof~\ref{pf:eq-general} we find 
    $
    \overline{D_M(\rho \rr \rhobar)}
    =
    \overline{\sum_x p_x \log p_x} - \sum_x \overline{p_x} \log q_x
    =
    \overline{H(q) - H(p)}.
    $    
    But from continuity (35) of \cite{schindler2023continuity}, with $s = \frac{1}{2}\| p-q\|_1$, 
    $
    \overline{H(q) - H(p)} \leq \overline{s \log m + g(s)} \leq \overline{s} \log m + g(\overline{s})
    $.
    The last inequality used concavity of $g(s)$ \cite{schindler2023continuity}. Finally, according to (13) of \cite{short2011equilibration}, $\overline{s} \leq \eps = m/4\sqrt{d_2(\rhobar)}$ (we assumed nondegenerate energies and energy gaps). Since $g(s)$ is monotonic, this implies $\overline{D_M(\rho \rr \rhobar)} \leq \eps \log m + g(\eps)$.
    \qed
\end{proofy}

\begin{proofy}[of Eq.~\eqref{eqn:thermo}]
\label{pf:thermo}
    Suppose \eqref{eqn:eq-eps} holds with small $\eps$ for a given $M,\tau$, and denote $S(t) = S^\tau_M(\rho(t))$. Proof~\ref{pf:eq-eps} showed that given an initial state with $S(t_0) \ll S(\tau)$, entropy is very likely to subsequently increase. And that that most of the time $S(t) \approx S(\tau)$ is near maximal.

    Now suppose the prior in question is $\tau = e^{-\beta H}/Z$, reflecting energy conservation, and the measurement is \mbox{$M_{E_A} \otimes M_{E_B}$}, a coarse measurement of local energies in two subsystems, where $H = H_A + H_B + H_{\rm int}$. The energy measurements are given by collection of projectors $\Pi_E$ onto microcanonical energy shells of width $\Delta E$ centered at $E$, w.r.t.~the local Hamiltonians.

    Consider a system with a definite macrostate, which implies $S^\tau_M(\rho) = \log V_{E_A,E_B}$. The volume reduces to $V_{E_A,E_B}  =  \tr \Pi_{E_A} \! \otimes \Pi_{E_B} \, e^{-\beta(H-\E)}$, where $\E = \tr \tau H$ is the total energy. Meanwhile, $W_{E_A} = \tr \Pi_{E_A}$ and $W_{E_B} = \tr \Pi_{E_B}$ are the traditional Boltzmann volumes, and $T^{-1}_A = \frac{d}{dE_A} \log W_{E_A}$ (likewise~$T^{-1}_B$) the Boltzmann temperatures. 

    This is the same setup as in \AppClassical.a (thermodynamic CGs), which can be consulted for clarifications.

    To obtain the result we make two assumptions. First, that the system is weakly coupled ($H_{\rm int} \approx 0$), so that $V_{E_A,E_B} \approx \tr\Pi_{E_A} e^{-\beta H_A} \tr\Pi_{E_B} e^{-\beta H_B} e^{\beta \E}$, and so that $E_A + E_B = \E$ for any occupied state. And second, that energy is approximately constant in each shell, so that $\tr \Pi_{E_A} e^{-\beta H_A} \approx W_{E_A} e^{-\beta E_A}$ (and likewise $B$). Together this gives $V_{E_A,E_B} \approx W_{E_A} W_{E_B}$. Such assumptions are common in textbook calculations~\cite{huang1987statistical}.

    With these assumptions $S^\tau_M(\rho) = S(E)$ is only a function of the energy $E_A = E$ (since $E_B = \E-E$), and 
    \begin{equation}
        S(E) = \log W_{E_A} + \log {W_{E_B}}.
    \end{equation}
    As discussed above, \eqref{eqn:eq-eps} implies the entropy is maximal at most times. Approximating the entropy as a smooth function, maximization implies $\frac{dS}{dE} = 0$. But
    \begin{equation}
        \frac{dS}{dE} = T_A^{-1} - T_B^{-1},
    \end{equation}
    so the Boltzmann temperatures are equal at most times. Integrating $\frac{dS}{dt} = \frac{dS}{dE} \frac{dE}{dt}$ over interval $\Delta t$ gives $\Delta S$ of~\eqref{eqn:thermo}. Then \eqref{eqn:eq-eps} implies that $\Delta S >0$ with high probability, if the initial entropy was low.

    If one considers instead $M_{E_A} \otimes \one_B$, then under the same assumptions one finds $S(E) = \log W_{E_A} - \beta (E - \E_A) + S(\tau_B)$, where $\E_A = \tr \tau H_A$. Thus the analogous calculation shows
    \begin{equation}
        \frac{dS}{dE} = T_A^{-1} - \beta,
    \end{equation}
    giving equilibrium condition $T_A = \beta^{-1}$. This $\beta$ is the one appearing in the prior $\tau \propto e^{-\beta H}$ (determined by total average energy), a global Gibbs temperature.
    \qed
\end{proofy}

\begin{proofy}[of Eq.~\eqref{eqn:generalized-thermo}]
\label{pf:generalized-thermo}
     We continue to assume that $\tau$ is fixed, with $q_x$ and $S(\tau)$ time-independent. Since $\sum_x p_x = 1$ it follows that  $\sum_x \frac{d p_x}{dt}  = 0$, and so $\frac{d}{dt} \sum_x p_x \log p_x = \sum_x \frac{d p_x}{dt} \log p_x$. One then evaluates the derivative of $S_M^\tau(\rho)$ finding 
    $\frac{d}{dt} S_M^{\tau}(\rho) 
    = - \frac{d}{dt} D_M(\rho \rr \tau)
    = \frac{d}{dt} \sum_x p_x (\log q_x - \log p_x)
    =  \sum_x \frac{d p_x}{dt} (\log q_x - \log p_x)
    $. Meanwhile, as already established, if \eqref{eqn:eq-eps} holds then $\frac{d}{dt} S_M^{\tau}(\rho) > 0$ with high probability, for low entropy initial states.
    \qed
\end{proofy}

\begin{proofy}[of Eq.~\eqref{eqn:rate-matrix}]
\label{pf:rate-matrix}
     Suppose $dp_x/dt = \sum_x R_{xx'} p_x'$. From $\sum_x p_x(t) = 1$ it follows that $\sum_x R_{xx'} = 0$ for all $x'$. Meanwhile since $p_x(t) \geq 0$,  $x \neq x'$ implies $R_{xx'} \geq 0$. Now consider $dS_M^\tau/dt = \sum_x (dp_x/dt) \log (q_x/p_x)$. Using the local detailed balance condition $q_x = R_{xx'}q_{x'}/R_{x'x}$ one finds $dS_M^\tau/dt = \sum_{x,x'} R_{xx'} p_{x'} \log (R_{xx'}q_{x'}/R_{x'x}p_x)$. Separating out the $\log q_{x'}$ term and using the normalization condition on $R_{xx'}$, one finds $\sum_{x,x'} R_{xx'} p_{x'} \log q_{x'} = \sum_{x,x'} R_{xx'} p_{x'} \log p_{x'}=0$, which allows one to rewrite $dS_M^\tau/dt = \sum_{x,x'} R_{xx'} p_{x'} \log (R_{xx'}p_{x'}/R_{x'x}p_x)$, which is the desired form. It remains to show non-negativity. To see this first observe that $\log (R_{xx'}p_{x'}/R_{x'x}p_x)$ is antisymmetric under $ x' \leftrightarrow x$. Therefore one can find $dS_M^\tau/dt = \frac{1}{2} \sum_{x,x'} (R_{xx'} p_{x'} - R_{x'x} p_{x}) \log (R_{xx'}p_{x'}/R_{x'x}p_x)$. But because $\log$ is monotonic, it follows that $(\log(x) - \log(y))$ and $(x-y)$ always have the same sign, and thus that $(x-y) \log (x/y) \geq 0$ for all $x,y$. Therefore each individual term in the sum is $\geq 0$, whence $dS_M^\tau / dt \geq 0$.
    \qed
\end{proofy}

\setcounter{propnum}{1}

\begin{proofy}[Sec.~\ref{sec:mathematical-properties}]
\label{pf:properties}

    For proofs that follow directly from the basic properties of REs (see \AppRE), we merely refer to the relevant property.

    \smallskip
    (\tpn) Proof~\ref{pf:bounds}.

    \smallskip
    (\tpn) RE joint convexity, concavity of $S(\rho)$. Bounded concavity and disjoint linearity follow as in \cite{schindler2023continuity,bonfill2023entropic}.
    
    \smallskip
    (\tpn) Classical RE monotonicity.
    
    \smallskip
    (\tpn) Classical RE chain rule.

    \smallskip
    (\tpn) The identity $D_M(\rho \rr \sigma) \leq D(\rho \rr \sigma)$ follows from RE monotonicity~\cite{hiai1991proper,vedral1997statistical}.

    \smallskip
    (\tpn) Let $\rho = \sum_k \lambda_k \rho_k$ and $P_{xk} = \lambda_k \tr(M_x \rho_k)$. The marginals are $P_x = \sum_k P_{xk} = \tr M_x \rho = p_x$ and $P_k = \sum_x P_{xk} = \lambda_k$.
    Then a straightforward expansion shows that  $S_M^\tau(\rho)
    = -\sum_{x} p_x \log \left(\frac{p_x}{V_x} \right)
    = -\sum_{xk} P_{xk} \log \left( \frac{p_x}{V_x} \frac{\lambda_k \tr(M_x \rho_k)}{\lambda_k \tr(M_x \rho_k)} \right)
    = -\sum_{xk} P_{xk} \log \left( \frac{\tr(M_x \rho_k)}{V_x} \frac{ p_x \lambda_k }{P_{xk})} \right).$
    Then one finds
    $S_M^\tau(\rho)
    = -\sum_{k} \lambda_k \sum_x \tr(M_x \rho_k) \log \left( \frac{\tr(M_x \rho_k)}{V_x}\right) + \sum_{xk} P_{xk} \log \left( \frac{ P_{xk} }{p_x \lambda_k)} \right)
    = \sum_{k} \lambda_k  S_M^\tau(\rho_k) + I(M\! : \! \mathcal{\E}) $. Meanwhile, Holevo's bound shows that $I(M\! : \! \mathcal{\E}) \leq S(\rho) - \sum_k \lambda_k S(\rho_k)$~\cite{wilde2011notes}.

    \smallskip
    (\tpn) This is a continuity bound for the case of a finite number of nonzero $V_x$. It assumes $\rho,\,\sigma$ each obey the constraint, $S(\sigma;\tau), \, S(\rho;\tau) \leq S(\tau) < \infty$. This ensures that $V_x = 0$ implies $\tr\rho M_x = \tr\sigma M_x = 0$, and that all $V_x<\infty$. Thus, only macrostates with \mbox{$0 < V_x < \infty$} contribute to the OE ($0 \log 0 = 0$), the rest can be discarded. Then there are effectively only $A$ outcomes of the measurement, each with $|\log V_x| \leq B < \infty$. The remainder of the proof is exactly as (5) of \cite{schindler2023continuity}.

    A more general result without the finite outcome restriction is expected but still under investigation. One should compare the broader realm of linearly constrained (energy-bounded) continuity relations~\cite{winter2016tight}.

    \smallskip
    (\tpn) See~\cite{schindler2023continuity}.

    \smallskip
    (vi,vii) We consider Petz recovery maps for the measurement channel $\Phi_M(\bullet) = \sum_x \tr(\bullet M_x) \ketbra{x}$~\cite{buscemi2022observational}, with prior $\tau$. This provides a simplification from the general case because the outputs of $\Phi_M$ are all diagonal in the same basis and mutually commute (a quantum-classical, or QC, map). 
    
    Coarse-grained states derived from the rotated Petz maps are defined by $\rcg^s = \mathcal{R}^{[s]}_{\tau, \Phi_M} \circ \Phi_M(\rho)$, with maps defined as in (61) of~\cite{sutter2017multivariate} (see also~\cite{wilde2011notes} for review). The trace dual of the measurement map evaluates to $\Phi_M^\dagger(\ketbra{x})=M_x$. Using this with fact that all outputs of the channel are diagonal in the $\ketbra{x}$ basis leads to $\rcg^s = \sum_x p_x \frac{\tau^{(1+is)/2} M_x \tau^{(1-is)/2}}{\tr\tau M_x}$. The standard Petz recovery $\rcg$ is the case $s=0$, and the smeared $\rcgt$ is defined by the integral given in the text.

    If $\tau$ commutes with all $M_x$ then clearly all $\rcg^s = \rcg$ as the imaginary exponent terms cancel out. Similarly, if $\tau = \Pi/W$ with $\Pi$ a projector, then $\tau^\alpha = \tau / W^\alpha$, and the imaginary exponents again cancel, giving $\rcg^s = \rcg$. In the case $\tau \propto \one$ one obtains $\rcg^s = \sum_x p_x M_x / \tr M_x$.
    
    We have used the quantum channel language here. Everything works equivalently in the classical case, in which case everything commutes and reduces to the classical recovery theory of \cite{li2018squashed,wilde2011notes}.

    \smallskip
    (\tpn) Let $\tilde{\tau}_M$ be the MaxEnt state consistent with measurement outcomes $\tr M_x \rho = p_x$. This is given by $\tilde{\tau}_M \propto e^{-\sum_x \lam_x M_x}$ for some appropriate $\lam_x$, assuming such a state exists that obeys the constraint. Then $S(\rho ; \tilde{\tau}_M) = S(\tilde{\tau}_M)$ defines a set containing all states with the given outcome probabilities. 

    In the case of a projective $M$, the POVM elements $M_x \to \Pi_x$ are a complete set of orthogonal projectors. Then one can write $\tilde{\tau}_M$ as the exponential of a diagonal matrix, leading to $\tilde{\tau}_M \propto \sum_x e^{-\lam_x} \Pi_x $. Evaluating outcome probabilities and comparing to the constraint establishes $\tilde{\tau}_M = \sum_x p_x \Pi_x / \tr \Pi_x $ for projective case.
    
    \smallskip
    (\tpn) For projective $M = (\Pi_x)_x$ with $\tau \propto \one$, reduction of all coarse-grained states to $\rcg = \sum_x p_x \Pi_x/\tr \Pi_x$ was already shown above. This state has eigenvalues $p_x/W_x$ with multiplicity $W_x$, so that $S(\rcg) = -\sum_x W_x (p_x/W_x) \log (p_x/W_x) = S_M(\rho)$.

    \smallskip
    (\tpn) The form for $\tilde{\tau}_M$ and $\tilde{\tau}_{M,X}$ follows immediately from the maximum entropy principle as stated in Sec.~\ref{sec:informational}, and can be checked by computing the cross entropy condition $S(\rho;\tau) \leq S(\tau)$ for states in the constrained set.

    \smallskip
    (\tpn) Follows from direct evaluation for the $\rcg$ states. That it is also $\tilde{\tau}_M$ can be checked from the cross entropy condition.
    
    \smallskip
    (\tpn) Each inequality follows from RE monotonicity under quantum channels (or the classical equivalent). The relevant channels are first the measurement map~$\Phi_M$, then the Petz recovery map, then $\Phi_M$ again. 

    In the case $\tau \propto \one$ with projective $M$, these four REs correspond to the four entropies in the later chain. In that case $\rho$ and $\rcg$ induce the same $p_x$ (see above), and so $S_M(\rcg) = S_M(\rho)$, which also equals $S(\rcg)$ as above.

    \smallskip
    (\tpn) The recovery bound is an application of strong RE monotonicity, which we use in the form of Sutter-Berta-Tomamichel (Cor.~4.2)~\cite{sutter2017multivariate}. The theorem is applied to the channel $\Phi_M(\bullet) = \sum_x \tr(\bullet M_x) \ketbra{x}$, which is the QC channel encoding the measurement~\cite{buscemi2022observational}. As discussed above, because it is QC, the outputs of $\Phi_M(\bullet)$ are all diagonal in the same basis and mutually commute. And the trace dual of the measurement map evaluates to $\Phi_M^\dagger(\ketbra{x})=M_x$. These properties simplify \cite{sutter2017multivariate} from the general case, giving
    \begin{equation}
        D(\rho \rr \sigma) - D_M(\rho \rr \sigma) \geq \sup_M D_{M}(\rho \rr \rcgt),
    \end{equation}
    with $\rcgt$ as earlier. In the classical case the $\sup$ saturates $D(\rho \rr \rcg)$ recovering \mbox{(17) of \cite{li2018squashed}}. For the conventional OE an additional simplification was possible because $\sigma \propto \one$ commutes with everything~\cite{buscemi2022observational}. Finally, to connect to the stated form we note that by virtue of the constraint~\eqref{eqn:cross-ent-constraint}, one has $S^\tau_M(\rho) - S(\rho) \geq D(\rho \rr \tau) - D_M(\rho \rr \tau)$.

    \smallskip
    (\tpn) $\sup_M D_M(\rho \rr \sigma) = 0$ if and only if $\rho = \sigma$, because the supremum contains informationally complete measurements~\cite{piani2009relative}.  For an informationally complete $M$ to yield exactly the same outcomes, $\rho,\sigma$ must have the same coefficients in a matrix basis.

    \smallskip
    (\tpn) Let $\mathcal{S}(\rho) = \inf_{M} S_M^{\tau}(\rho)$ where the infimum is over all POVMs. Let $\Phi$ be a channel such that $\Phi(\tau)=\tau$. Then $\mathcal{S}\big(\Phi(\rho)\big) = S(\tau) - \sup_M D_M(\Phi(\rho) \rr \tau) = S(\tau) - \sup_M D_M(\Phi(\rho) \rr \Phi(\tau) )$ using the $\tau$-preserving property. But $\sup_M D_M(\Phi(\rho) \rr \Phi(\tau) ) = \sup_M D_{\Phi^{\dagger}M}(\rho \rr \tau ) = \sup_{N=\Phi^{\dagger}M} D_{N}(\rho \rr \tau ) \geq \sup_{M} D_{M}(\rho \rr \tau )$. Here $\Phi^{\dagger}M$ is the POVM obtained by applying the trace dual of $\Phi$ to each POVM element, and the inequality is due to the infimum being taken over a smaller set of POVMs. Combining the above, $\mathcal{S}\big(\Phi(\rho)\big) \leq \mathcal{S}\big(\rho\big)$. Note that measured RE does not have monotonicity under quantum channels. The proof comes not from RE monotonicity, but from absorbing the channel into the measurement set.

    \smallskip
    (\tpn) As an example let $\Omega$ be the set of separable POVMs (POVM elements are convex sums of tensor products of positive operators), and let  $\mathcal{S}(\rho) = \inf_{M \in \Omega} S_M^{\tau}(\rho)$. Let $\Phi$ be a separable (admits product Kraus operators) channel. If $M \in \Omega$ is separable then $\Phi^{\dagger} M \in \Omega$ is also separable, as seen by calculating in terms of Kraus operators. If also $\Phi(\tau) = \tau$ then by the same type of argument as above, $\mathcal{S}\big(\Phi(\rho)\big) \leq \mathcal{S}\big(\rho\big)$. Thus $\mathcal{S}$ is a monotone of separable $\tau$-preserving maps.
    \qed
\end{proofy}

\clearpage

\bookmark[dest=table2,level=1]{Table II}

\begin{table*}[h]
    \centerline{ 
    \begin{tabular}{P{50mm}L{5mm}L{140mm}} 
         \toprule \addlinespace[2pt]
        {\hypertarget{table2}{}}
         \textsc{Quantity} & &
         \centerline{$S_M^\tau(\rho)$ \textsc{Interpretation}} 
    \\[-4pt] \midrule \addlinespace[10pt]
    Jaynes' max entropy $S(\tau)$~\cite{jaynes1957informationI,jaynes1957informationII} & &
    Trivial or uninformative $M$. Upper bound $S_M^\tau(\rho) \leq S(\tau)$. Equilibrium entropy. See \eqref{eqn:upper-bound}.
    \\[6pt]
    von Neumann entropy $S(\rho)$ & &
    $S(\rho)$ quantifies information in the state $\rho$, while $S_M^\tau(\rho)$ quantifies info extractable by observations. Bound $S_M^\tau(\rho) \geq S(\rho)$ shows observations extract no more info than available in the state. See \eqref{eqn:lower-bound}.
    \\[6pt]
    Traditional OE $S_M(\rho)$~\cite{safranek2021brief,strasberg2020first} & &
    The case of no constraint, corresponding to trivial prior $\tau \propto \one$.
    \\[6pt]
    Boltzmann entropy~\cite{boltzmann1909further,goldstein2020gibbs} & &
    The case when the system in a definite macrostate (when only one $p_x$ is nonzero).
    \\[6pt]
    Shannon entropy of measurement~\cite{meier2024emergence} & &
    The case (up to a constant) when macrostate volumes $V_x$ are all equal.
    \\[16pt]
    diagonal entropy~\cite{polkovnikov2011microscopic,giraud2016average,oliveira2024thermodynamic} & &
    The case $M = (\ketbra{E})_E$ of a measurement in the energy basis. Also $S(\rhobar)$, the entropy of the tightest stationary prior as in \eqref{eqn:time-average}, assuming nondegenerate energies and energy gaps.
    \\[6pt]
    canonical entropy~\cite{matty2017comparison} & &
    The case of trivial $M$ and a time-dependent average energy constraint, giving $S(\tau(t))$.
    \\[6pt]
    entanglement entropy~\cite{schindler2020correlation,eisert2010area} & &
    Minimum OE for local measurements: $S_{\rm ent}(\ket{\psi_{AB}}) = \inf\limits_{ M_A, M_B} S_{M_A \otimes M_B}(\ket{\psi_{AB}})$. See~\cite{schindler2020correlation,rossetti2025observational}.
    \\[6pt]
    stochastic thermodynamics entropy~\cite{strasberg2022book,seifert2018stochastic} & &
    The case of a weakly coupled system/bath with global $\tau = \Pi_E/W_E$ (microcanonical energy constraint) and $M = M_S \otimes \one_B$ with projective measurement $M_S$ on the system.
    \\[6pt]
    surface/volume entropies~\cite{dunkel2014consistent,frenkel2015gibbs} & &
    The cases  $\tau \propto \Pi_{E \approx E_0}$ and $\tau \propto \Pi_{E \leq E_0}$, respectively.
    \\[6pt]
    Wehrl entropy~\cite{wehrl1978general,lieb1978proof} & &
    The case $M = \big(\frac{\ketbra{z}}{\pi}\big)_z$ (a POVM) where $\ket{z}$ are the over-complete basis of coherent states.
    \\[6pt]
    generalized OE~\cite{bai2023observational} & &
    Equivalent to $S(\rho;\tau) - D_M(\rho \rr \tau)$, which equals $S^\tau_M(\rho)$ for equality constraints.
    \\[6pt]
    Helmholtz free energy~\cite{huang1987statistical} & &
    The case of $M=M_S \otimes \one_B$ with $\tau = e^{-\beta H_{SB}}/Z$ for a system weakly coupled to a bath, where one finds total entropy is a constant minus the free energy.
    \\[6pt]
    R{\'e}nyi/Tsallis/etc entropies & &
    The natural R{\'e}nyi extension is given in the text. Similarly, other alternatives to the standard algebraic form can be incorporated by modifying $D_M(\rho \rr \tau)$ to an appropriate divergence.
    \\[6pt]
    HEP coarse- and fine- \linebreak grained entropies~\cite{polchinski2017blackhole,almheiri2020entropy} & &
    $S(\tau)$ and $S(\rho)$, respectively.
    \\[16pt]
    thermodynamic entropies & &
    Related to energy measurements/constraints, precise meaning depends on context. See text.
    \\[6pt]
    entropy production~\cite{potts2019introduction,varizi2024entropy} & &
    Can often be rewritten as $\Delta S_M^{\tau} = S_{M_t}^{\tau_t}\big(\rho(t)\big) - S_{M_0}^{\tau_0}\big(\rho(0)\big)$. See text.
    \\[6pt]    
    entropy production as \linebreak in (26--40) of~\cite{potts2019introduction} & &
    The prior is $\tau(t)= \one_S/d_s\otimes \tau_B(t)$ where $\tau_B(t)=e^{-\beta(t) H_B}/Z_B$ is a canonical state determined by the bath energy, and $M(t)=M_S(t) \otimes \one_B$ where $M_S(t)$ is an optimal measurement on the system.
    \\[6pt]
    entropy from Boltzmann's \linebreak
    H-theorems~\cite{boltzmann1909further} & &
    The case of $\tau \propto e^{-\beta H}$ and either $M_{P(E)}$ or $M_{P(\vec{x},\vec{p})}$ as in (\ref{eqn:empirical-boltzmann}--\ref{eqn:empirical-boltzmann-energy}). See text.
    \\[16pt]
    entropy from Gibbs' \linebreak H-theorem~\cite{gibbs1902book,ehrenfest1912conceptual} & &
    The case where $M$ divides phase space in a grid of equal cells and $\tau \propto \one$. See text.
    \\[16pt]
    entropy from von Neumann's \linebreak H-theorem~\cite{vonNeumann1929proof} & &
    The case where $\tau$ is a coarse version of the time-averaged state and $M$ is his ``quantum phase cell'' measurement. Interpreted this way his theorem becomes an instance of \eqref{eqn:eq-eps}. See text.
    \\[6pt]
    micro-/grand-/canonical entropies & &
    See Table \ref{tab:constraints-and-priors}.
    \\[6pt]
    Sackur-Tetrode \linebreak ideal gas entropy~\cite{huang1987statistical} & &
    Equilibrium entropy $S(\tau)$ in Figs.~\ref{fig:classical-example}--\ref{fig:ideal-gas-extra}. See \hyperref[fig:ideal-gas-1]{App.~A}.
    \\ \addlinespace[6pt]
    \bottomrule
    \end{tabular}
    }
    \caption{Observational entropy (OE) interpretation of other entropic quantities.}
    \label{tab:special-cases-and-limits}
\end{table*}

\end{document}